\documentclass[pra,twocolumn,preprintnumbers,amsmath,amssymb,showpacs,
nofootinbib,floatfix]{revtex4}

\usepackage{graphicx,bm}

\makeatletter
\def\graphicscale{\twocolumn@sw{0.33}{0.4}}
\def\graphicthreescale{\twocolumn@sw{0.33}{0.4}}
\makeatother

\renewcommand\Re{\mathop{\rm Re}\nolimits}

\renewcommand\mod{\mathop{\rm mod}\nolimits}

\newcommand\txf[2]{{\textstyle{#1\over#2}}}
\newcommand\half{\txf12}

\begin{document}

\title{Quantum critical behavior and trap-size scaling of trapped bosons 
   in a one-dimensional optical lattice}


 
 \author{Massimo Campostrini and Ettore Vicari} \affiliation{Dipartimento di
   Fisica dell'Universit\`a di Pisa and I.N.F.N., Sezione di Pisa, Largo Bruno
   Pontecorvo 2, I-56127 Pisa, Italy} 
\date{March 17, 2010}

\begin{abstract}
  
  We study the quantum (zero-temperature) critical behaviors of confined
  particle systems described by the one-dimensional (1D) Bose-Hubbard model in
  the presence of a confining potential, at the Mott insulator to superfluid
  transitions, and within the gapless superfluid phase.  Specifically, we
  consider the hard-core limit of the model, which allows us to study the
  effects of the confining potential by exact and very accurate numerical
  results.  We analyze the quantum critical behaviors in the large trap-size
  limit within the framework of the trap-size scaling (TSS) theory, which
  introduces a new trap exponent $\theta$ to describe the dependence on the
  trap size. This study is relevant for experiments of confined quasi 1D cold
  atom systems in optical lattices.
  
  At the low-density Mott transition TSS can be shown analytically within the
  spinless fermion representation of the hard-core limit. The trap-size
  dependence turns out to be more subtle in the other critical regions, when
  the corresponding homogeneous system has a nonzero filling $f$, showing an
  infinite number of level crossings of the lowest states when increasing the
  trap size.  At the $n=1$ Mott transition this gives rise to a modulated TSS:
  the TSS is still controlled by the trap-size exponent $\theta$, but it gets
  modulated by periodic functions of the trap size. Modulations of the
  asymptotic power-law behavior is also found in the gapless superfluid
  region, with additional multiscaling behaviors.

\end{abstract}


\maketitle

\section{Introduction}
\label{intro}

The impressive progress in the experimental manipulation of cold atoms in
optical lattices (see, e.g., Ref.~\cite{BDZ-08} and references therein) have
provided a great opportunity to investigate the interplay between quantum and
statistical behaviors in particle systems. Cold atoms in optical lattices can
be used to study many-body phenomena in dilute gases, such as quantum
Mott-Hubbard transitions for bosonic atoms, see, e.g.,
Refs.~\cite{GBMHS-02,SMSKE-04,PWMMFCSHB-04,KWW-05,HSBBD-06,FWMGB-06,
SPP-07,CFFFI-09}.  An important feature of these experiments is the presence
of a confining potential which traps the particles within a limited spatial
region of the optical lattice created by laser-induced standing waves.  The
theoretical framework~\cite{JBCGZ-98} is based on the Bose-Hubbard (BH)
model~\cite{FWGF-89} in the presence of a confining potential coupled to the
particle density, i.e.,
\begin{eqnarray}
H_{\rm BH} &=& -{J\over 2}
\sum_{\langle ij\rangle}(b_i^\dagger b_j+b_j^\dagger b_i)
+ {U\over2} \sum_i n_i(n_i-1)
\nonumber \\&+&
\mu \sum_i n_i + \sum_i V(r_i) n_i ,
\label{bhm}
\end{eqnarray}
where $\langle ij\rangle$ is the set of nearest-neighbor sites,
$n_i\equiv b_i^\dagger b_i$ is the particle density operator.

We consider a power-law trapping potential
\begin{equation}
V(r) = v^p r^p \equiv (r/l)^p,
\label{potential}
\end{equation}
where $r\equiv |\vec{x}|$, $v$ and $p$ are positive constants and $l\equiv
1/v$ is the trap size.  Experiments are usually set up with a harmonic
potential, i.e., $p=2$.  Far from the origin the potential $V(r)$ diverges,
therefore $\langle n_i\rangle$ vanishes and the particles are trapped.  The
inhomogeneity due to the trapping potential strongly affects the phenomenology
of quantum transitions in homogeneous systems.

The homogeneous BH model without trap undergoes Mott insulator to superfluid
quantum transitions driven by the chemical potential $\mu$, whose low-energy
properties are described by a {\em nonrelativistic\/} U(1)-symmetric bosonic
field theory~\cite{FWGF-89}, which is characterized by the dynamic exponent
$z=2$.\footnote{The special transitions at fixed integer density
belong to a different universality class~\cite{FWGF-89}, described by
a {\em relativistic} U(1)-symmetric bosonic field theory, which is the
$(d+1)$-dimensional XY universality class~\cite{PV-02}. Thus its
dynamic exponent is $z=1$.}
In the presence of a confining potential, theoretical
and experimental results have shown the coexistence of Mott insulator and
superfluid regions when varying the total occupancy of the lattice, see, e.g.,
Refs.~\cite{FWMGB-06,JBCGZ-98,BRSRMDT-02,KPS-02,KSDZ-04,PRHD-04,WATB-04,
RM-04,DLVW-05,GKTWB-06,ULR-06,RBRS-09}.
However, at fixed trap size, the system does not develop a critical behavior
with a diverging length scale~\cite{BRSRMDT-02,WATB-04}.

Criticality can be recovered only in the limit of large trap size.  As put
forward in Refs.~\cite{CV-10,CV-09}, this critical regime can be described in
the framework of the trap-size scaling (TSS) theory, where the critical
behavior is cast in the form a TSS with a nontrivial trap exponent $\theta$,
which determines how the length scale of the critical modes at the critical
point diverges with increasing trap size, i.e., $\xi\sim l^{\theta}$.  For
example, let us consider a standard scenario (see, e.g.,
Ref.~\cite{Sachdev-book}), in which the quantum $T=0$ transition of the
the homogeneous $d$-dimensional system has one relevant parameter $\mu$, with
critical value $\mu_c$. The simplest TSS Ansatz~\cite{CV-10} for the
asymptotic behavior of the free-energy density in the presence of a confining
potential (\ref{potential}) is
\begin{equation}
F(\mu,T,l,x) = l^{-\theta (d+z)} 
{\cal F}(\bar{\mu} l^{\theta/\nu},Tl^{\theta z},xl^{-\theta}),
\label{freee}
\end{equation}
where $x$ is the distance from the middle of the trap, $\bar{\mu}\equiv
\mu-\mu_c$, $z$ is the dynamic exponent and $\nu\equiv 1/y_\mu$ where $y_\mu$
is the renormalization-group (RG) dimension of $\mu$.  Moreover, any
low-energy scale at $T=0$, and specifically the gap, is expected to behave as
\begin{equation}
\Delta = l^{-\theta z} {\cal D}(\bar{\mu} l^{\theta/\nu}).
\label{gapscalgen}
\end{equation}
The above TSS has been verified by analytical and accurate numerical
calculations~\cite{CV-10} within the quantum XY chain in a space-dependent
transverse external field, which acts as a trap for the spinless fermions of
its quadratic Hamiltonian representation, which can be obtained by a 
Jordan-Wigner transformation.

The general features of the TSS at the Mott to superfluid transitions in
$d$-dimensional BH models have been discussed in Ref.~\cite{CV-10}.  Beside
the dynamic critical exponent $z$ and the RG dimension $y_\mu$ of $\mu$, which
control the critical behavior of the homogeneous system and can be determined
from the corresponding continuum theory~\cite{FWGF-89,Sachdev-book}, the TSS
requires the trap exponent $\theta$, which can be derived by a RG analysis of
the corresponding perturbation~\cite{CV-10}.  For one- and two-dimensional
systems, the critical exponents entering the scaling formulas (\ref{freee})
and (\ref{gapscalgen}) are
\begin{equation}
z=2,\qquad y_\mu\equiv 1/\nu=2,\qquad \theta= p/(p+2),
\label{critexps}
\end{equation}
where $p$ is the power of the confining potential (\ref{potential}).

In this paper we investigate the quantum critical behaviors of the
one-dimensional (1D) BH model at zero temperature in the presence of a
confining potential, at the Mott to superfluid transitions, and within the
gapless superfluid phase.  The 1D BH model in the presence of a confining
potential is of experimental relevance in optical lattices, where quasi
1D confined particle systems have been realized, see, e.g.,
Refs.~\cite{BDZ-08,PWMMFCSHB-04,KWW-05,CFFFI-09}.

Specifically, we consider the hard-core limit, $U\to\infty$, of the 1D BH
model, which allows us to study the effects of the confining potential 
by exact and very accurate numerical results. The hard-core
limit implies that the particle number $n_i$ per site is restricted to the
values $n_i=0,1$.  In this limit the model can be mapped into the XX
chain model with a space-dependent transverse external field,
\begin{eqnarray}
H_{\rm XX} &=& - J \sum_i \left( S^x_i S^x_{i+1} + S^y_i S^y_{i+1} \right)
\nonumber \\&-&
\sum_i [\mu+V(x_i)] S^z_i,
\label{XX}
\end{eqnarray}
where $S^a_i=\sigma^a_i/2$ and $\sigma^a$ are the Pauli matrices, which are
related to the boson operators $b_i$ by $\sigma^x_i = b_i^\dagger + b_i$,
$\sigma^y_i = i(b_i^\dagger - b_i)$, $\sigma^z_i = 1-2b_i^\dagger b_i$.  In
the following we fix $J=1$. Then, by a Jordan-Wigner transformation, one can
further map it into a model of spinless fermions, see, e.g.,
Ref.~\cite{Sachdev-book}.

In the absence of the trap, the 1D hard-core BH model has three phases: two
Mott insulator phases, for $\mu>1$ with $\langle n_i\rangle=0$ and for
$\mu<-1$ with $\langle n_i\rangle=1$, separated by a gapless superfluid phase
for $|\mu|<1$.  Therefore, there are two Mott insulator to superfluid
transitions at $\mu=1$ and $\mu=-1$. At both transitions the exponents
controlling the critical behavior are those reported in Eq.\
(\ref{critexps}). The gapless superfluid phase is instead described by a free
massless bosonic field theory with dynamic exponent $z=1$, see, e.g.,
Ref.~\cite{Tsvelik-book}.

The effects of the confining potential at the low-density transition ($\mu=1$)
has been already studied in Ref.~\cite{CV-10}, where TSS has been shown to
emerge by analytical calculations exploiting the spinless fermion formulation
of the 1D hard-core BH model.

In this paper we extend the study of the trap-size dependence to all critical
regions of its phase diagram. We present results for the trap-size dependence
in the gapless superfluid region and at the $n=1$ Mott insulator to superfluid
transition.

For this purpose we exploit the free spinless fermion representation of the 1D
hard-core BH model, which allows us to perform computations for very large
systems, since they only require the diagonalization of a $L\times L$ matrix
where $L$ is the number of lattice sites.  We obtain numerical results for
chains of size $L$, with a trap of size $l$ centered in the middle of the
chain (unless explicitly stated, we consider odd $L$s so that the middle of the
trap coincides with the middle site of the chain); we choose $L$ large enough
to have negligible finite-$L$ effects; we are able to obtain results correct
to machine precision for $l$ up to $O(10^3)$.  We then analyze the quantum
critical behaviors in the presence of the trap within the framework of the TSS
theory.

The trap-size dependence shows subtle effects in the parameter region where
the homogeneous model without trap has a nonzero filling $f$, i.e., for
$\mu<1$, therefore in the superfluid region and at the $n=1$ Mott transition.
This is essentially related to the presence of level crossings at finite trap
size.  They arise because the particle number is conserved, i.e., the particle
number operator $\hat{N}=\sum_i n_i$ commutes with the BH Hamiltonian
(\ref{bhm}) even in the presence of the trapping potential; thus the
eigenvectors do not depend on $\mu$, even though the eigenvalues do.  In the
presence of the trapping potential (\ref{potential}), the particle number
$N\equiv\langle\hat N\rangle$ is finite and increases as $N\sim l$ with
increasing the trap size $l$.  Therefore, as $l\to\infty$, there is an infinite
number of ground-state level crossings where $N$ jumps by 1 and the gap
$\Delta$ vanishes.

As we shall see, this phenomenon gives rise to a new interesting scenario at
the $n=1$ Mott transition, requiring a revision of the TSS Ansatz
(\ref{freee}) and (\ref{gapscalgen}) into a {\em modulated} TSS: the TSS is
still controlled by the trap-size exponent $\theta= p/(p+2)$, as in the case
of the low-density Mott transition, but it gets modulated by periodic
functions of the trap size.

We provide numerical evidence of universality at the low-density and $n=1$
Mott transitions by considering a more general 1D hard-core BH model with
nearest-neighbor density-density interactions, corresponding to the so-called
XXZ model.  Since the corresponding fermion representation is not longer
quadratic, the numerical results are obtained by using density matrix
renormalization group (DMRG) methods.

We also show that the trap-size dependence in the gapless superfluid phase is
characterized by power-law asymptotic behaviors which are modulated by
periodic functions of the trap size. Moreover, it shows a multiscale behavior
characterized by different length scales diverging with different power laws
with increasing trap size, associated with the smooth modes and with the
modes at the Fermi momentum $k_F=\pi f$.

The paper is organized as follows.  In Sec.~\ref{tssmu1} we discuss the TSS at
the low-density Mott transition in the 1D hard-core BH model.  We extend the
analysis of Ref.~\cite{CV-10}, providing the TSS of several observables and
checking its universality within the XXZ model.  In Sec.~\ref{ldasec} we
discuss the local density approximation to determine the particle density in
the presence of the confining potential.  In Sec.~\ref{muint} we study the
trap-size dependence within the gapless superfluid phase, presenting results
for the gap, the particle density and its correlators, the one-particle
density matrix and the von Neumann entanglement entropy.  In Sec.~\ref{mum1}
we consider the $n=1$ Mott insulator to superfluid transition at $\mu=-1$.  We
present results at fixed trap size for the XX and XXZ models, and show that
their trap-size dependence is described by a modulated TSS.
Finally, in Sec.~\ref{conclusions} we draw our conclusions.  In
App.~\ref{Sec:numerics} we report some details on the numerical calculations
presented in paper.  App.~\ref{muintpinf} reports some results for the
homogeneous 1D hard-core BH model with open boundary conditions, showing that
modulated scaling behaviors of the gap, the particle density, and the
subleading corrections of the entanglement entropy, are already present in the
finite size behavior of the homogeneous model within the superfluid region; we
provide exact formulae for the total particle number and for the particle
density at the middle of the chain.

\section{Trap-size scaling at the low-density Mott transition}
\label{tssmu1}

\subsection{The TSS limit}
\label{tsslmu1}

In order to show the existence of a nontrivial TSS limit around
$\mu=1$, i.e., at the transition between a low-density superfluid and the
empty vacuum state (which may be named $n=0$ Mott phase), we exploit the exact
mapping of the XX model into a model of spinless fermions,
by the Jordan-Wigner transformation
\begin{eqnarray}
&& \sigma^{x}_i = \prod_{j<i}(1-2c_j^\dagger c_j) (c_i^\dagger+c_i),
\label{JWtra}\\
&& \sigma^{y}_i = i \prod_{j<i}(1-2c_j^\dagger c_j) (c_i^\dagger-c_i),
\nonumber \\
&& \sigma^{z}_i = 1-2c_i^\dagger c_i, 
\nonumber 
\end{eqnarray}
which leads to the Hamiltonian
\begin{eqnarray}
&&H_c = \sum_{ij} c_{i}^\dagger h_{ij} c_{j},\label{fermmod}\\
&&h_{ij} = \delta_{ij} - {1\over 2} \delta_{i,j-1} - {1\over 2} \delta_{i,j+1}
+ [\bar{\mu} + V(x_i)] \delta_{ij}, 
\nonumber
\end{eqnarray}
with $\bar{\mu}\equiv \mu-1$.
In the fermion representation the Hamiltonian can be easily diagonalized by
introducing new canonical fermionic variables $\eta_k=\sum_i \phi_{ki} c_i$,
where $\phi$ satisfies the equation
\begin{equation}
h_{ij}\phi_{kj} = \omega_k \phi_{ki},
\label{aeq}
\end{equation}
obtaining
\begin{equation}
H_c = \sum_k \omega_k \eta_k^\dagger \eta_k.
\label{hdiag}
\end{equation}
The ground state contains all $\eta$-fermions with $\omega_k<0$; the
number of filled energy levels is $N$.  The energy gap is
\begin{equation}
\Delta = \min_k |\omega_k|.
\label{gapres}
\end{equation}
Since $\phi_{ki}$ is an orthogonal matrix, the expectation values of the
$c$-operators can be obtained by using the inverse relation $c_i=\sum_k
\phi_{ki} \eta_k$.

The above equations have a nontrivial TSS limit around
\begin{equation}
\bar{\mu}\equiv\mu-1=0, 
\label{mubardef}
\end{equation}
i.e., at the transition between a low-density superfluid and the empty vacuum
state.  We consider the continuum limit of Eq.\ (\ref{aeq}), by rewriting the
discrete differences in terms of derivative expansions.  Near the critical
point $\bar{\mu}=0$ and for sufficiently small values of $k$ (this is required
by the smoothness hypothesis underlying the continuum limit), we obtain
\begin{equation}
\left[ \bar{\mu} + (x/l)^p - 
{1\over 2} {d^2 \over dx^2} - {1\over 24} {d^4 \over dx^4} + ...
\right]\phi_k(x) =
\omega_k \phi_k(x),
\label{trapscaleqxx00}
\end{equation}
where $\phi_k(x)\equiv \phi_{kx}$.
Then, by replacing
\begin{eqnarray}
&&\displaystyle x = l^{p/(2+p)} X, \nonumber \\
&&\displaystyle \bar{\mu} = l^{-2p/(2+p)} \mu_r, 
\nonumber \\
&&\displaystyle \omega_k = l^{-2p/(2+p)} \Omega_k, 
\label{resca}
\end{eqnarray}
and neglecting terms which are suppressed in the large-$l$ limit, we
obtain
\begin{equation}
\left( - {1\over 2} {d^2 \over dX^2} + X^p \right)\varphi_k(X) =
(\Omega_k - \mu_r) \varphi_k(X) \equiv \bar\Omega_k \varphi_k(X),
\label{trapscaleqxx}
\end{equation}
where $\varphi_k(X)\equiv \phi_k(l^{p/(2+p)} X)$; note that
$\bar\Omega_k$ is independent of $\mu_r$. Recalling that $z=2$ and
$y_\mu=2$ at this transition, we infer $\theta=p/(2+p)$, as also
obtained by RG arguments~\cite{CV-10}.  Therefore, the solutions of
Eq.\ (\ref{trapscaleqxx}) determine the TSS in the limit
$\bar{\mu}\equiv \mu-1\to 0$ and $l\to \infty$, keeping 
$\mu_r\equiv l^{2\theta} \bar{\mu}$ and $X\equiv l^{-\theta}x$ fixed.
On the other hand, subleading terms, such as the $d^4/dx^4$ term
in the l.h.s.\ of Eq.\ (\ref{trapscaleqxx00}), give rise to
$O(l^{-2\theta})$ corrections in the TSS limit (\ref{resca}) of the
Eq.\ (\ref{aeq}).  Thus, we expect that the TSS of any quantity is
approached with $O(l^{-2\theta})$ scaling corrections.

For $p=2$, we obtain
\begin{eqnarray}
&&\bar\Omega_k \equiv \Omega_k - \mu_r
= 2^{1/2}(k + 1/2), \quad k\ge 0, \label{eq:p2eig}\\
&&\varphi_k(X) = {2^{1/8} H_k(2^{1/4}X)\over
 \pi^{1/4} 2^{k/2} (k!)^{1/2}} \, \exp(-X^2/\sqrt{2}), 
 \nonumber 
\end{eqnarray}
where $X\equiv x/l^{1/2}$ and $H_k(x)$ are Hermite's polynomials $H_k(x) =
(-1)^k e^{x^2} d^ke^{-x^2}/dx^k$.

For $p=4$, we can solve numerically Eq.\ (\ref{trapscaleqxx}) by Numerov's
method; the resulting energy levels are $\bar\Omega_0=0.667986$,
$\bar\Omega_1=2.39364$, $\bar\Omega_2=4.69680$, $\bar\Omega_3=7.33573$,
$\bar\Omega_4=10.2443$, $\bar\Omega_5=13.3793$, etc.; Bohr-Sommerfield
quantization formula gives
\begin{eqnarray}
\bar\Omega_k &\approx& [\gamma_4(k+1/2)]^{4/3},
\nonumber \\
\gamma_4 &=& \sqrt{\pi/2}\;\Gamma(7/4)/\Gamma(5/4)\cong1.27082,
\label{eq:p4eig}
\end{eqnarray}
which is accurate to $0.1\%$ already for $\bar\Omega_5$.
For large $k$ and generic $p$, the semiclassical limit gives
\begin{equation}
\bar\Omega_k\sim k^{2p/(2+p)}.
\label{largekome}
\end{equation}

For $p\to\infty$, Eq.\ (\ref{trapscaleqxx}) becomes equivalent to the
Schr\"odinger equation of a free particle in a box of size $L=2l$ with
boundary conditions $\varphi(-1)=\varphi(1)=0$, leading to
\begin{eqnarray}
&&\bar\Omega_k = {\pi^2\over 8} (k+1)^2, \qquad k\ge 0,
\label{eq:pinfeig}\\
&&\varphi_k(X) = \sin\left[{\pi\over 2} (k+1) (X+1)\right],
\nonumber
\end{eqnarray}
where $X\equiv x/l$.

\subsection{TSS of observables}
\label{tssobsmu1}

\subsubsection{The energy gap}
\label{egap}

\begin{figure}[tbp]
\includegraphics*[scale=\graphicscale]{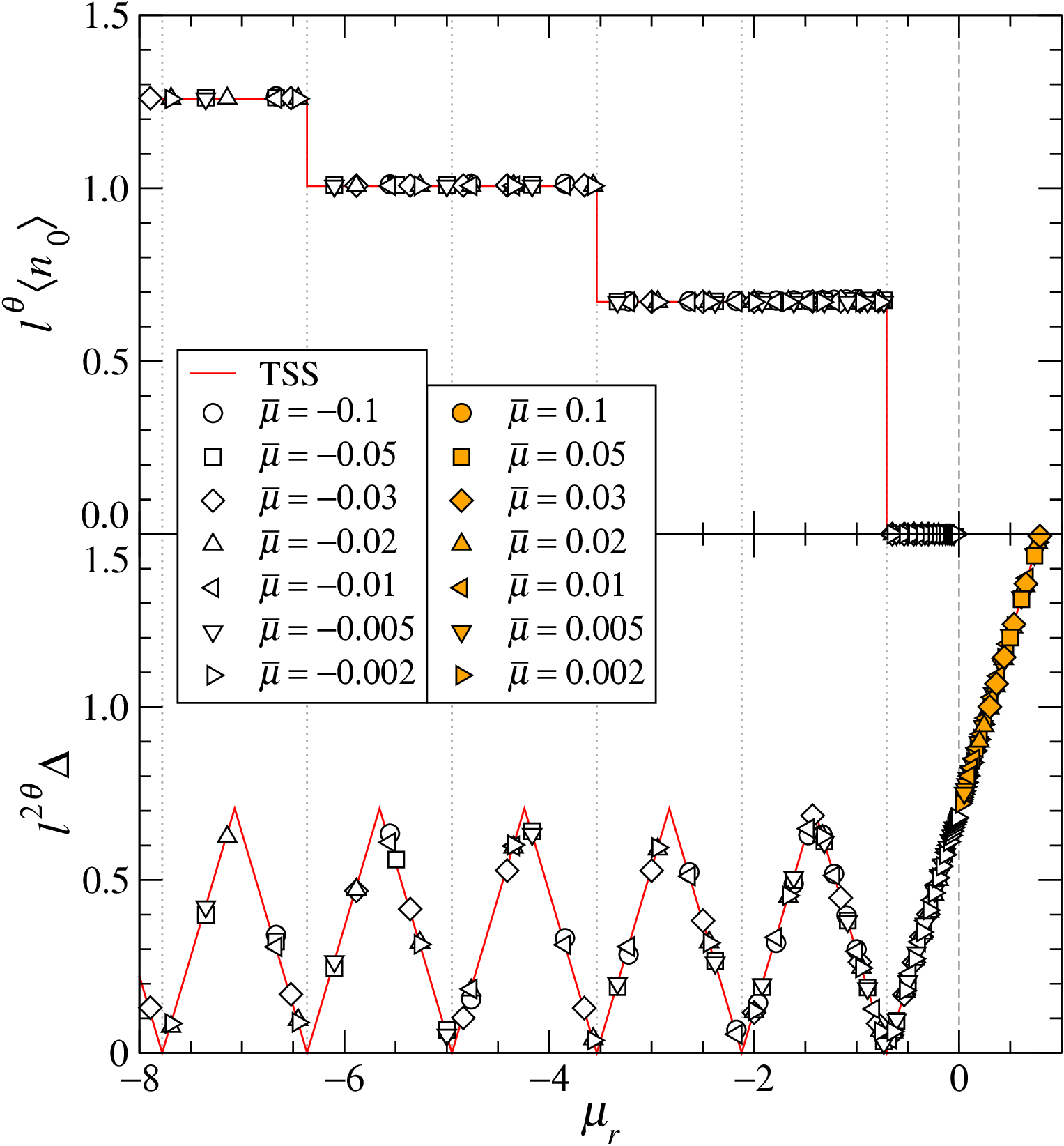}
\caption{
  The rescaled energy gap $l^{2\theta}\Delta$ (below) and the rescaled
  particle density $l^{\theta}\langle n_0\rangle$ in the middle of the trap 
  (above) for $p=2$ ($\theta=1/2$), compared with the predictions of TSS.  We
  report results for several values of $\bar{\mu}$ and $l\gtrsim 10$.  They
  are perfectly consistent with the analytical TSS.}
\label{fig:Drho0,p=2}
\end{figure}
\begin{figure}[tbp]
\includegraphics*[scale=\graphicscale]{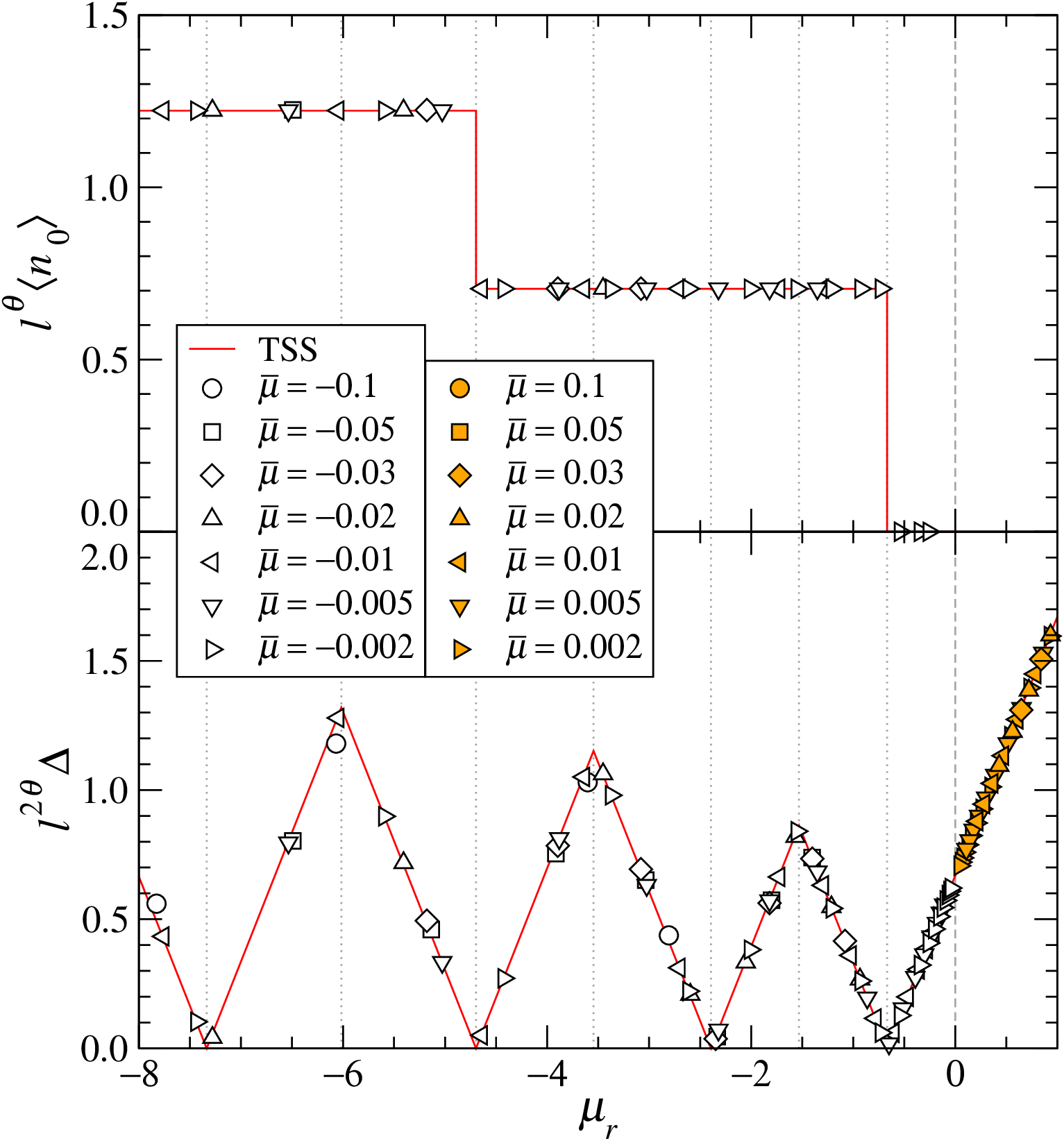}
\caption{The rescaled energy gap $l^{2\theta}\Delta$ (below)
  and the rescaled particle density $l^{\theta}\langle n_0\rangle$ in the
  middle of the trap (above) for $p=4$ ($\theta=3/4$), compared with the
  predictions of TSS.  We report results for several values of $\bar{\mu}$ and
  $l\gtrsim 10$.}
\label{fig:Drho0,p=4}
\end{figure}
\begin{figure}[tbp]
\includegraphics*[scale=\graphicscale]{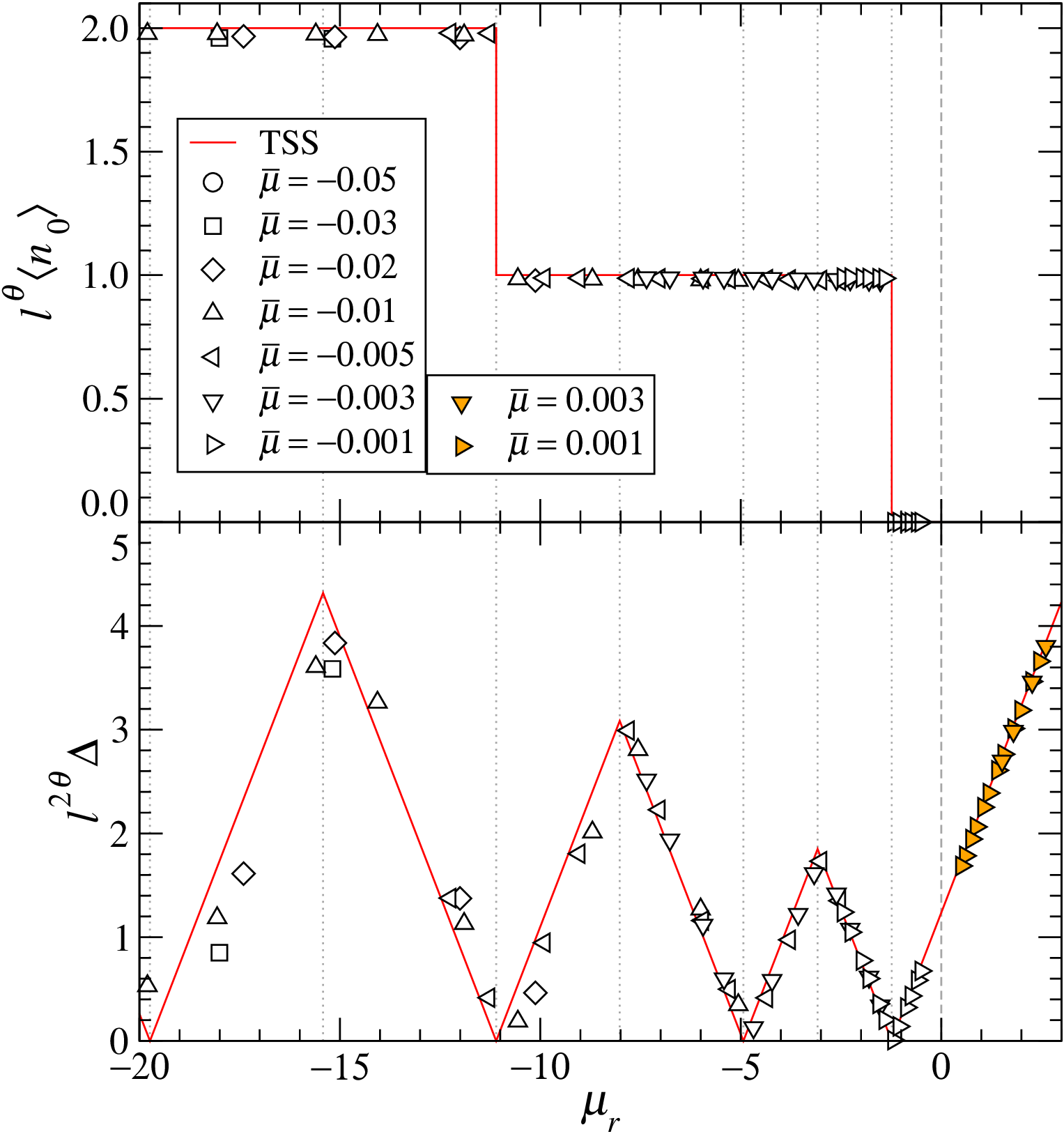}
\caption{The rescaled energy gap $l^{2\theta}\Delta$ (below)
  and the rescaled particle density $l^{\theta}\langle n_0\rangle$ in the
  middle of the trap (above) for $p\to\infty$ ($\theta=1$), compared with the
  predictions of TSS.  We report results for several values of $\bar{\mu}$ and
  $l\gtrsim 10$.  They approach the analytical TSS results with increasing
  $l$.  }
\label{fig:Drho0,p=inf}
\end{figure}

The existence of the TSS limit implies that any low energy scale behaves as
$l^{-2\theta} {\cal E}(\mu_r)$, where ${\cal E}(\mu_r)$ is a scaling function.
Specifically, in the case of the gap $\Delta=E_1-E_0$ we have
\begin{equation}
\Delta \approx l^{-2\theta} {\cal E}_\Delta(\mu_r),\qquad
{\cal E}_\Delta(\mu_r) = \min_k |\Omega_k|,
\label{eq:gap}
\end{equation}
which can be easily computed from the solution of Eq.\ (\ref{trapscaleqxx}).
The results are shown in Figs.\ \ref{fig:Drho0,p=2}, \ref{fig:Drho0,p=4}, and
\ref{fig:Drho0,p=inf}, respectively for $p=2,\,4$ and in the limit
$p\to\infty$. For any $p$, the scaling function ${\cal
E}_\Delta(\mu_r)$ shows a triangular structure for $\mu_r\le0$ and it
is linear for $\mu_r \ge -\bar\Omega_0 < 0$.  
In Figs.\ \ref{fig:Drho0,p=2}, \ref{fig:Drho0,p=4}, and
\ref{fig:Drho0,p=inf} we also show results obtained by numerical
diagonalization at fixed $l$.  They
clearly approach the analytical TSS computations. 
Corrections to scaling turn out to be very small for $p=2$ and $p=4$.

\subsubsection{The particle density}
\label{padebmu1}

We now consider the expectation value and correlators of the particle density
\begin{equation}
n_x\equiv b_x^\dagger b_x = {1-\sigma^z_x\over 2} = c_x^\dagger c_x =
\sum_{k_1,k_2} \phi_{k_1\,x}^* \phi_{k_2\,x} \,\eta_{k_1}^+ \eta_{k_2}.
\label{nx}
\end{equation}
$\langle n_x\rangle$ is zero for $\bar\mu\ge0$, but it can be nonzero for
$\bar\mu<0$.  Since the RG dimension of the particle density is given by
$y_n=d+z-y_\mu=1$, we expect
\begin{equation}
\langle n_x \rangle =  l^{-\theta} {\cal D}(\mu_r,X).
\label{nxsc}
\end{equation}
This is confirmed by the analytical results of the previous section, which
lead to
\begin{equation}
{\cal D}(\mu_r,X) = \sum_{k:\;\Omega_k<0} \varphi^2_k(X),
\label{eq:rho}
\end{equation}
where we used the fact that $\langle \eta_k^+ \eta_k\rangle=1$ if $\Omega_k<0$
and zero otherwise.  $\varphi_k(X)$ are the normalized eigenfunctions of Eq.\ 
(\ref{trapscaleqxx}).  Note that ${\cal D}$ depends on $\mu_r$ only through
the number of negative energy levels $N$, i.e., the number of levels included
in the sum of Eq.\ (\ref{eq:rho}); therefore, it vanishes for
$\mu_r\ge-\bar\Omega_0$ and it falls on a discrete set of curves as a function
of $X$ (with jumps at $\Omega_k=0$, i.e., zeroes of $\Delta$).  Since
$\varphi_k(X)=(-1)^k\varphi_k(-X)$, only even $k$s contribute to ${\cal
  D}(\mu_r,0)$.

Specifically, for $p=2$ we obtain
\begin{eqnarray}
&&{\cal D}(\mu_r,0) =
2^{1/4} \pi^{-1/2} \sum_{j=0}^{j_{\rm max}} {[(2j-1)!!]^2\over(2j)!},
\nonumber \\
&&j_{\rm max}= \max\left( 0, 
\left\lfloor |\mu_r|/\sqrt{8} - 1/4 \right\rfloor \right),
\label{n0xx}
\end{eqnarray}
where $\lfloor x \rfloor\equiv{\mathop{\rm floor}}(x)$ is the largest
integer not greater than $x$.  For $p\to\infty$ we have
\begin{equation}
{\cal D}(\mu_r,0) = \max\left( 0, 
\left\lfloor \sqrt{2|\mu_r|}/\pi + 1/2\right\rfloor\right).
\label{n0xxpinfty}
\end{equation}

\begin{figure}[tbp]
\includegraphics*[scale=\graphicthreescale]{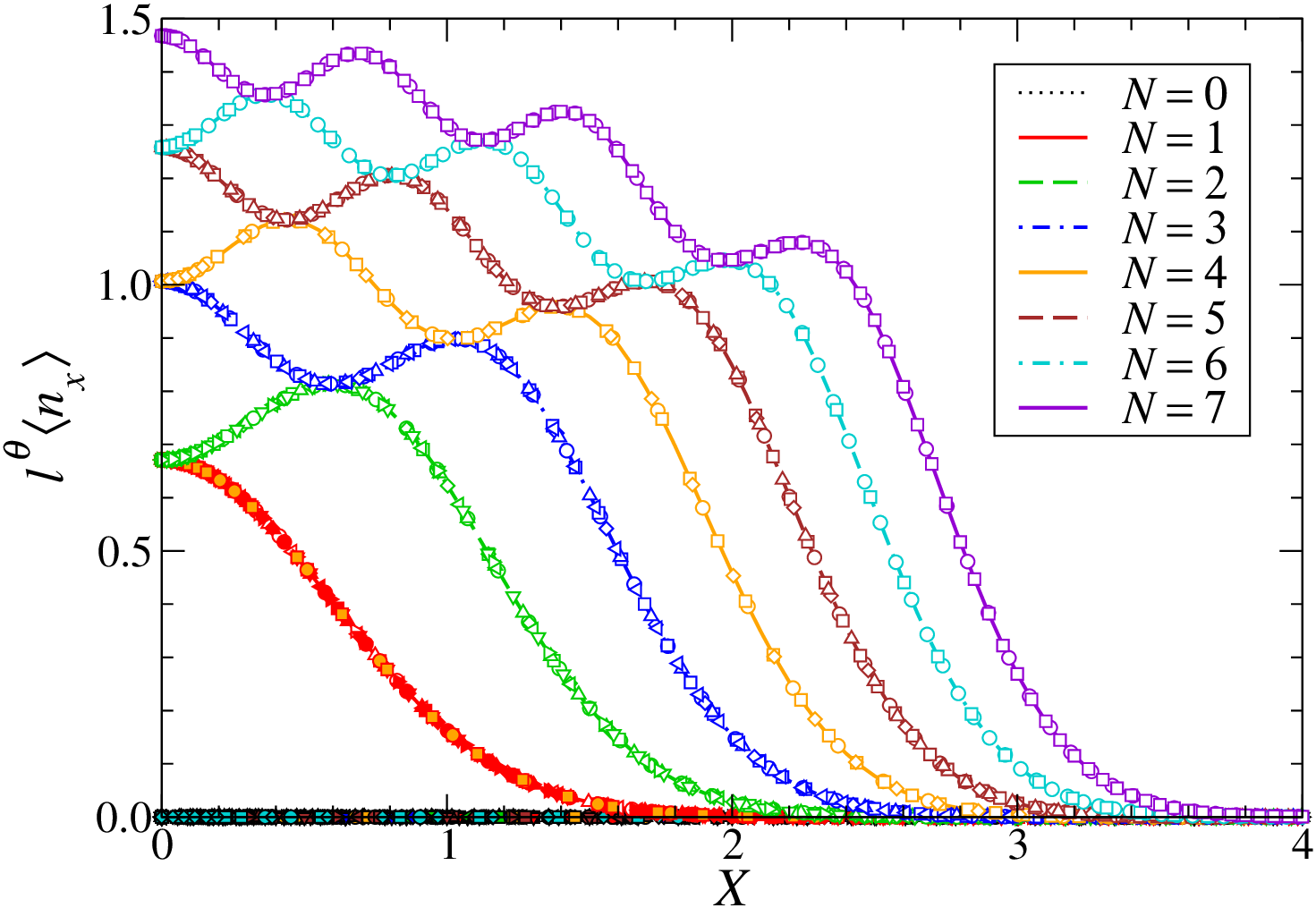}
\includegraphics*[scale=\graphicthreescale]{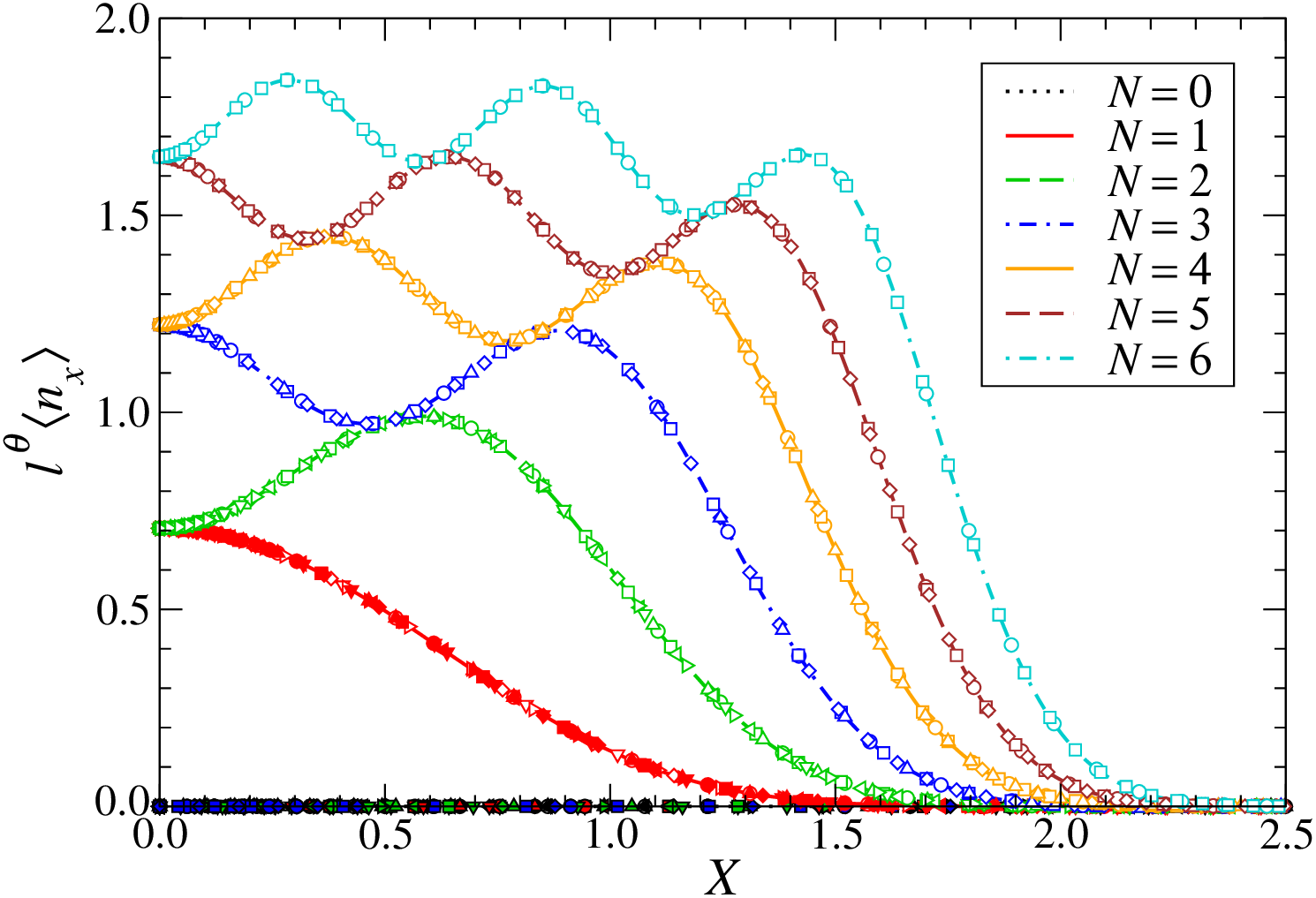}
\includegraphics*[scale=\graphicthreescale]{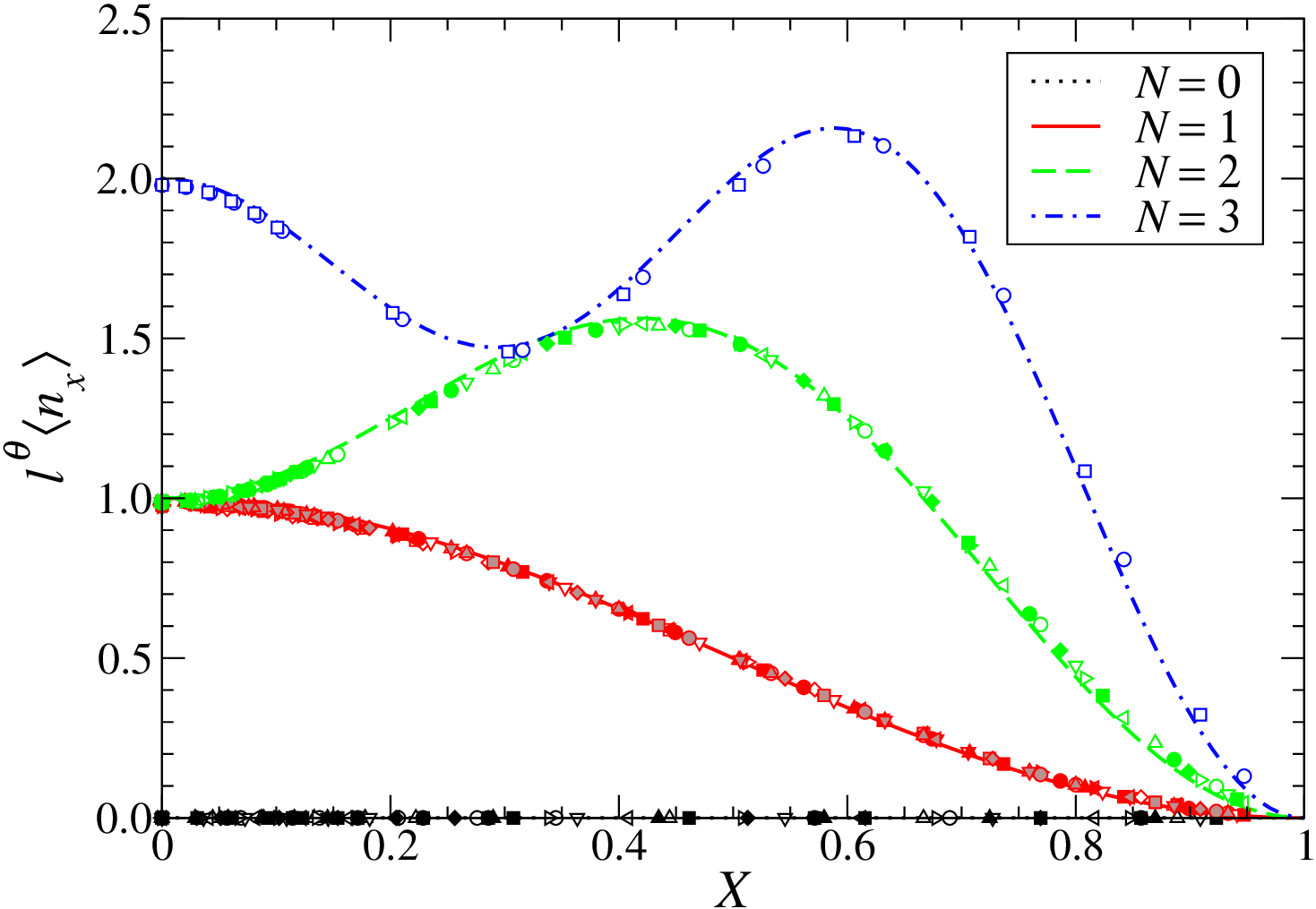}
\caption{
  The rescaled density $l^\theta \langle n_x\rangle$ for $p=2$ (above),
  $p=4$ (middle), and $p\to\infty$ (below), compared with Eq.\ (\ref{eq:rho}).
  $N$ is the number of negative energy levels with $\Omega_k<0$.  Here and in
  the following, set of parameters $(\mu,l)$ are differentiated by symbol
  color (matching $N$), shape, and filling color.  The results approach the
  analytical $N$-dependent TSS results, cf.\ Eq.~(\ref{eq:rho}), with
  increasing $l$.}
\label{fig:rho}
\end{figure}

Numerical results for the particle density at the origin are shown in Figs.\ 
\ref{fig:Drho0,p=2}, \ref{fig:Drho0,p=4}, and \ref{fig:Drho0,p=inf}, for
$p=2,4,\infty$ respectively; they fully support TSS.  Note the peculiar
plateaus and the discontinuities in the particle density at negative values of
the scaling variable $\mu_r$.  Moreover, asymptotically for $\mu_r\to-\infty$,
$\langle n_0\rangle \sim |\bar\mu|^{1/2}$, which matches the critical behavior
for $\bar{\mu}<0$ in the absence of the trap~\cite{Sachdev-book}.  Numerical
results for $\langle n_x \rangle$, showing the dependence on the distance form
the middle of the trap, are shown in Fig.\ \ref{fig:rho}, for $-0.01\le
\bar\mu<0$ and $l\ge10$.  They show that the quantity $l^{\theta} \langle n_x
\rangle$ approaches the analytical functions obtained using Eq.~(\ref{eq:rho}).

\subsubsection{The particle density correlator}
\label{padecobmu1}

Let us consider the particle density correlation
\begin{equation}
G_n(x)\equiv \langle n_0 n_x\rangle-\langle n_0\rangle \langle n_x\rangle =
{1\over 4} \langle \sigma_0^z \sigma_x^z\rangle_c.
\label{gnx}
\end{equation}
Like the particle density, it vanishes for $\bar{\mu}\ge 0$, and it can be
nonzero for $\bar{\mu}<0$.  In the TSS limit we expect the scaling behavior
\begin{equation}
G_n(x) \approx l^{-2\theta} {\cal G}_n(\mu_r,X).
\label{gnxsc}
\end{equation}
Straightforward  calculations, i.e., writing $G_n$ in terms of $\eta$-operators
and then using the Wick theorem to compute the resulting ground-state
expectation values, show that the scaling function ${\cal G}_n(\mu_r,X)$ can
be written in terms of the eigensolutions of Eq.\ (\ref{trapscaleqxx}):
\begin{eqnarray}
&&{\cal G}_n(\mu_r,X) =  \\
&&\quad
\Bigl[\textstyle\sum_{k:\;\Omega_k<0} \varphi_k(X) \varphi_k(0)\Bigr]
\Bigl[\textstyle\sum_{k:\;\Omega_k>0} \varphi_k(X) \varphi_k(0)\Bigr];
\nonumber
\label{eq:GnTSSe}
\end{eqnarray}
then, using the completeness relation
$\sum_k \varphi_k(X) \varphi_k(0) = \delta(X)$, we obtain
\begin{equation}
{\cal G}_n(\mu_r,X) = 
-\Bigl[\textstyle\sum_{k:\;\Omega_k<0} \varphi_k(X) \varphi_k(0)\Bigr]^2;
\label{eq:GnTSS}
\end{equation}
note that odd $k$s do not contribute to the sum.  Like the particle density,
${\cal G}_n(\mu_r,X)$ is nonzero only for $\mu_r < -\bar\Omega_0 < 0$, and
depends on $\mu_r$ only through $N$ (actually, only
through $\lfloor(N+1)/2\rfloor$). 

\begin{figure}[tbp]
\includegraphics*[scale=\graphicscale]{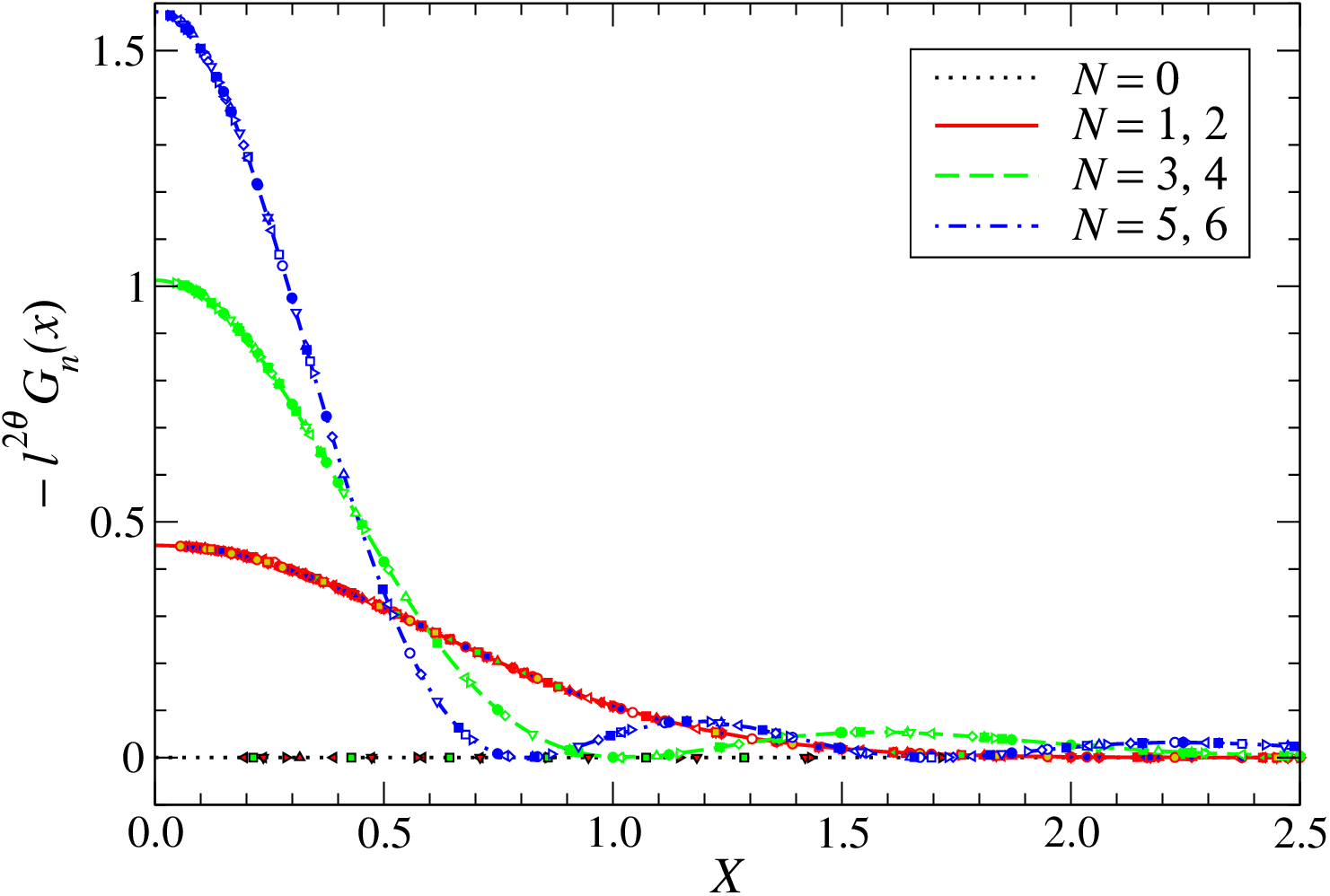}
\caption{
 The rescaled connected correlation $l^{2\theta} G_n(x)$ for $p=2$,
  $l\ge10$, $\mu\ge0.95$, and $N\le6$.  The results approach the analytical
  $N$-dependent TSS results, cf.\ Eq.~(\ref{eq:GnTSS}), with increasing $l$.}
\label{fig:rhocc}
\end{figure}

From the numerical data, we find that $l^{2\theta} G_n(x)$ as a function of
$X$ approaches rapidly ${\cal G}_n(\mu_r,X)$.  The results for $p=2$ are shown
in Fig.\ \ref{fig:rhocc}.

\subsubsection{The one-particle density matrix}
\label{onepdbmu1}

\begin{figure}[tbp]
\includegraphics*[scale=\graphicscale]{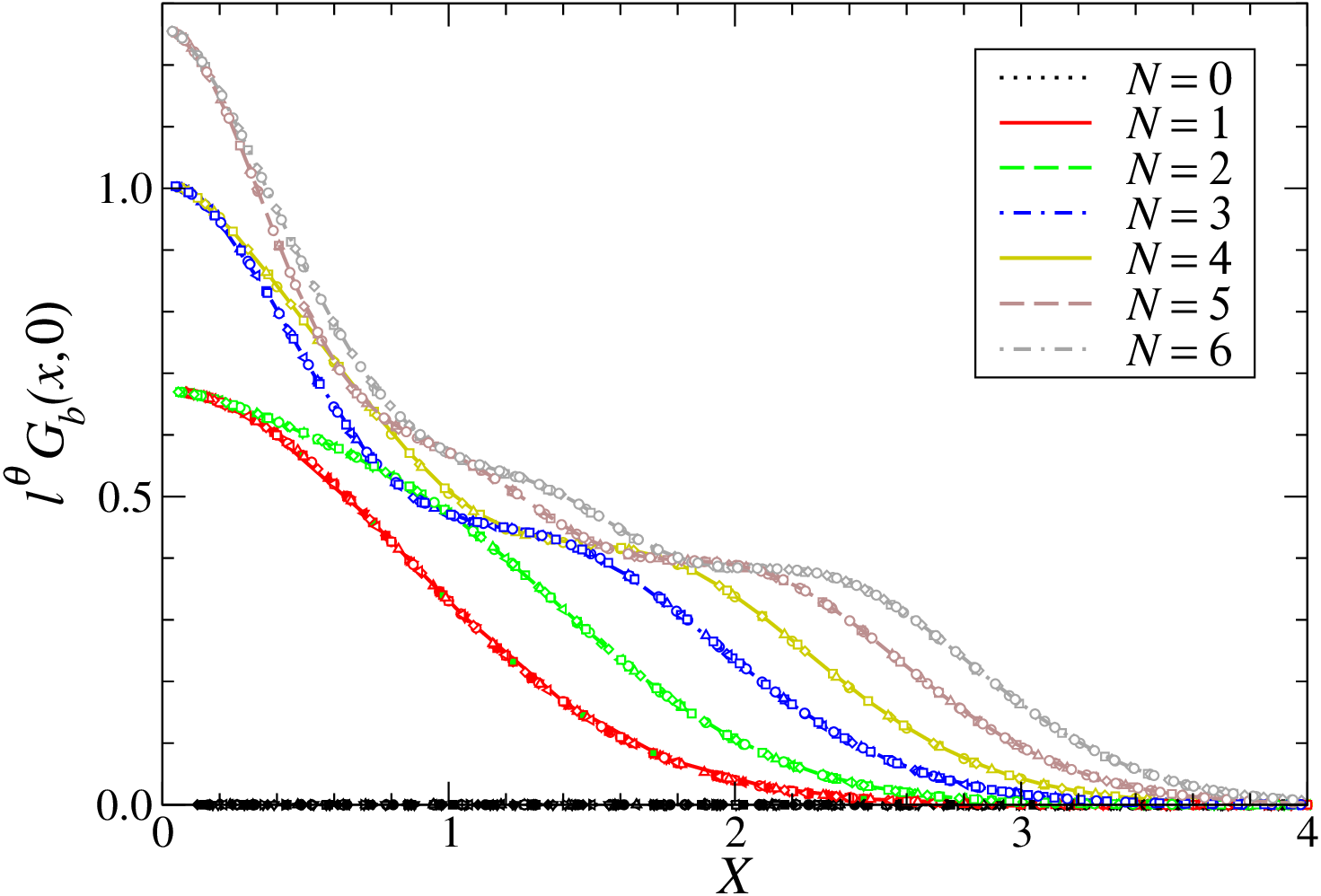}
\caption{The rescaled $\sigma_x$ correlation
$l^{\theta} G_b(x,0)$ for $p=2$, $l\ge10$, $\mu\ge0.95$, and $N\le6$.
With increasing $l$, they converge toward $N$-dependent TSS curves.
}
\label{fig:cxx}
\end{figure}

We now consider the one-particle density matrix defined as
\begin{equation}
G_b(x_i,x_j) \equiv \langle b_{i}^\dagger b_{j} \rangle
= \langle \sigma^-_{i} \sigma^+_{j} \rangle,
\label{gij}
\end{equation}
where $\sigma^\pm = (\sigma^x\pm i \sigma^y)/2$. 
For $\bar{\mu}\ge 0$, since the ground state is empty, $G_b(x_i,x_j)=0$.  But,
like the particle density (note that $G_b(x,x)=\langle n_x \rangle$), it can
be nonzero for $\bar{\mu}<0$. Its scaling behavior is also determined by the
RG dimension $y_b$ of the field associated with the boson operator $b_i$,
which is~\cite{Sachdev-book} $y_b=1/2$.  Considering specifically the
correlations with one boson operator at the origin, i.e., $G_b(x,0)$, we expect
\begin{equation}
G_b(x,0) \approx l^{-\theta} {\cal M}(\mu_r,X).
\label{gbscal1}
\end{equation}
Numerical results can be obtained using the method of Ref.~\cite{YR-96}.  They
confirm the scaling behavior (\ref{gbscal1}), i.e., $l^{\theta} G_b(x,0)$
appears to approach a function of $\mu_r$ and $X$ in the large-trap limit.
Fig.\ \ref{fig:cxx} shows results for $p=2$.  Again, the dependence on $\mu_r$
is only through 
the number $N$ of negative energy levels.

\subsubsection{The von Neumann entanglement entropy}
\label{hlentmu1}

\begin{figure}[tbp]
\includegraphics*[scale=\graphicscale]{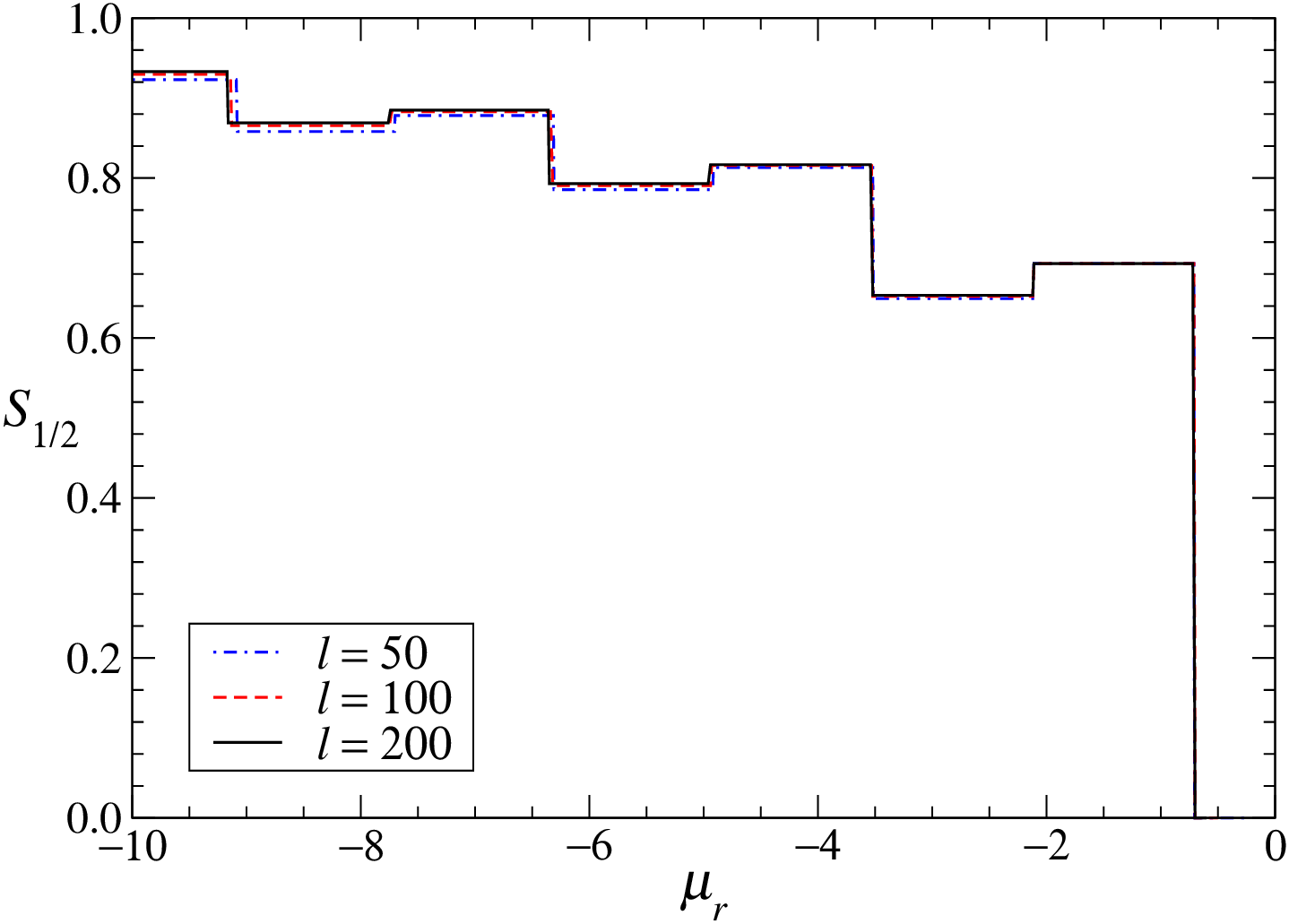}
\caption{The half-lattice von Neumann entanglement
entropy $S_{1/2}$, cf.\ Eq.~(\ref{sdef1o2}), for $p=2$.}
\label{fig:Se}
\end{figure}

We consider the von Neumann entanglement entropy $S(l_A;L)$ in the presence of
the confining potential, defined by dividing the chain in two parts of length
$l_A$ and $L-l_A$.  The entanglement entropy trivially vanishes for
$\bar{\mu}\ge 0$, but it can be nonzero for $\bar{\mu}<0$ \cite{JK-04}.  It
can be computed using the techniques of Ref.~\cite{Peschel-03}.  We consider
the half-lattice von Neumann entanglement entropy $S(L/2;L)$ for even $L$ and
open boundary conditions in the presence of the trap of size $l$ (in this case
the trap is centered between the two central sites of the chains). Its
large-$L$ limit,
\begin{equation}
  S_{1/2}\equiv \lim_{L\to\infty} S(L/2;L),
\label{sdef1o2}
\end{equation}
depends on the trap size $l$ only.  The TSS limit of $S_{1/2}$ turns out to
depend only on $N$, and therefore it is a function of the scaling quantity
$\mu_r$, as shown in Fig.\ \ref{fig:Se} for the harmonic potential.

\subsection{Universality of the TSS}
\label{tssunivmu1}

In order to check the universality of the TSS, we consider the XXZ chain model
\begin{eqnarray}
H_{\rm XXZ} = &&-\sum_{i=1}
    (S^x_i S^x_{i+1} + S^y_i S^y_{i+1}
     + j_z S^z_i S^z_{i+1})
\nonumber \\ &&-
    \sum_{i=1} [\mu + V(x_i)] S^z_i.
\label{eq:H-XXZ}
\end{eqnarray}
In terms of bosonic operators $b_i$, the $j_z$ term corresponds to
nearest-neighbor density-density interactions, i.e.,
\begin{equation}
j_z\sum_i (n_i-1/2)(n_{i+1}-1/2),
\label{jzterm}
\end{equation}
where $n_i=b_i^\dagger b_i$.

For $|j_z|<1$, the homogeneous XXZ chain model (i.e., with $V=0$) undergoes
two Mott insulator to superfluid transitions at $\mu_{c}^\pm=\pm(1+j_z)$ in
the same universality class of those of the XX model at $\mu=\pm 1$.  Indeed,
the $j_z$ term is irrelevant at these transitions: the RG dimension of the
coupling $j_z$ is $y_{j_z}=-1$~\cite{Sachdev-book}.  We therefore expect that
the $j_z$ term remains irrelevant also in the TSS limit. The main difference
concerns the dominant scaling corrections, which are expected to be
$O(\xi^{-1})$, thus $O(l^{-\theta})$, when this term is present. Therefore
they are expected to be larger than those of the XX model, where they are
$O(l^{-2\theta})$.

\begin{figure}[tbp]
\includegraphics*[scale=\graphicscale]{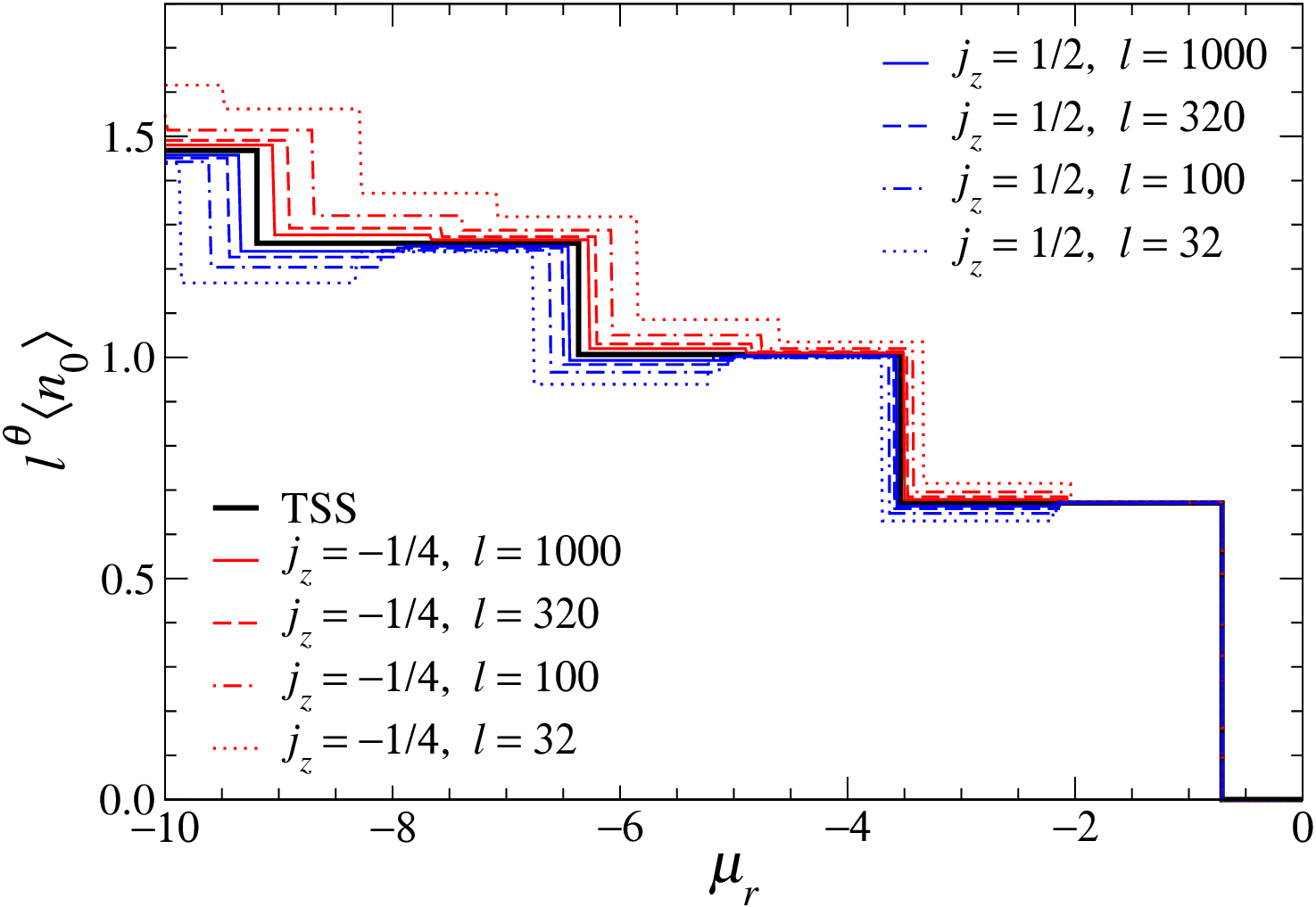}
\includegraphics*[scale=\graphicscale]{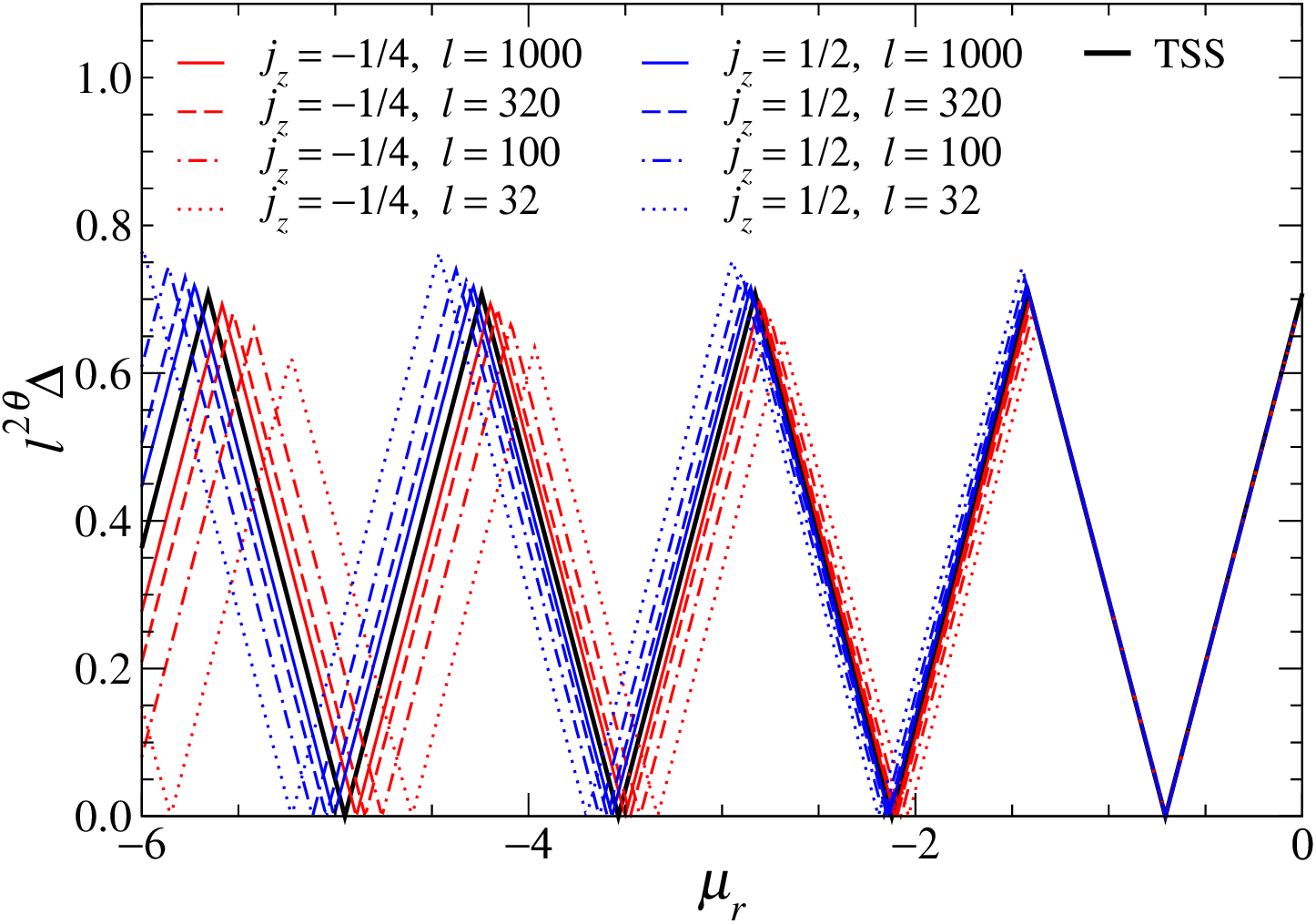}
\caption{Scaling plot of the energy gap (below) and of the
  particle density at the origin (above) for the XXZ model with $p=2$ and
  $j_z=-1/4$, $1/2$.  The data at $j_z=-1/4$ and $1/2$ approach the TSS
  functions, with increasing $l$, from opposite sides.}
\label{fig:XXZ}
\end{figure}

The Hamiltonian (\ref{eq:H-XXZ}) is no longer equivalent to a quadratic
fermionic Hamiltonian.  The properties of the model can be studied numerically
via DMRG.  Results for the gap and the particle density in the middle of the
trap at the low-density Mott transition, around $\mu_{c}^+$ (we define
$\bar{\mu}=\mu-\mu_{c}^+$), are shown in Fig.\ \ref{fig:XXZ} for $p=2$.  They
clearly support universality of TSS.  As expected, corrections to scaling are
larger for $j_z\ne0$ than for $j_z=0$.

It is interesting to notice that energy differences and expectation values
over states with $z$-component of the total spin $M=L/2$ and $M=L/2-1$ depend
on $\mu$ and $j_z$ only through $\mu_{c}^+$.  This implies that, in the
scaling region, vacuum expectation values (as functions of $\mu_r$) are
independent on $j_z$ for $\mu_r\gtrsim-\bar\Omega_2$; the energy gap $\Delta$
is independent on $j_z$ for $\mu_r\gtrsim-(\bar\Omega_1+\bar\Omega_2)/2$.
This is clearly shown in Fig.\ \ref{fig:XXZ}.

\section{The local density approximation of the particle density}
\label{ldasec}

The homogeneous 1D hard-core BH model, i.e., the model (\ref{bhm}) for
$U\to\infty$ and $V(x)=0$, has a nonzero filling below the low-density Mott
transition, i.e., for $\mu<1$.  In the infinite-chain limit $L\to\infty$, the
filling $f$ for $|\mu|\le1$ is given by~\cite{Sachdev-book}
\begin{equation}
f \equiv \langle n_i\rangle = {N\over L} = 
{1\over\pi}\arccos\mu, \qquad \mu=\cos\pi f.
\label{eq:fvsmu}
\end{equation}
The corresponding Fermi momentum is $k_F=\pi f$. For $\mu\le -1$,
$f=1$ independently of $\mu$.  In the following, $f$ will always denote the
value for the infinite homogeneous chain.

In the presence of a space-dependent confining potential, the so-called
local density approximation (LDA) estimates the spatial dependence of the
particle density by taking the value of the particle density of the
homogeneous system at the effective chemical potential
\begin{equation}
\mu_{\rm eff}(x) \equiv \mu + (x/l)^p .
\label{mueff}
\end{equation}
The LDA has been widely used to get quantitative information on the
behavior of BH models in a confining potential, and, more generally,
of inhomogeneous systems, 
see, e.g., Refs.~\cite{BHR-04,PBS-08,BRSRMDT-02,PRHD-04,WATB-04}.  

\begin{figure}[tbp]
\begin{center}
\leavevmode \includegraphics*[scale=\graphicscale]{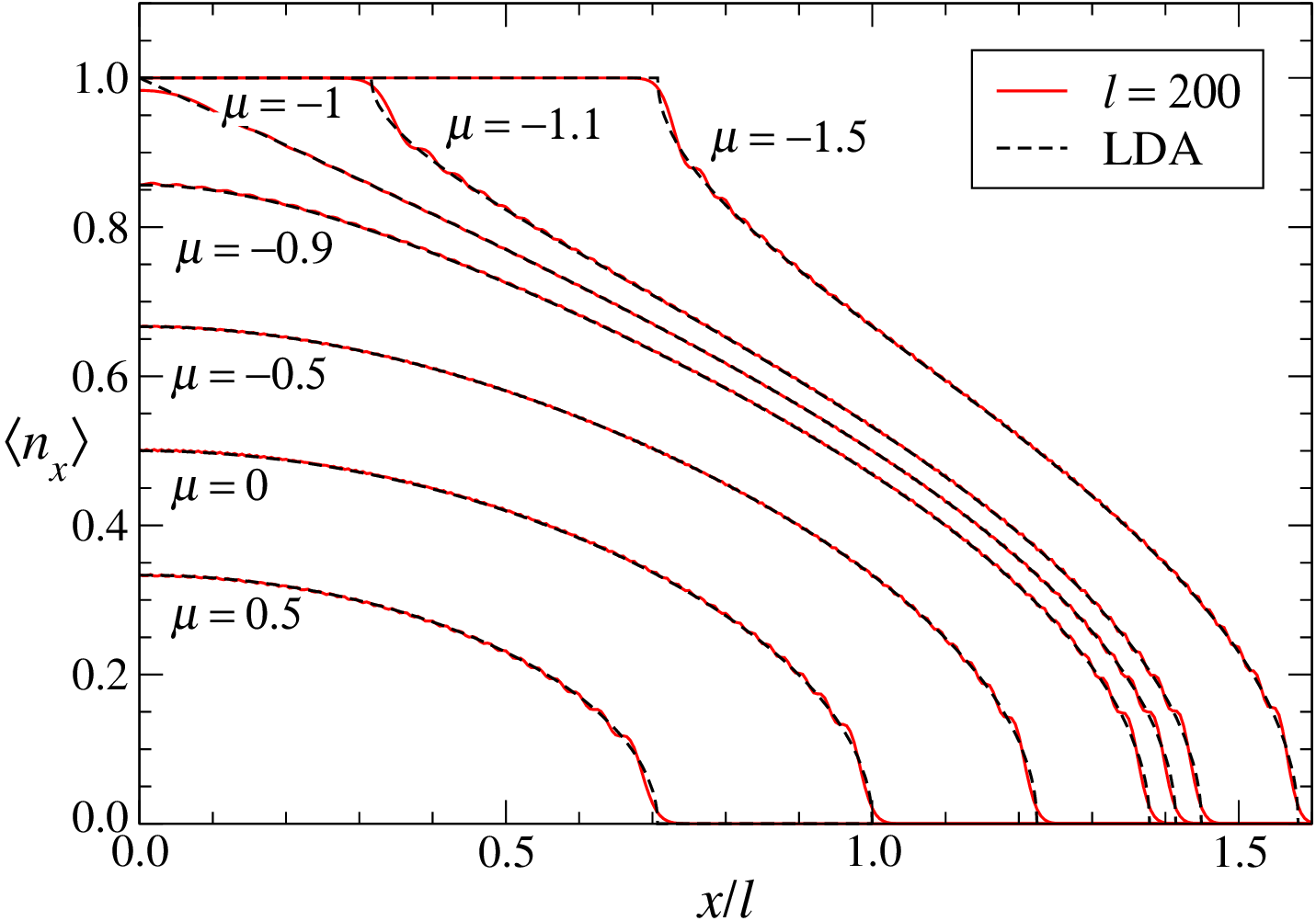}
\caption{The particle density vs.\ $x/l$ for $p=2$ and
several values of $\mu$. We show results from the LDA, cf.\
Eq.~(\ref{nxlda}), and from numerical calculations on a large chain
with $l=200$.  The differences are hardly visible in the figure.}
\label{XXpartdensnum1}
\end{center}
\end{figure}

The LDA of the particle density reads
\begin{eqnarray}
&&\langle n_x \rangle_{\rm lda} \equiv \rho_{\rm lda}(x/l) = \label{nxlda} \\
&&\quad\left\{
\begin{array}{l@{\ \ }l@{\ \ }l}
0 & {\rm for} & \mu_{\rm eff}(x) > 1, \\
(1/\pi)\arccos\mu_{\rm eff}(x) &
    {\rm for} & -1 \le \mu_{\rm eff}(x) \le 1, \\
1 & {\rm for} & \mu_{\rm eff}(x) < -1. \\
\end{array} \right. \nonumber
\end{eqnarray}
This would imply the presence of a plateau at $n=1$ when 
$\mu_{\rm eff} \le -1$, for 
\begin{equation}
x/l\le (-1-\mu)^{1/p},
\label{mott1ph}
\end{equation}
and a vanishing particle density when $\mu_{\rm eff} \ge 1$, for
\begin{equation}
x/l \ge (1-\mu)^{1/p}.
\label{mott0ph}
\end{equation}
In Fig.\ \ref{XXpartdensnum1} we compare the LDA of the particle
density with numerical
results for $p=2$ and the trap size $l=200$.  Note the flat regions
related to the $n=0,1$ Mott phases, already observed in experimental
and numerical works, see, e.g., Refs.~\cite{JBCGZ-98,BRSRMDT-02,FWMGB-06}.
Analogous results are found for other powers of the confining
potential.  The LDA provides a good approximation of the particle
density, which improves with increasing trap size.  The
differences of the trap-size dependence from the LDA results show a
nontrivial scaling behavior; they will be considered in the next sections.

Using Eq.~(\ref{nxlda}) we can also obtain the LDA of the total particle
number:
\begin{equation}
N_{\rm lda} = \sum_x \langle n_x \rangle_{\rm lda} \approx
2 l \int_0^\infty \rho_{\rm lda}(y) \,dy \equiv c(\mu) l,
\label{ntlda}
\end{equation}
where $c(\mu)$ is a finite function of $\mu$, which can be easily computed by
integrating $\rho_{\rm lda}(y)$.  Comparing with the numerical results, we
find excellent agreement, i.e., 
\begin{equation}
|c(\mu) l-N_{\rm exact}|<1
\label{cmunex}
\end{equation}
 (note that the
$N_{\rm exact}$ is an integer number, while the LDA is a continuous linear
function).  Eq.\ (\ref{cmunex}) implies that,
in the large-$l$ limit at fixed $\mu$,
the total particle number increases as 
\begin{equation}
N\sim l.
\label{thermlim}
\end{equation}

\section{Trap-size dependence in the superfluid phase}
\label{muint}

We now discuss the trap-size dependence in the gapless superfluid phase for
$|\mu|<1$.  In the continuum limit, the gapless superfluid phase of the
homogeneous system is described by a free massless bosonic field theory with
dynamic exponent $z=1$, corresponding to a conformal field theory with central
charge $c=1$, see, e.g., Ref.~\cite{Tsvelik-book}.

\subsection{Level crossings in the presence of the trap}
\label{levcro}

In the gapless superfluid phase, and more generally for
$\bar\mu\equiv\mu-1<0$, the ground state contains all the $\eta$-fermions with
$\omega_k<0$, cf.\ Eq.\ (\ref{hdiag}). In the presence of the trapping
potential (\ref{potential}), level crossings of the lowest states occur in the
$\bar\mu$-$l$ plane separating the regions with $N=k$
and $N=k+1$. This is essentially related to the
fact that the particle number is conserved, i.e., the particle number operator
$\hat{N}=\sum_i b_i^\dagger b_i$ commutes with the BH Hamiltonian (\ref{bhm});
thus the eigenvectors do not depend on $\mu$, even though the eigenvalues do.
Since, for $\mu<1$ and in the absence of the trap potential, the ground state
has a finite {\em density\/} $N/L>0$, and in the presence of the trap
$N$ is finite and increases as $N\sim l$,  the lowest states 
show an infinite number of level crossings as $l\to\infty$ (after
$L\to\infty$) where the gap $\Delta$ vanishes. Note that the hard-core limit,
$U\to \infty$ in Eq.~(\ref{bhm}), does not play any special role, so we
expect that level crossings at finite trap size are a general feature of the
BH model in the presence of a confining potential, when the homogeneous limit
of infinite trap size has a finite particle density.

In the following of this section we present results for the trap-size
dependence of several observables in the superfluid phase, for $|\mu|<1$.  As
we shall see, the above-mentioned level crossings and the competition of
smooth modes and modes at the Fermi momentum $k_F=\pi f$ 
give rise to peculiar modulated trap-size dependencies.

\subsection{Analytical results for the energy gap at small $\bm{|\bar\mu}|$}
\label{sect:superfluid,smallmu}

We can infer some information on the trap-size dependence of the energy gap at
$\bar{\mu}<0$ and small $|\bar\mu|$ by using the analytical calculations of
Sec.~\ref{tsslmu1}.  Let us introduce a few definitions.  $E^{(k)}_0(l)$ is
the lowest energy in the sector with $N=k$. $l_0^{(k)}$ is the value of $l$
such that $E^{(k)}_0(l)=E^{(k+1)}_0(l)$; Eq.\ (\ref{hdiag}) implies that this
is also the doubly-degenerate ground-state energy.  The gap shows peaks at
$l_{\rm peak}^{(k)}$ with $l_0^{(k)}<l_{\rm peak}^{(k)}<l_0^{(k+1)}$; $l_{\rm
  peak}^{(k)}$ is the value of $l$ such that $E^{(k-1)}_0(l)=E^{(k+1)}_0(l)$;
this is the energy of the doubly-degenerate first excited state in the sector
with $N=k$, where the energy of the ground state is $E^{(k)}_0$.

For $p=2$ Eq.\ (\ref{eq:p2eig}) tells us that, for large trap size, the gap
vanishes at
\begin{equation}
l_0^{(k)} \approx {\sqrt{2}\over |\bar{\mu}|}  (k+1/2),
\label{l0xxm}
\end{equation}
and therefore the interval between two zeroes is constant, i.e.,
\begin{equation}
l_0^{(k+1)}-l_0^{(k)} \approx {\sqrt{2}\over |\bar{\mu}|}.
\label{intpert}
\end{equation}
Moreover, the gap $\Delta = \min_k |\omega_k|$ has peaks at
\begin{equation}
l_{\rm peak}^{(k)} \approx {\sqrt{2}\over |\bar{\mu}|} (k+1),
\label{peakvalues}
\end{equation}
where the gap decreases as
\begin{equation}
\Delta_{{\rm peak}} \sim l^{-1}.
\label{deltapeakan}
\end{equation}
The product between the trap size and the gap, i.e., $l \Delta$, has a
periodic asymptotic behavior. Indeed, defining
\begin{equation}
\phi = {l-l_0^{(k)}\over l_0^{(k+1)}-l_0^{(k)}}, 
\quad {\rm for}\quad l_0^{(k)}\le l < l_0^{(k+1)},
\label{phildefmu0pert}
\end{equation}
thus $0\le \phi < 1$, we obtain a simple triangle-like form
\begin{eqnarray}
l \Delta = a \,t(\phi), \label{pertresde}
\end{eqnarray}
where $t(\phi)$ is the triangle function
\begin{eqnarray}
t(\phi)= 1/2 - |\phi-1/2|
\label{pertresde2}
\end{eqnarray}
and $a=\sqrt{2}$. Note that $l_{\rm peak}^{(k)}$ corresponds 
asymptotically to the value $\phi=1/2$.

More generally, for any $p$, the asymptotic behavior
(\ref{largekome}) for large $k$ of the eigensolutions of Eq.\
(\ref{trapscaleqxx}) implies
\begin{equation}
l_0^{(k)} \sim k |\bar{\mu}|^{-\rho}, \quad \rho = {p+2\over 2p}.
\label{l0xxm2}
\end{equation}
Thus, for any power $p$, the location of the zeroes increases linearly
with $k$, and therefore the intervals between subsequent zeroes
$l_0^{(k+1)}-l_0^{(k)}$ approaches a constant as in the $p=2$
case. This implies again that the peak of the gap decreases as
$l^{-1}$, and that the product $l\Delta$ is a periodic function of
$l$, and, specifically, of the corresponding phase-like variable 
$\phi$, cf.\ Eq.\ (\ref{phildefmu0pert}).  Moreover, asymptotically, $l
\Delta$ must have a triangle-like form for any $p$, analogous to
Eq.~(\ref{pertresde}). Notable values are
\begin{eqnarray}
&l_0^{(k)} \approx \gamma_4 |\bar{\mu}|^{-3/4} (k+1/2) 
\quad&(p=4),\nonumber \\
&l_0^{(k)} \approx \pi 2^{-3/2}|\bar{\mu}|^{-1/2} (k+1) \quad&(p=\infty),
\label{eq:l0-p4,inf}
\end{eqnarray}
and
\begin{eqnarray}
&a = \txf43 \gamma_4 |\bar{\mu}|^{1/4} \quad&(p=4),\nonumber \\
&a = \pi\sqrt{|\bar{\mu}|/2} \quad&(p=\infty).
\label{eq:a-p4,inf}
\end{eqnarray}

The above results suggest that the gap vanishes with a global power scaling
$\Delta\sim l^{-1}$. However, its amplitude in not a constant, but a periodic
function of the trap size.  This scenario will be confirmed by
the results from numerical diagonalization at fixed $\mu<1$.

\subsection{Results from numerical diagonalization}
\label{muintp24}

We now present numerical results at fixed $l$ and $\mu$, obtained by
numerical diagonalization of the quadratic Hamiltonian
(\ref{fermmod}), see App.~\ref{Sec:numerics}.  The results at fixed
trap size are essentially correct up to machine precision.

\subsubsection{Interval between level crossings}
\label{intlc}

The numerical results show that the interval between two level crossing
approaches a constant value in the large trap-size limit. They turn out to 
fit the simple Ansatz
\begin{equation}
Q_l^{(k)} \equiv l_0^{(k+1)}-l_0^{(k)} = Q_l^* + c_2 l^{-2} + c_3 l^{-3} + ...,
\label{ql}
\end{equation}
for sufficiently large trap size, $l\gtrsim 10$, which allows us to get
accurate estimates of the large-$l$ limit $Q_l^*$.  We report a
selection of results in Table \ref{tab:someres}.
\begin{table}
\caption{Numerical results in the superfluid region
for the XX model and several values of $\mu$ and $p$, for the
asymptotic interval $Q_l^*$ between level crossings,
the constant $a$ entering the asymptotic
behavior (\ref{eq:Delta,mu0}) of the gap,
the constant $b$ appearing in Eq.~(\ref{eq:rho0,mu0}) 
of the particle density in the middle of trap, and the
amplitude of the entanglement length scale $a_e$, cf.\ Eq.~(\ref{xie}).
The reported estimates are obtained by extrapolating data at finite trap size;
their uncertainty is at most one on the last figure.}
\label{tab:someres}
\begin{ruledtabular}
\begin{tabular}{clllll}
$\mu$& $p$  &
\multicolumn{1}{c}{$Q_l^*$}&
\multicolumn{1}{c}{$a$}&
\multicolumn{1}{c}{$b$}&
\multicolumn{1}{c}{$a_e$}\\
\colrule
$-1/2$ & 2 & 0.8265187 & 1.03010681  & 0.378618  & 2.64118  \\
$-1/2$ & 4 & 0.8021659 & 1.19898955  & 0.440691  & 2.26916  \\
   0   & 2 & 1.3110287 & 1.198140234 & 0.3813798 & 2.622058 \\
   0   & 4 & 1.1635926 & 1.349953898 & 0.4297036 & 2.327185 \\
  1/2  & 2 & 2.7337528 & 1.3177760   & 0.484352  & 2.06462  \\
  1/2  & 4 & 2.0551051 & 1.29539790  & 0.476126  & 2.10028  \\
\end{tabular}
\end{ruledtabular}
\end{table}

Note that the results for $p=2$ are quite close to, and clearly approach, the
small-$\bar\mu$ estimate (\ref{intpert}), i.e.,
$Q_l^*\approx\sqrt{2}/|\bar{\mu}|$.  In the limit $p\to\infty$ the value of
$Q_l^*$ should converge toward the corresponding value in the finite-size
behavior of the homogeneous system with open boundary conditions, see
App.~\ref{muintpinf}. Thus
\begin{equation}
\lim_{p\to\infty} Q_l^* = {1\over 2f},
\label{pinfql}
\end{equation} 
where $f$ is the filling factor given in Eq.~(\ref{eq:fvsmu}).

The asymptotic values $Q_l^*$ can also be computed using the LDA of the total
particle number, cf.\ Eq.~(\ref{ntlda}).  Since the level crossings occur at
the boundary between the regions with $N=k$ and $N=k+1$, asymptotically we
have $k\approx N_{\rm lda} = c(\mu) l$, thus
\begin{equation}
Q_l^*=1/c(\mu), 
\label{qlstar}
\end{equation}
whose numerical values coincide with the estimates of $Q_l^*$ obtained above,
see Table \ref{tab:someres}.  This is actually a further evidence, beside the
results of Sect.\ \ref{ldasec}, that the LDA of the particle number is
asymptotically exact.

\subsubsection{Trap-size dependence of the gap}
\label{gapsufl}

For any $|\mu|<1$ and power of the confining potential,
the results for the gap show the behavior
\begin{eqnarray}
&&\Delta = l^{-1} A_\Delta(\phi) + O(l^{-2}),\label{eq:Delta,mu0}\\
&&A_\Delta(\phi)=a\,t(\phi), \nonumber
\end{eqnarray}
with $t(\phi)$ given by Eq.\ (\ref{pertresde2}), and the
coefficient $a$ depends on $p$ and $\mu$. We determine the single parameter
$a$ by fitting the peak values to the form
\begin{equation}
\Delta_{\rm peak} = \half a l^{-1} + c l^{-3}
\label{eq:Delta,p=2,mu=0}
\end{equation}
(the term $l^{-2}$ is absent just for this quantity); the fit quality is
usually excellent for $l\gtrsim 10$.  We report some results for the constant
$a$ in Table \ref{tab:someres}.  Note that the values for $p=2$ appear to
approach the small-$|\bar\mu|$ estimate $a=\sqrt{2}$ with increasing $\mu$.
The value of $\phi$ at the peak of the gap converges as 
$\phi_{\rm peak} = 1/2 + c l^{-1} + O(l^{-2})$ with $c\ll 1$, of the
order of $10^{-2}$.

We note that the modulated trap-size dependence of the gap within the
superfluid region $|\mu|<1$ appears to be largely universal, being
substantially independent of $\mu$ and $p$, apart from a trivial normalization.

This is also confirmed by the finite-size behavior of the homogeneous XX chain
corresponding to the $p\to\infty$ limit of the confining potential, where the
model becomes equivalent to a homogeneous chain of size $L=2l$ with open
boundary conditions (more precisely, the $p\to\infty$ limit corresponds to a
chain with $L=2\lfloor l\rfloor +1$ when the center of the trap coincides with
the middle site of the chain, and $L=2\lfloor l\rfloor$ when the center is in
the middle between two sites).  The results of App.~\ref{muintpinf} show that
the modulated trap-size behavior of the gap found at finite values of $p$
persists in the $p=\infty$ limit.

\subsubsection{The particle density}
\label{pdsufl}

\begin{figure}[tbp]
\includegraphics*[scale=\graphicscale]{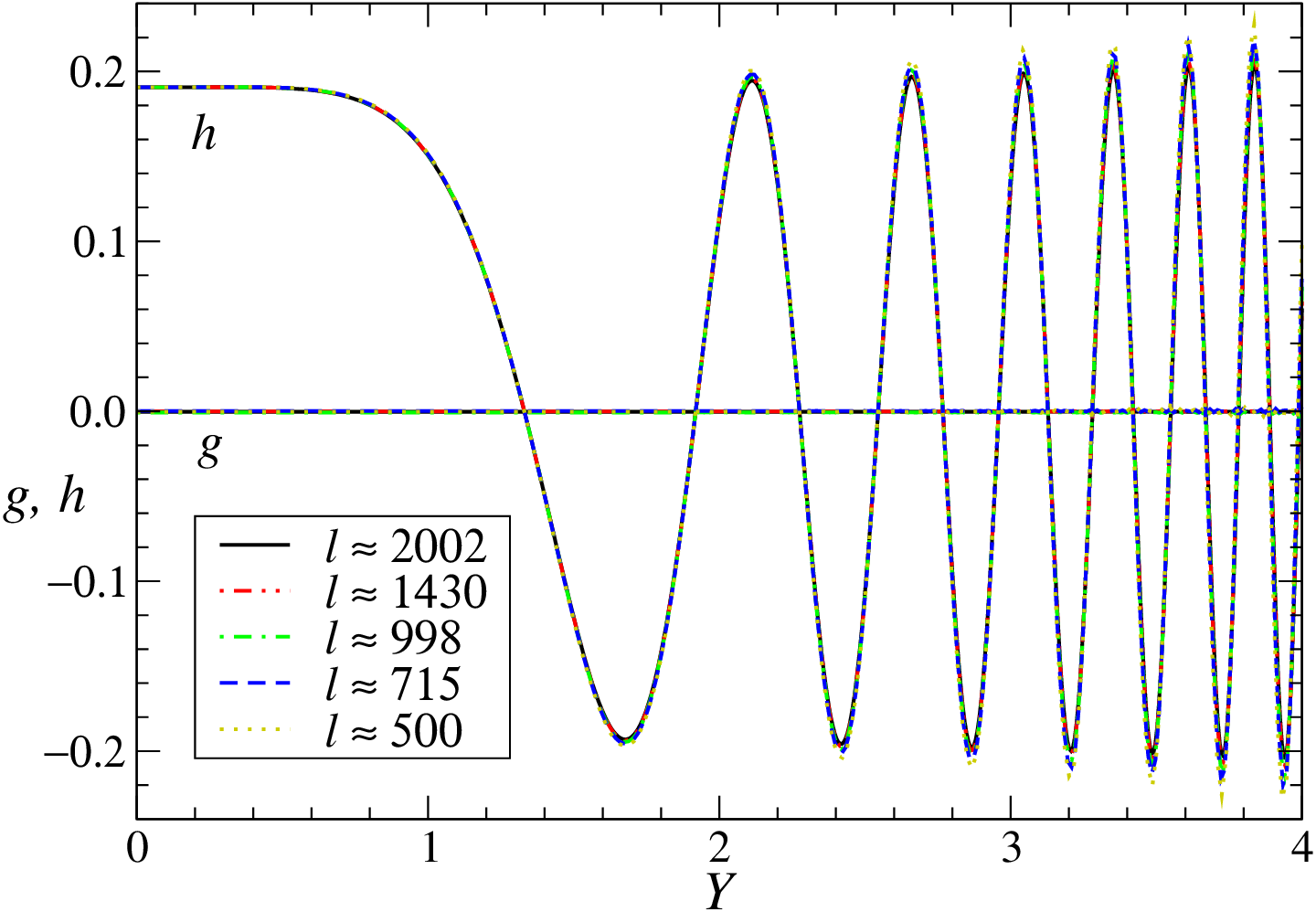}
\includegraphics*[scale=\graphicscale]{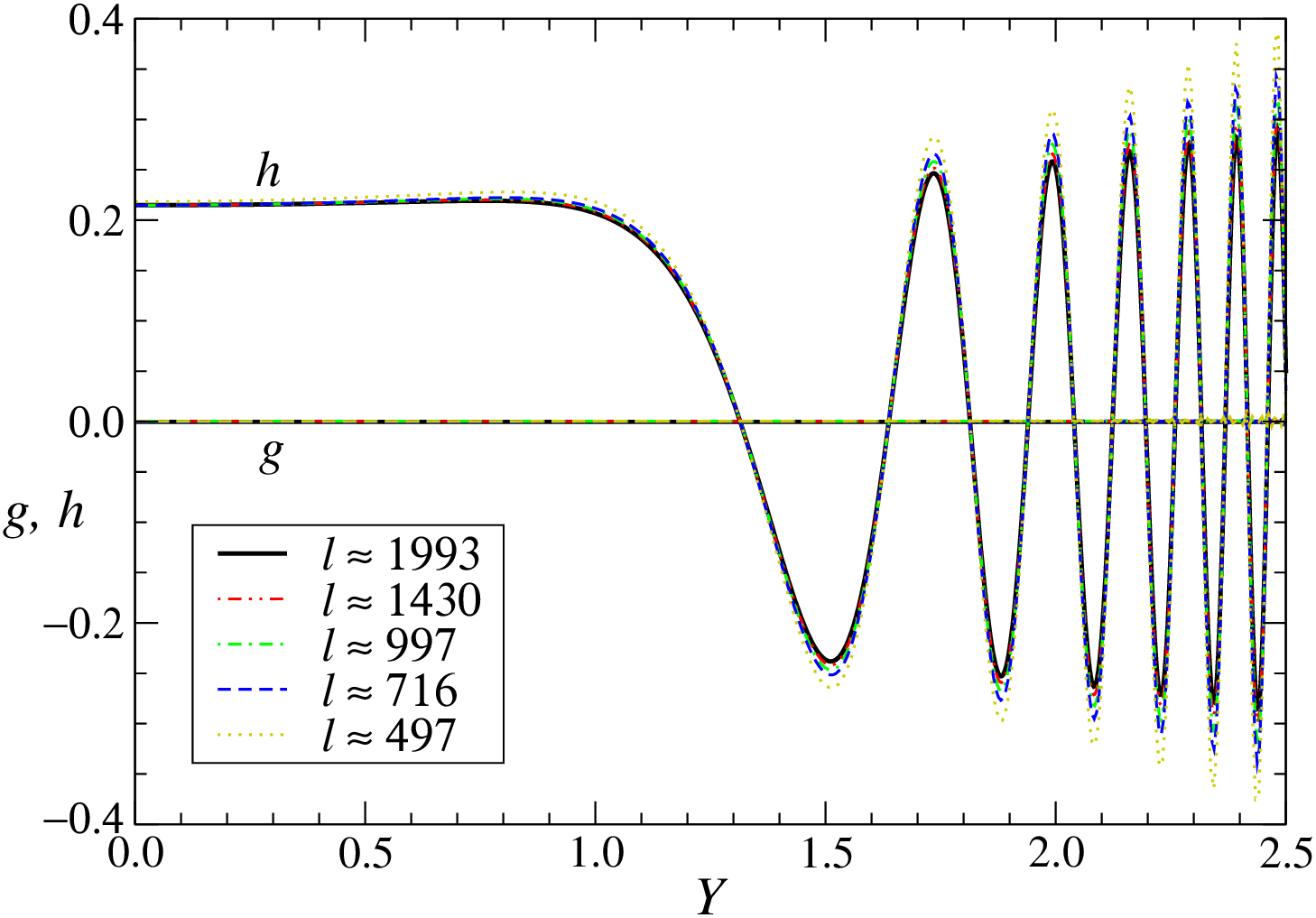}
\caption{
  The functions $h$ and $g$, cf.\ Eq.~(\ref{eq:rhox,sf}), vs.\
  $Y=l^{-p/(p+1)} x$ at odd peaks of $\Delta$ at $\mu=0$, for $p=2$ (above)
  and $p=4$ (below).  Results for $p=6$ are of the same quality.  $h$ is real,
  since $k_F=\pi/2$.  $g$ at peaks is compatible with 0.  $h$ at even peaks is
  opposite in sign to $h$ at odd peaks.  The data at different trap sizes
  clearly approach TSS functions when increasing $l$.  Here and in the
  following, in the legends $l$ is rounded to the nearest integer for the sake
  of presentation.}
\label{fig:rho,mu=0,p}
\end{figure}

\begin{figure}[tbp]
\includegraphics*[scale=\graphicscale]{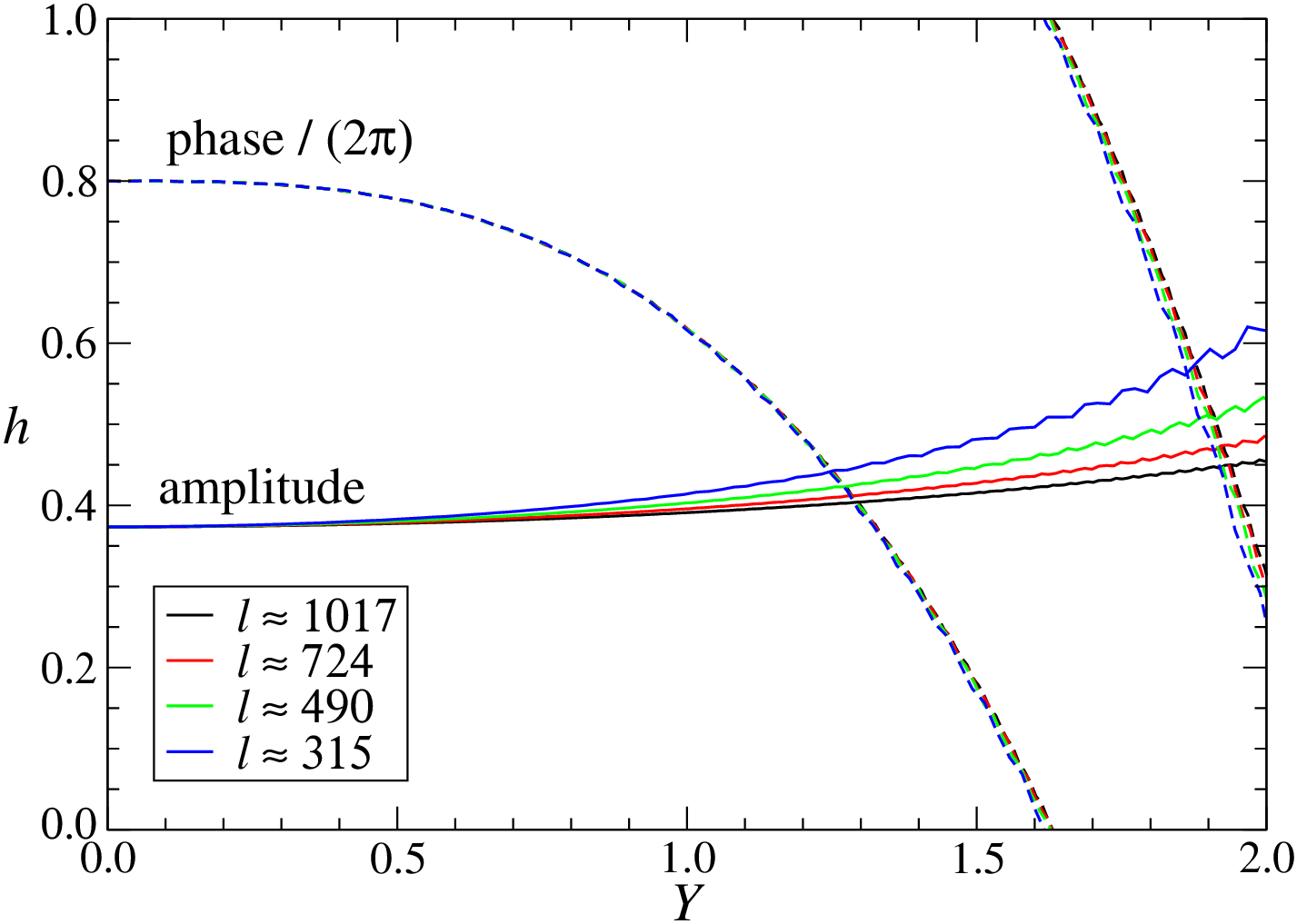}
\caption{
  The functions $h$ (solid lines) and $g$ (dashed lines), cf.\ 
  Eq.~(\ref{eq:rhox,sf}), vs.\ $Y=l^{-2/3} x$ at odd peaks of $\Delta$ for
  $p=2$ and $\mu=\cos\pi/5$ ($f=1/5$).  $g$ at peaks is compatible with 0.
  $h$ at even peaks is opposite in sign to $h$ at odd peaks.  The ``noise''
  more evident at smaller $l$ is due to the ambiguity of disentangling a
  ``fast'' oscillation $e^{2ik_Fx}$ from a ``slow'' oscillation $h(Y)/|h(Y)|$
  in Eq.\ (\ref{eq:rhox,sf}).}
\label{fig:rho,mu=0.809017,p=2}
\end{figure}

The particle density $\langle n_0\rangle$ at the origin shows a modulated
asymptotic behavior as well, but with a period twice the period of $\Delta$;
therefore it is useful to define
\begin{equation}
\bar\phi = 2 {l-l_0^{(2k)}\over  l_0^{(2k+2)} - l_0^{(2k)}}
\quad {\rm for}\ l_0^{(2k)}\le l < l_0^{(2k+2)};
\label{barphildef}
\end{equation}
thus $0\le \bar\phi < 2$ (note that either $\bar\phi=\phi$ or
$\bar\phi=\phi+1$ for large $l$).  The results of the numerical
diagonalization show that $\langle n_0\rangle$ satisfies the asymptotic
behavior
\begin{eqnarray}
&&\langle n_0\rangle - f = l^{-1} A_n(\bar\phi) + O(l^{-2}),
  \label{eq:rho0,mu0} \\
&&A_n(\bar\phi) = b(1-\bar\phi),
\nonumber
\end{eqnarray}
where $f$ is the filling factor of the homogeneous model without trap, cf.\
Eq.~(\ref{eq:fvsmu}).  Some results for the constant $b$ are reported in Table
\ref{tab:someres}.  Note that the $O(l^{-1})$ term has zero average over the
period.

An analogous behavior of the particle density in the middle of trap is found
in the limit $p\to\infty$, as shown by the results of App.~\ref{muintpinf} for
the finite-size behavior of the homogeneous XX chain with open boundary
conditions.

The particle density as a function of the distance $x$ from the middle of the
trap turns out to behave as
\begin{eqnarray}
&&\langle n_x\rangle \approx 
\rho_{\rm lda}(X) + l^{-1} \Re\Bigl\{h(Y,\bar\phi) e^{2ik_Fx} +
g(Y,\bar\phi)\Bigr\} \nonumber \\
&&X = x/l, \qquad Y=x l^{-p/(p+1)}, \label{eq:rhox,sf}
\end{eqnarray}
where $k_F=\pi f=\arccos\mu$, $g$ is real, and terms suppressed by higher
powers of $l^{-1}$ are neglected.  $g$ and $h$ are discontinuous at
$\bar\phi=1$.  Note that two scaling variables $X$ and $Y$ appear in the above
equation, distinguishing the scaling behavior of the two terms. Note also that
the leading term depending on $X$ is the LDA of the particle
density, cf.\ Eq.\ (\ref{nxlda}).  Some results are plotted in Figs.\ 
\ref{fig:rho,mu=0,p} and \ref{fig:rho,mu=0.809017,p=2}.  We find that
$g(Y)=0$ at peaks of $\Delta$ ($\phi=1/2,\,3/2$).

\subsubsection{The particle density correlation}
\label{dcsufl}

\begin{figure}[tbp]
\includegraphics*[scale=\graphicscale]{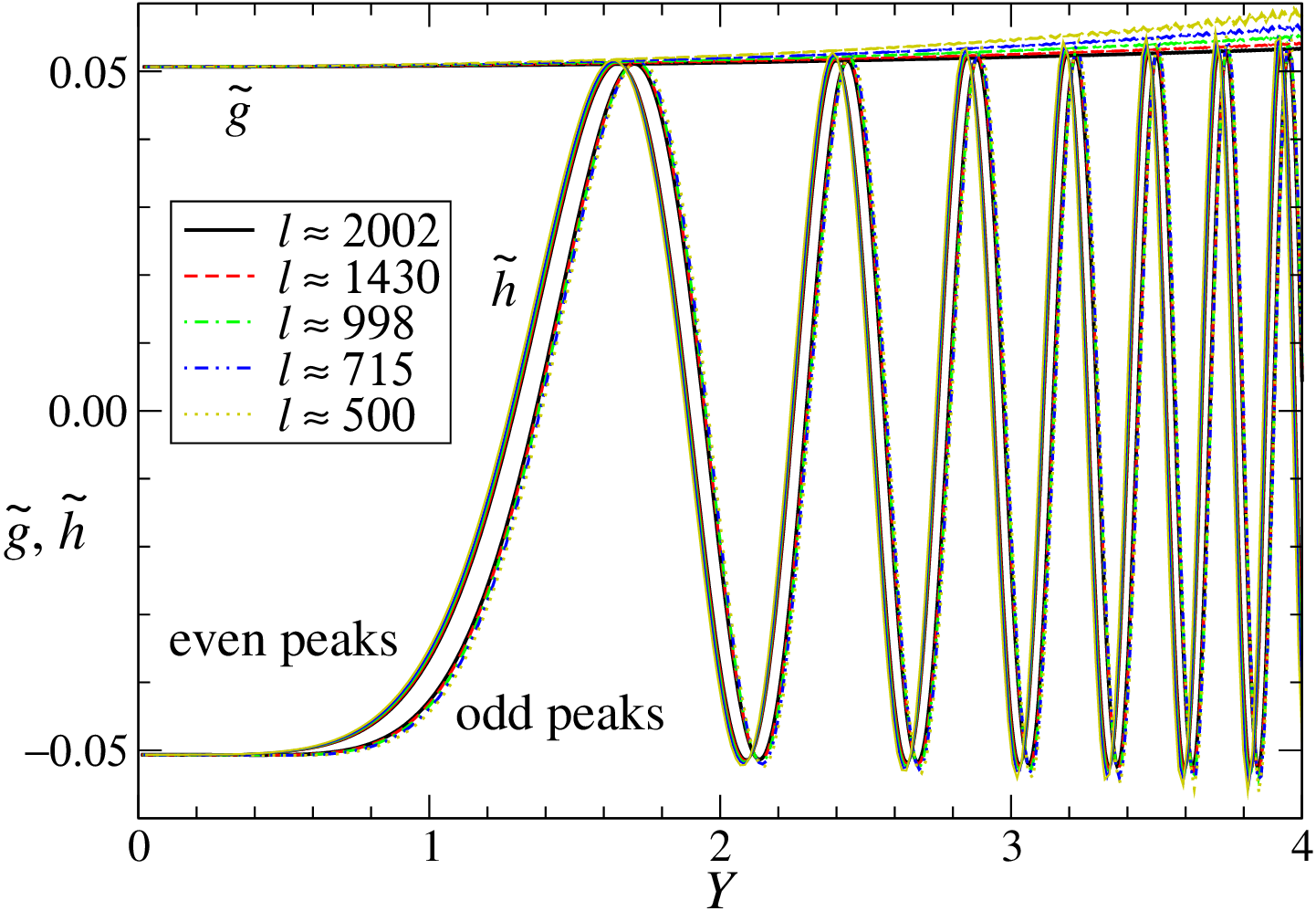}
\caption{
The functions $\tilde h$ and $\tilde g$, cf.\ Eq.~(\ref{eq:Gn,mu=0}),
  vs.\ $Y=l^{-2/3} x$ at peaks of $\Delta$ for $p=2$ and $\mu=0$; $\tilde h$
  is real, since $k_F=\pi/2$.  The data at different trap sizes clearly
  approach TSS functions when increasing $l$.}
\label{fig:rhocc,mu=0,p=2}
\end{figure}

\begin{figure}[tbp]
\includegraphics*[scale=\graphicscale]{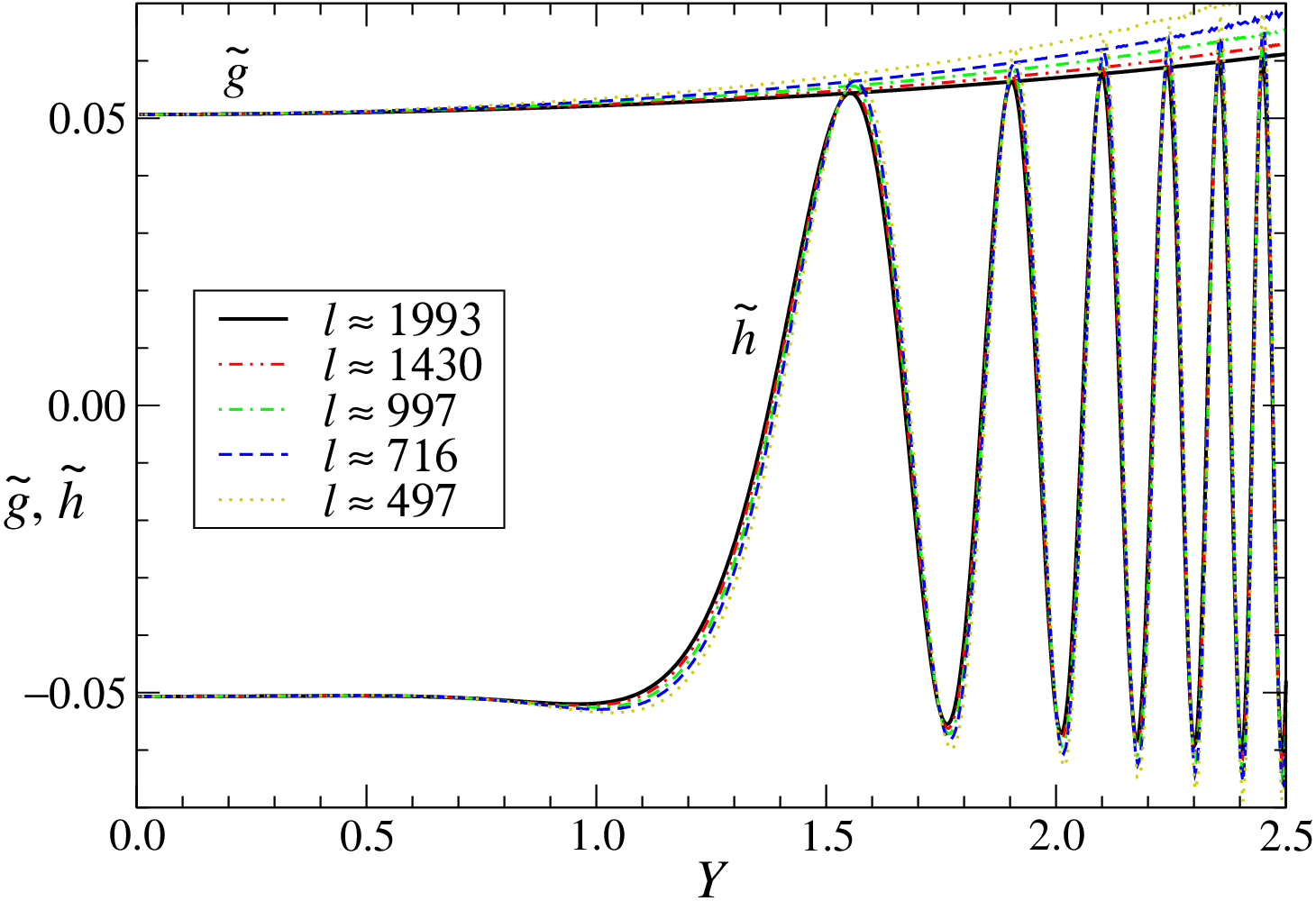}
\caption{The functions
$\tilde h$ (solid lines) and $\tilde g$ (dashed lines),
cf.\ Eq.~(\ref{eq:Gn,mu=0}), 
vs.\ $Y=l^{-p/(p+1)} x$ at $\mu=0$ and at odd
peaks of $\Delta$ for $p=4$.
Results for $p=6$ are of the same quality.
}
\label{fig:rhocc,mu=0,p=4and6}
\end{figure}

\begin{figure}[tbp]
\includegraphics*[scale=\graphicscale]{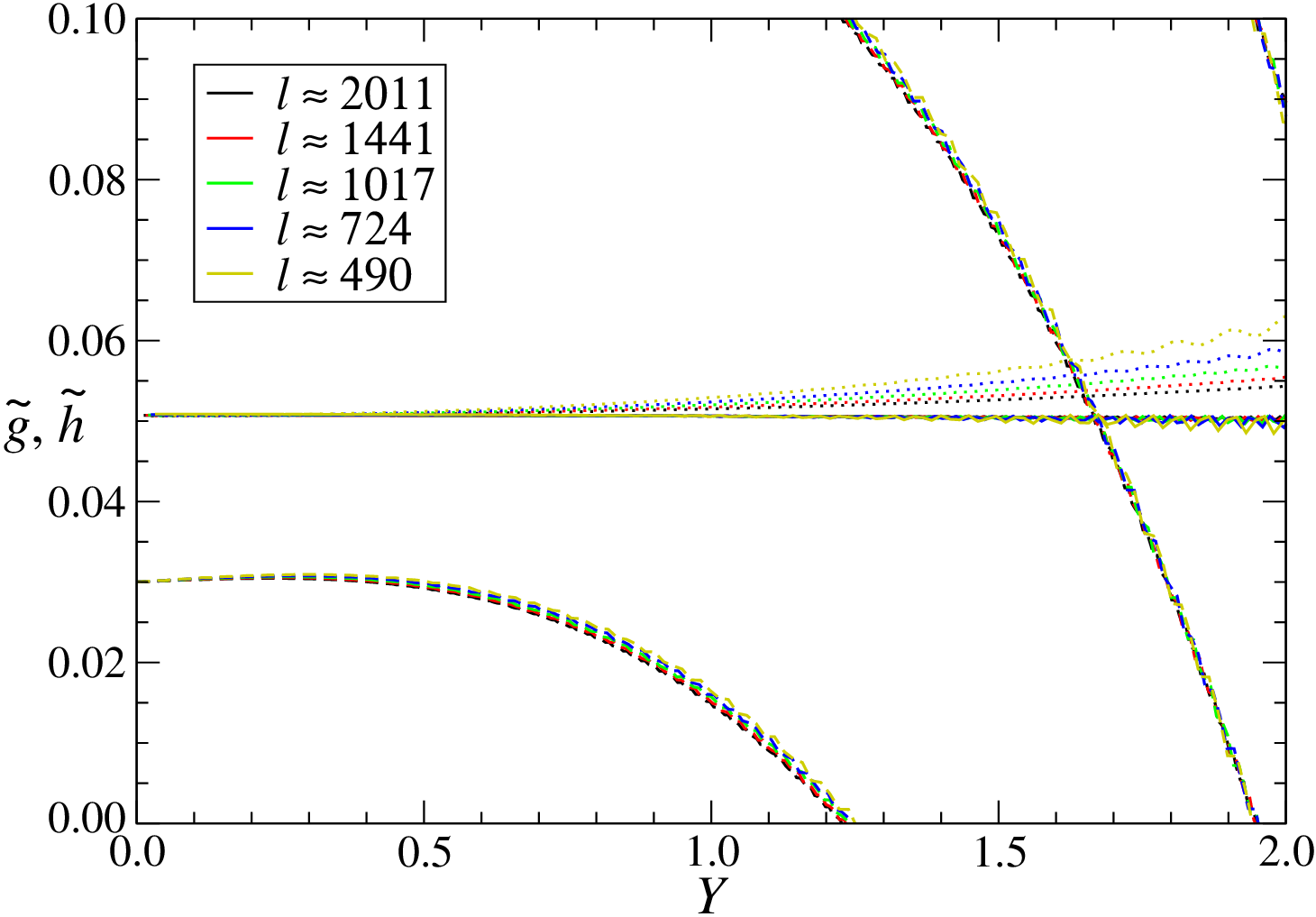}
\caption{
The functions $\tilde h$ (solid lines: amplitude; dashed lines:
phase$/(20\pi)$) and $\tilde g$ (dotted lines) vs.\ $Y=l^{-2/3} x$ at odd peaks
of $\Delta$ for $p=2$ and $\mu=\cos\pi/5$ ($f=1/5$).
}
\label{fig:rhocc,mu=0.809017,p=2}
\end{figure}

Results for $p=2$, $4$, and $6$ show also that the connected density
correlation scales as
\begin{eqnarray}
G_n(x,0)\approx l^{-2p/(p+1)} 
\Re\Bigl\{\tilde h(Y,\bar\phi) e^{2ik_Fx} + \tilde g(Y,\bar\phi)\Bigr\},
\label{eq:Gn,mu=0}
\end{eqnarray}
where $Y = x l^{-p/(p+1)}$, and $\tilde g$ is real.  $\tilde h$ and 
$\tilde g$ are discontinuous at $\bar\phi=1$.  Results are shown in Figs.\
\ref{fig:rhocc,mu=0,p=2} and \ref{fig:rhocc,mu=0,p=4and6} at $\mu=0$ and
$p=2,\,4$ respectively, and in Fig.~\ref{fig:rhocc,mu=0.809017,p=2} at
$\mu={\rm cos}\pi/5$ and $p=2$, at the peaks of the gap.  We find that
$\tilde g(Y)=|\tilde h(Y)|$ at peaks of $\Delta$ ($\phi=1/2,\,3/2$).

\subsubsection{The one-particle density matrix}
\label{opdmsufl}

Another interesting quantity is the one-particle density matrix (\ref{gij}).
In the homogeneous system without trap 
\begin{equation}
G_b(x_i,x_j) \sim |x_i-x_j|^{-1/2},
\label{gmu0}
\end{equation}
from which we can read the RG dimension of the bosonic operator $b$ in the
superfluid phase, i.e., $y_b=1/4$.  This behavior is also observed in the
presence of the trap for sufficiently small distances~\cite{RM-04}.
Fig.~\ref{fig:cxx,p=2} shows results for $p=2$ and
$\mu=0$: the data of $x^{1/2}G_b(x,0)$ for different $l$s appear to collapse
to a unique curve when plotted versus $x/l$, apart from small oscillations
with a decreasing amplitude for $l\to\infty$. Only the oscillations depend on
$\phi$.  For $x\gtrsim l$, $G_b(x,0)$ is exponentially suppressed.  These
results indicate an asymptotic scaling behavior given by
\begin{equation}
G_b(x,0)\approx l^{-1/2} g(x/l).
\label{gbx02}
\end{equation}
\begin{figure}[tbp]
\begin{center}
\leavevmode
\includegraphics*[scale=\graphicscale]{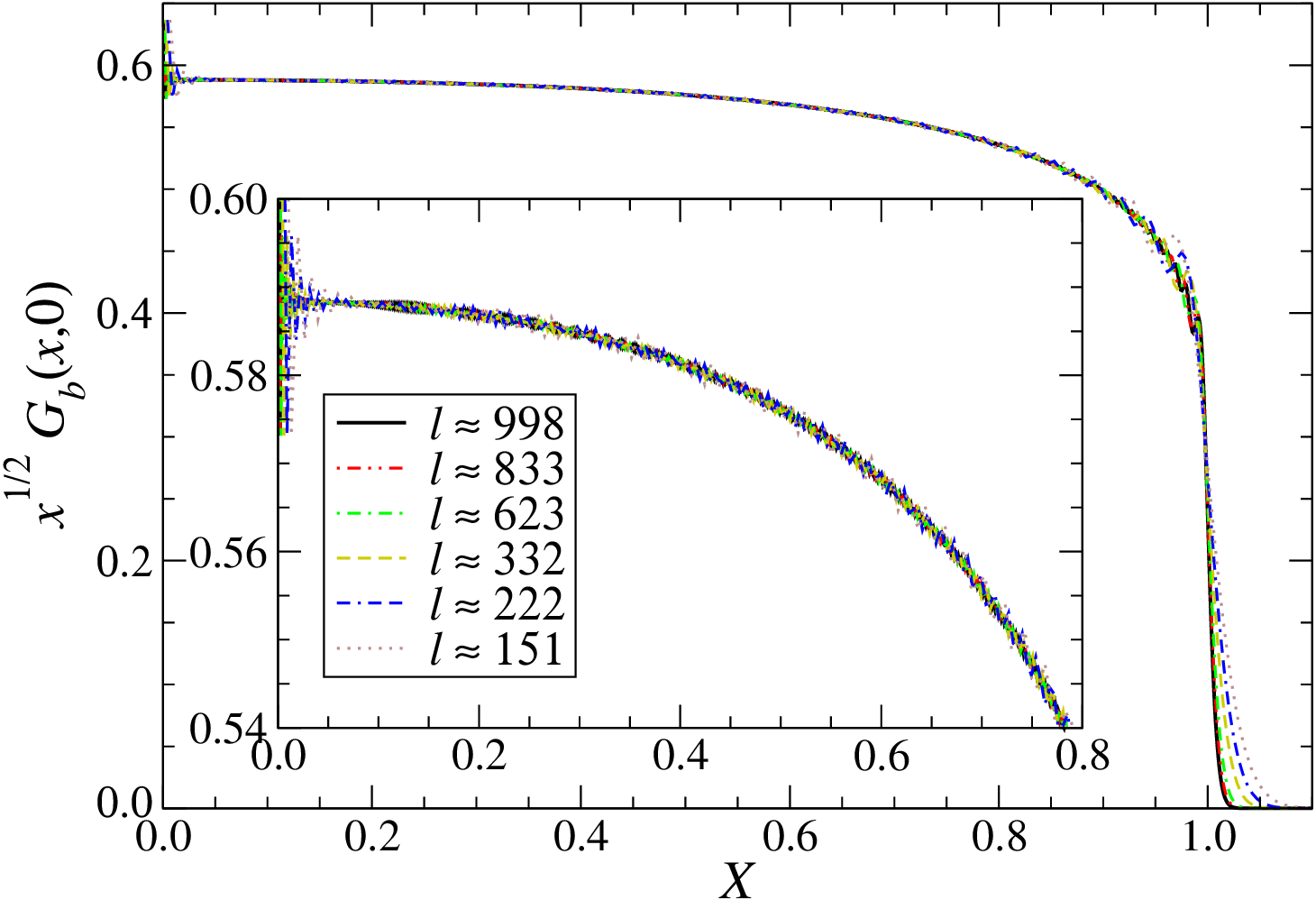}
\caption{
$x^{1/2}G_b(x,0)$ at peaks of $\Delta$ vs.\ $x/l$ for $p=2$
and $\mu=0$.
}
\label{fig:cxx,p=2}
\end{center}
\end{figure}

The region where $G_b(x,0)$ appears to rapidly vanish, i.e., for $x/l\approx1$
in Fig.~\ref{fig:cxx,p=2}, corresponds to the region where $\mu_{\rm eff}
\approx 1$, cf.\ Eq.~(\ref{mueff}), which is the value of the chemical
potential corresponding to the superfluid to empty state transition, where the
particle density of the ground state vanishes.  We thus expect that, for
generic values of $\mu$ and $p$, the region around $x=x_c=l(1-\mu)^{1/p}$,
where $\mu_{\rm eff}(x_c)=1$, develops critical modes related to a low-density
Mott transition.  The effective chemical potential can be expanded around
$x_c$ as
\begin{equation}
\mu_{\rm eff}=\mu+(x/l)^p 
= 1 + (1-\mu)p{x-x_c\over l} + O[(x-x_c)^2].
\label{linV}
\end{equation}
Thus, the behavior around $x_c$ is essentially analogous to that arising at
the low-density Mott transition $\mu=1$ in the presence of a linear potential
$V_l\sim r/l$.  Around $x_c$, critical modes should appear with length scale
$\xi\sim l^\sigma$, where $\sigma$ is the exponent associated with a linear
external potential.  The value of $\sigma$ can be inferred by RG arguments
analogous to those leading the determination of the trap exponent $\theta$ at
the low-density Mott transition~\cite{CV-10}, which give
$\sigma=1/3$.\footnote{ The exponent $\sigma$ can be determined by a RG
  analysis of the perturbation corresponding to a linear potential
  $V_l(x)=ux$, i.e., $\int d^dx\,dt\, V_l(x) |\phi(x)|^2$, at the fixed point
  of the continuous theory describing the Mott transition~\cite{FWGF-89}.  The
  exponent $\sigma$ is related to the RG dimension $y_u$ of the parameter $u$,
  which can be obtained from the relations $y_u - 1 = d+z-y_{|\phi|^2} =
  y_\mu=2$, thus $y_u=3$, and therefore $\sigma\equiv 1/y_u=1/3$ for $d=1$ and
  $d=2$.}  We thus expect that the transition region around $x=x_c$ enlarges
as
\begin{equation}
\Delta x \sim l^{1/3}, 
\label{enleq}
\end{equation}
independently of the power-law $p$ of the confining potential.  We study this
phenomenon numerically by computing $G_b(x,0)$ for $l$s corresponding to odd
peaks of $\Delta$; we take 
\begin{equation}
\Delta x\equiv l-x_{\rm max}, 
\label{deltaxmax}
\end{equation}
where $x_{\rm max}$ is the abscissa of the rightmost maximum of
$x^{1/2}G_b(x,0)$; the results agree very well with $\Delta x \sim l^{1/3}$,
see Fig.~\ref{fig:Deltax,mu=0}.
\begin{figure}[tbp]        
\begin{center}
\leavevmode
\includegraphics*[scale=\graphicscale]{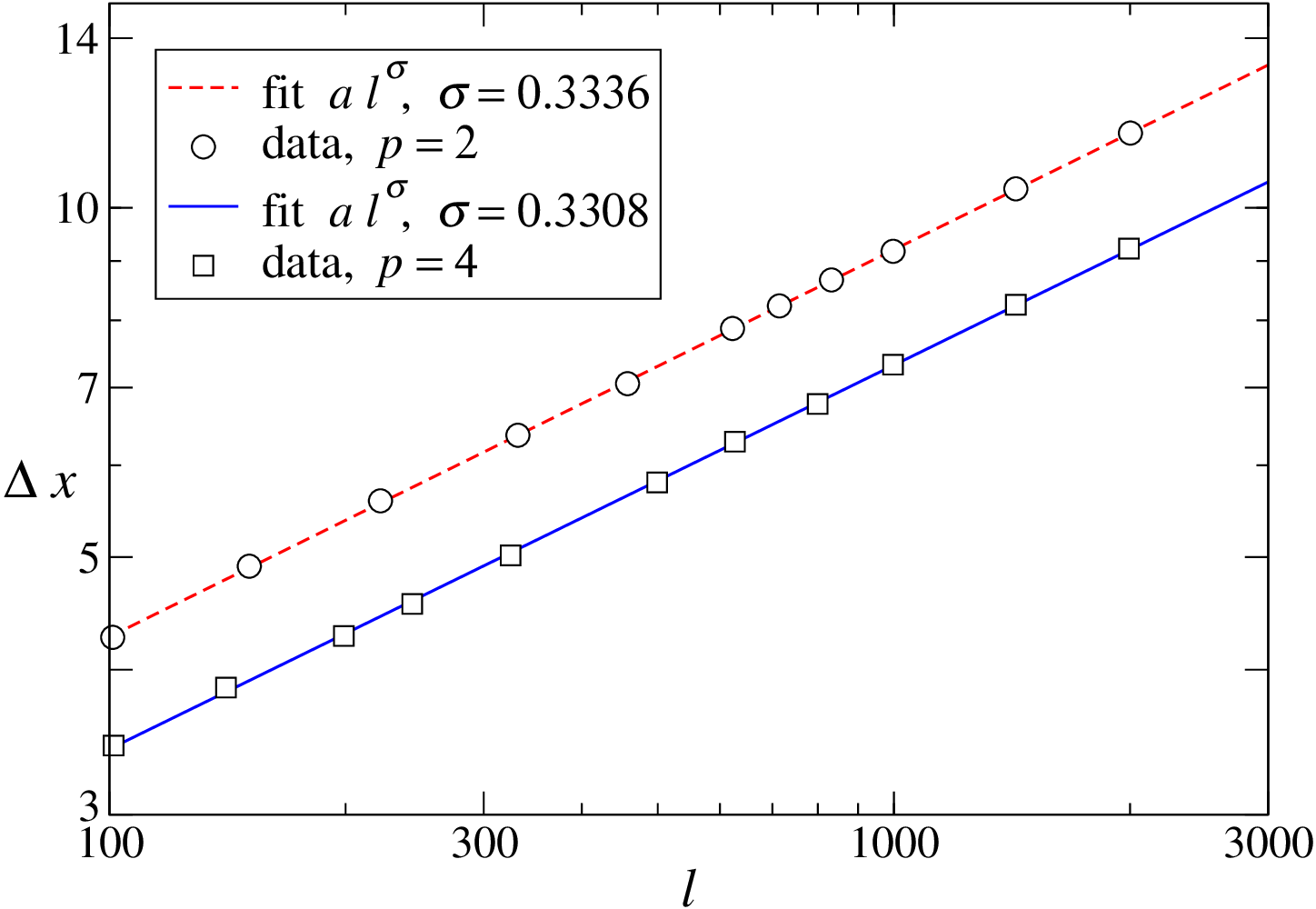}
\caption{$\Delta x$, defined by Eq.~(\ref{deltaxmax}), vs.\ $l$
  for $p=2,\,4$ and $\mu=0$.  The full lines shows a fit to $\Delta x = a
  l^{\sigma}$ leaving free the exponent $\sigma$.  The resulting estimates of
  $\sigma$ are in perfect agreement with the RG prediction $\sigma=1/3$
  independently of $p$.}
\label{fig:Deltax,mu=0}
\end{center}
\end{figure}

\subsection{Quantum entanglement in the superfluid phase}
\label{entanglementsuperfluid}

We divide the chain in two parts of length $l_A$ and $L-l_A$, and consider the
von Neumann entanglement entropy $S(l_A;L)$ for open boundary conditions.  In
the absence of the trap and for open boundary conditions, the von Neumann
entanglement entropy is essentially determined by the conformal field theory
which describes its continuum limit~\cite{CC-04}, 
\begin{equation}
S(l_A;L) \approx {c\over 6} \ln \left[L 
\sin( \pi l_A/L)\right] + E(\mu),
\label{ccfo}
\end{equation}
where $c=1$ is the central charge corresponding to a relativistic free boson
field theory. The $O(1)$ term, $E(\mu)$, depends on $\mu$; it is exactly
known, see Eq.~(\ref{emufunc}) (notably, $E(0)=0.287769699...$).  We
consider specifically the half-lattice von Neumann entanglement entropy, 
\begin{equation}
S(L/2;L) = {1\over6} \ln L + E(\mu) + O(1/L).
\label{eq:Se-E2}
\end{equation}
As shown in App.~\ref{muintpinf}, the amplitude of the $O(1/L)$ correction is
generally modulated by a function of $\bar\phi \equiv 2\{[(L+1)f+1]/2\}$
(where $\{x\}\equiv x - \lfloor x\rfloor$ is the fractional part of $x$),
giving rise to peculiar oscillations.

In the presence of a trapping potential, and for $L\to\infty$, our numerical
results show the behavior
\begin{eqnarray}
S_{1/2}&= &\lim_{L\to\infty} S(L/2;L) \label{ccfot}\\
&=&{1\over 6} \ln [a_e(\mu) \,l]  + E(\mu) + A_o(\mu,\bar\phi)/l  +
O(1/l^2)
\nonumber
\end{eqnarray}
for any $p$, where $E(\mu)$ is the same constant in the absence of the trap,
cf.\ Eq.~(\ref{eq:Se-E2}).  Eq.\ (\ref{ccfot}) defines an entanglement length
scale~\cite{CV-10}
\begin{equation}
\xi_e\equiv a_e \,l.
\label{xie}
\end{equation}
Some results for the amplitude $a_e$ are reported in Table \ref{tab:someres}.
For $p\to\infty$ we recover Eq.~(\ref{eq:Se-E2}) with $L=2l$, thus
\begin{equation}
\lim_{p\to\infty} a_e(\mu) = 2.
\label{aepinf}
\end{equation}
The amplitude $A_o$ of the $O(1/l)$ term turns out to be a periodic function
of the trap size, through the dependence on $\bar\phi$, i.e., the phase-like
variable $\bar\phi$ defined in Eq.~(\ref{barphildef}).  An analogous behavior
is found in the limit $p\to\infty$, i.e., in the case of a homogeneous system
of size $L$ with open boundary conditions, see App.~\ref{muintpinf}.

We also mention that similar subleading oscillations are
observed in the half-lattice entanglement entropy of the XX model with
gradients~\cite{EIP-09}, i.e., in the presence of a linear external field.

\subsection{Some notable relations}
\label{pecmu0}

We have already shown that in the gapless superfluid phase the asymptotic
modulated power-law behavior of the gap and the particle density in the middle
of the trap is largely universal, being independent of $\mu$ and $p$, apart
from trivial normalizations.  In the following we show that also the
amplitudes for different values of $p$ are strictly related; they can be
derived from the $p=\infty$ limit which corresponds to the homogeneous system
with open boundary conditions, see App.~\ref{muintpinf}.

We note that, using the entanglement definition (\ref{xie}) of length scale,
for any $|\mu|<1$ and any $p$ including $p\to\infty$, the asymptotic behavior
of the gap and the particle density can be written as
\begin{eqnarray}
&&\Delta \approx {\pi\sqrt{1-\mu^2}\over  \xi_e} t(\phi),
\label{deltamu0anyp}\\
&&\langle n_0\rangle - f \approx {1-\bar\phi\over \xi_e},
\label{n0mu0anyp}
\end{eqnarray}
with $\phi$ and $\bar\phi$ defined in Eqs.~(\ref{phildefmu0pert}) and
(\ref{barphildef}) respectively.  Indeed, one can check that, within the high
accuracy of our numerical estimates, the results reported in Table
\ref{tab:someres} satisfy the relations
\begin{eqnarray}
&&a(\mu;p)={\pi\sqrt{1-\mu^2}\over a_e(\mu;p)}, \label{aabrel}\\
&&b(\mu;p)={1\over a_e(\mu;p)}. 
\end{eqnarray}
Moreover, Eqs.~(\ref{deltamu0anyp}) and (\ref{n0mu0anyp}) reproduce the
results for the homogeneous system with open boundary conditions, see
App.~\ref{muintpinf}, by replacing the entanglement length scale $\xi_e$ with
$L+1$, with $\phi$ and $\bar\phi$ given by the corresponding expressions
(\ref{phideffss}) and (\ref{barphifss}) respectively.

These results provide a strong numerical evidence of the following statement:
in the superfluid phase the asymptotic trap-size dependence of smooth
observables can be obtained by replacing $L$ (or, more precisely $L+1$) with
$\xi_e$ in the asymptotic behavior of the homogeneous system of size $L$ with
open boundary conditions.

Another notable numerical relation is found at $\mu=0$: the numerical data of
the half-lattice entanglement $S_{1/2}$ in the presence of the trap indicates
that the entanglement length scale $\xi_e$, cf.\ Eq.~(\ref{xie}), is exactly
given by
\begin{equation}
\xi_e = 2 Q_l^* l, 
\label{xieql}
\end{equation}
for any $p$, where $Q_l^*$ is the asymptotic periodicity of the level
crossings, cf.\ Eq.~(\ref{ql}), which is exactly derived from the LDA of the
total particle number, i.e., from the relation
\begin{equation}
1/Q_l^* = {2\over\pi} \int_0^1 dx\,\arccos x^p= 
{2 p \Gamma\left({1+p\over 2p}\right) \over  
 \sqrt{\pi}\Gamma\left({1\over 2p}\right)} .
\label{qlm1}
\end{equation}
We are quite confident that Eq.~(\ref{xieql}) holds, since the numbers
reported in Table~\ref{tab:someres} show that it is verified within the
numerical accuracy of our estimates of the amplitude $a_e$ for $p=2$ and
$p=4$, which is $O(10^{-7})$. Moreover, it correctly reproduces the
$p\to\infty$ limit $\xi_e=2l$.

In addition, we find that at $\mu=0$ the behavior of the half-lattice
entanglement $S_{1/2}$ in the presence of the trap turns out to be consistent
with the following formula
\begin{equation}
S_{1/2} =  {1\over 6} \ln \xi_e  + E(0) + {\pi\over 4\xi_e} A_o(\bar\phi)
+O(1/\xi_e^2) 
\label{se2lguess}
\end{equation}
for any $p$, where
\begin{equation}
A_o(\mu;\bar\phi) = \Biggl\{
\begin{array}{cl}
1 + c (\bar\phi-1/2) &
\quad {\rm for} \quad 0 < \bar\phi < 1, \\
-1 - c (3/2-\bar\phi) &
\quad {\rm for} \quad 1 < \bar\phi < 2. \\
\end{array}
\label{eq:Ae2l}
\end{equation}
with $c\approx -0.150$ for $p=2$ and $c\approx -0.044$ for $p=4$.  Note that
the behavior of the homogeneous system with open boundary condition, cf.\ 
Eq.~(\ref{eq:Se-E2mu0}), is obtained by replacing $\xi_e\to L+1$ for the
values $\bar\phi=1/2,\,3/2$, which are the only possible values for the
homogeneous system in a chain with even $L$, corresponding to odd and even
$L/2$ respectively.

\subsection{Discussion}
\label{RGint}

The trap-size dependence of the half-lattice entanglement shows that the
confining potential induces a length scale which behaves as $\xi\sim l$ for
any power of the potential, at least for smooth observables. In the framework
of the TSS, this would imply that the trap exponent is $\theta=1$
independently of $p$. This value of $\theta$ is also obtained from the
trap-size dependence of the gap, which scales as $\Delta\sim l^{-1}$, apart
from a periodic dependence on $l$ of its amplitude, cf.\ 
Eq.~(\ref{eq:Delta,mu0}).  Indeed, the exponent of the power behavior is
expected to be $z \theta$, and in the superfluid region we have $z=1$.  A
consistent scaling is also observed in the case of the one-particle density
matrix $G_b(x_i,x_j)$,
see Sec.~\ref{opdmsufl}. On the other hand, the density correlations show
clearly a coupling with modes at the Fermi momentum $k_F=\pi f$, which are
apparently characterized by a different length scale $\xi_{\rm f}$, scaling as
$\xi_{\rm f}\sim l^{\zeta}$ with $\zeta=p/(p+1)$.

These results may be explained by the nontrivial coupling of the confining
potential with the free bosonic field $\phi(x)$ of the continuum theory, due
to the fact that in the continuum limit the spin operator $\sigma^z_i$ can be
written as a sum of a slow contribution proportional to $\partial_x \phi$ and
a rapidly oscillating contribution proportional to $e^{2ik_Fx}\phi$, see,
e.g., Refs.~\cite{Sachdev-book,Tsvelik-book}.  Assuming that the latter term
is suppressed for smooth or global quantities, such as the half-lattice
entanglement entropy and the gap, where its effects should get averaged out,
we may argue that $\theta=1$ is indeed the expected trap exponent. A heuristic
argument may be obtained by noting that the perturbation $\int
dx\,V(x)\,(d\phi(x)/dx)$ can be rewritten as $\int dx\,(dV(x)/dx)\,\phi(x)$ by
integration by parts, whose first-order perturbation vanishes because $V(x)$
is even in $x$.  We then expect that the leading contribution comes from
next-to-leading terms, like $\int dx\,V(x)\,(d^2\phi(x)/dx^2)$. RG arguments
applied to this perturbation, taking into account that $\phi$ is a free Bose
field, lead to $\theta=1$ independently of $p$, which is the result emerging
from the numerical diagonalization of the Hamiltonian.  On the other hand,
correlators which are nontrivially coupled to the modes at $k_F$ may show a
different length scale due to the coupling of the confining potential with the
staggered term in $\sigma^z$.  This is indeed what we observe in the
correlation of density operators which are directly related to the operator
$\sigma^z$.

\section{Modulated TSS at the $\bm{n=1}$ Mott insulator to superfluid
transition}
\label{mum1}

In the section we study the effects of the trap at the $n=1$ Mott insulator to
superfluid transition, i.e., at $\mu_c=-1$, where the filling factor of the
homogeneous system is $f=1$.  The confining potential $V(x)$ gives rise to a
change of the particle density from $\langle n \rangle\approx 1$ in the middle
of the trap to $\langle n \rangle=0$ at large distance, passing through the
gapless superfluid phase, see Fig.\ \ref{XXpartdensnum1}.  Specifically, for
$\mu=-1$, the particle density appears to vanish when $\mu_{\rm eff}\gtrsim 1$,
cf.\ Eq.~(\ref{mueff}), thus, $x/l \gtrsim 2^{1/p}$.

We recall that the behavior around $\mu=-1$ of the homogeneous BH model
without trap is essentially analogous to that at $\mu=1$, because of the
invariance under the particle-hole exchange.  At the $n=1$ Mott insulator to
superfluid quantum transition, the critical exponents $z$ and $y_\mu$ and the
trap-size exponent $\theta$ are the same as those at $\mu=1$, i.e., $z=2$,
$y_\mu=2$ and $\theta=p/(2+p)$.  However, the particle-hole symmetry does not
hold in the presence of the trapping potential, and the asymptotic trap-size
dependence appears more complicated at the $n=1$ Mott transition.  This is
essentially related to the presence of level crossings at finite values of the
trap size, where the gap vanishes, as already found in the superfluid region,
for $|\mu|<1$. As we shall see, the resulting trap-size dependence can be cast
in the form of a modulated TSS, that is a TSS controlled by the same exponents
as those at the low-density Mott transition, but modulated by periodic
functions of the trap size.

\subsection{Modulated TSS of the gap}
\label{mtssgap}

\begin{figure}[tbp]
\includegraphics*[scale=\graphicscale]{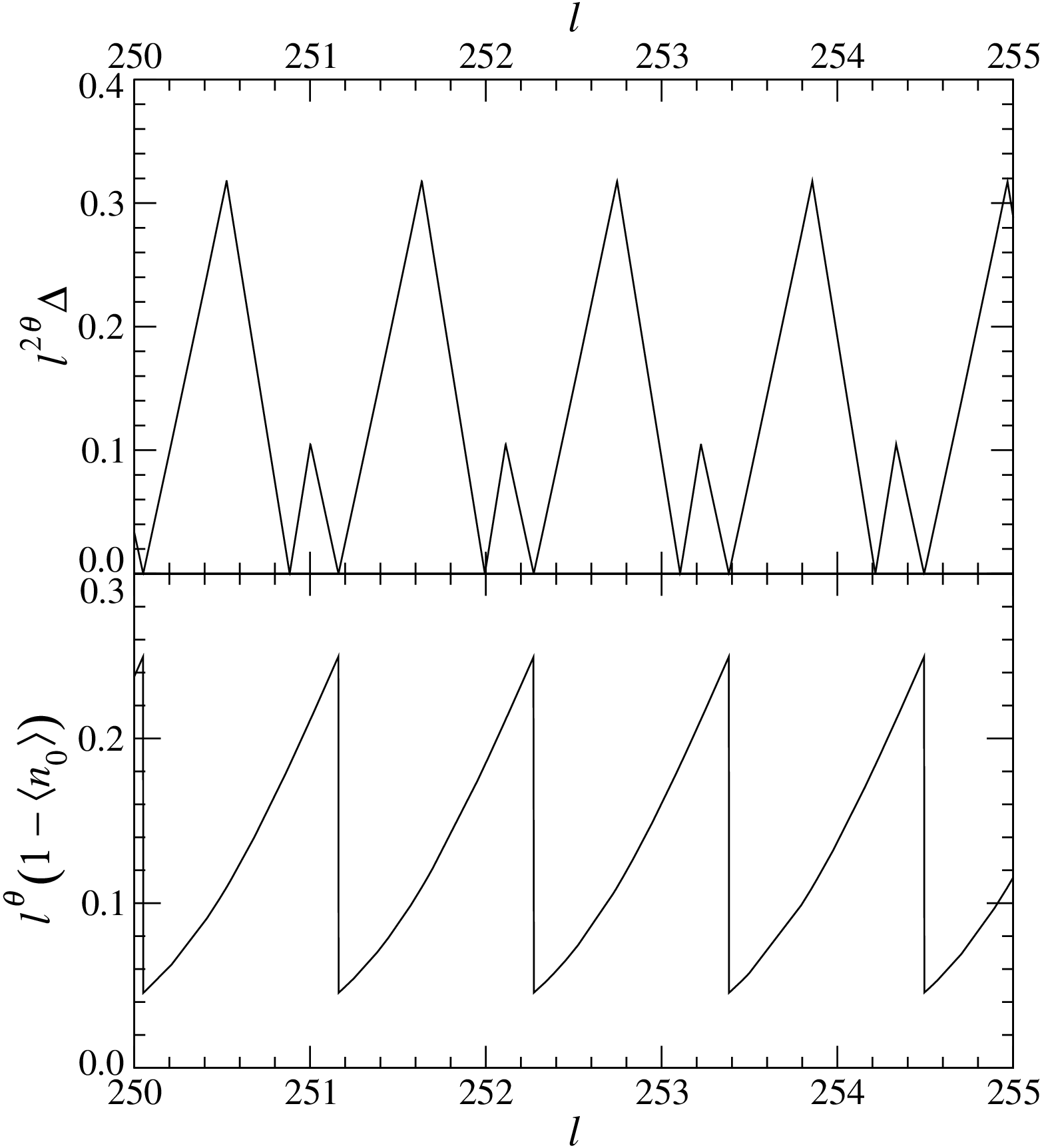} 
\caption{
The rescaled energy gap $l^{2\theta}\Delta$ (above) and the
rescaled particle density in the middle of the trap
$l^{\theta}(1-\langle n_0\rangle)$ (below) vs.\ $l$ for 
$\mu=-1$ and $p=2$, whose trap exponent is $\theta=1/2$.
Results for $p=4$ and $p=6$ are similar.}
\label{gapanddensityp2}
\end{figure}

Results for the gap and the particle density at the middle of the trap are
shown in Fig.\ \ref{gapanddensityp2}.
They suggest periodic asymptotic behaviors of the scaling quantities
$l^{2\theta}\Delta$ and $l^{\theta}(1-\langle n_0\rangle)$ as functions of
$l$, with a period given by the interval between two even (or odd) zeroes of
the gap, and a marked difference between even- and odd-numbered crossings and
peaks.

\begin{figure}[tbp]
\begin{center}
\leavevmode
\includegraphics*[scale=\graphicscale]{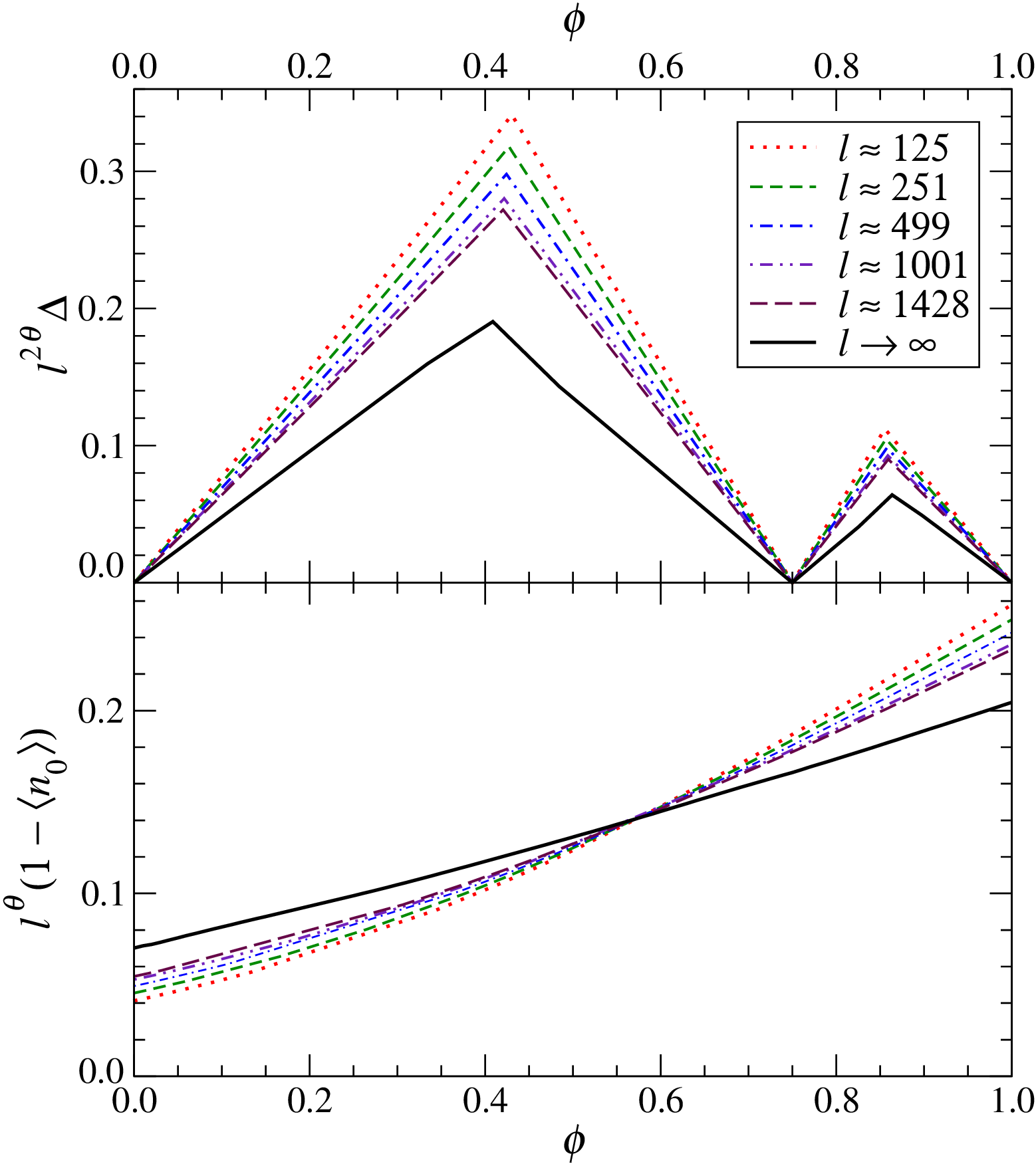}
\caption{
$l^{2\theta}\Delta$ (above) and $l^{\theta}(1-\langle n_0\rangle)$
  (below) vs.\ $\phi$, defined in Eq.~(\ref{phildef}), for $p=2$
  ($\theta=1/2$).  The extrapolation to $l\to\infty$ is obtained by assuming
  $O(l^{-\theta/2})$ leading corrections.  In the legends $l$ is
  rounded to the nearest integer for the sake of presentation.}
\label{gapplp2}
\end{center}
\end{figure}
\begin{figure}[tbp]
\begin{center}
\leavevmode
\includegraphics*[scale=\graphicscale]{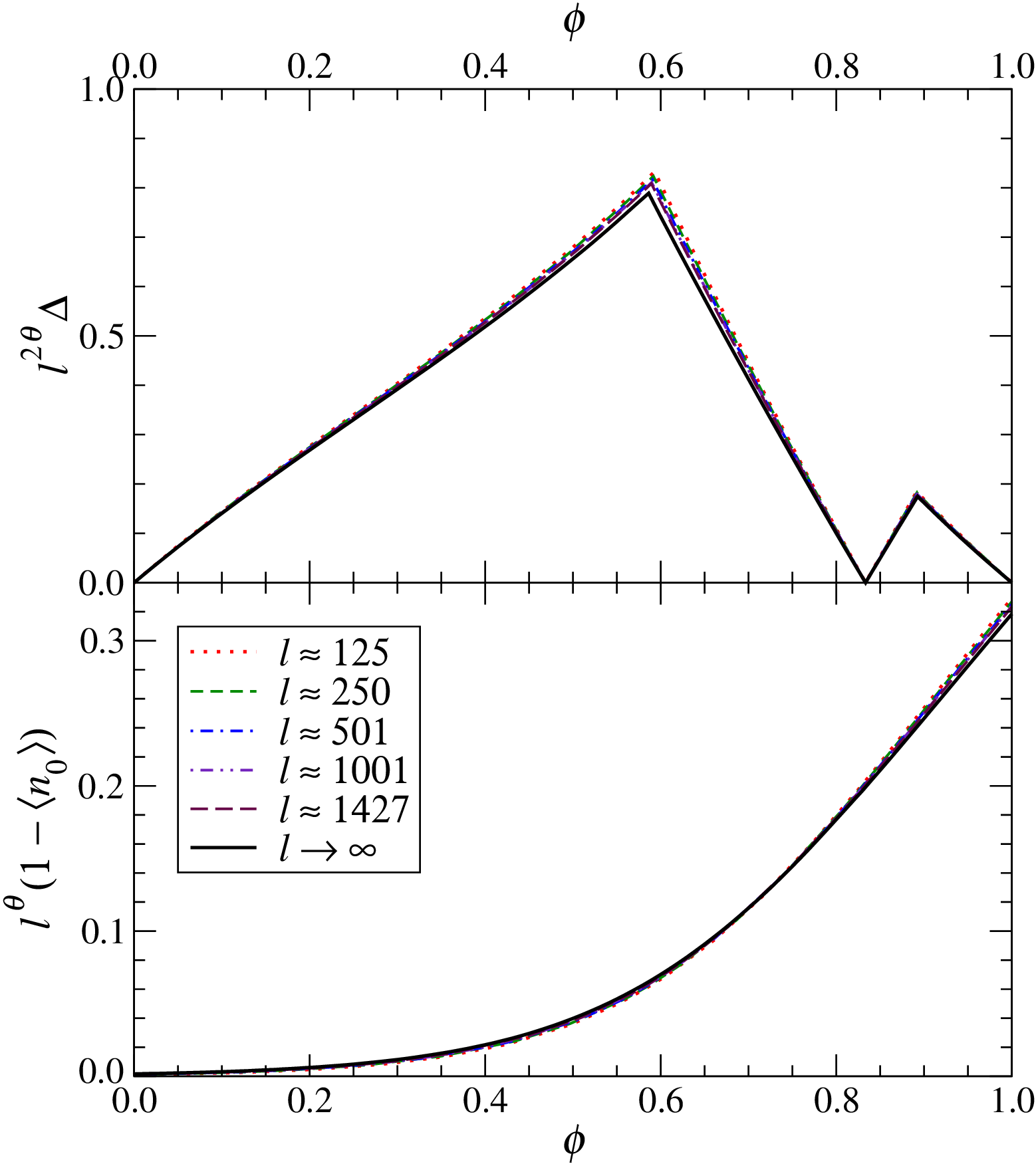}
\caption{
 $l^{2\theta}\Delta$ (above) and $l^{\theta}(1-\langle n_0\rangle)$
 (below) vs.\ $\phi$, defined in Eq.~(\ref{phildef}), 
 for $p=4$ ($\theta=2/3$).}
\label{gapplp4}
\end{center}
\end{figure}
\begin{figure}[tbp]
\begin{center}
\leavevmode
\includegraphics*[scale=\graphicscale]{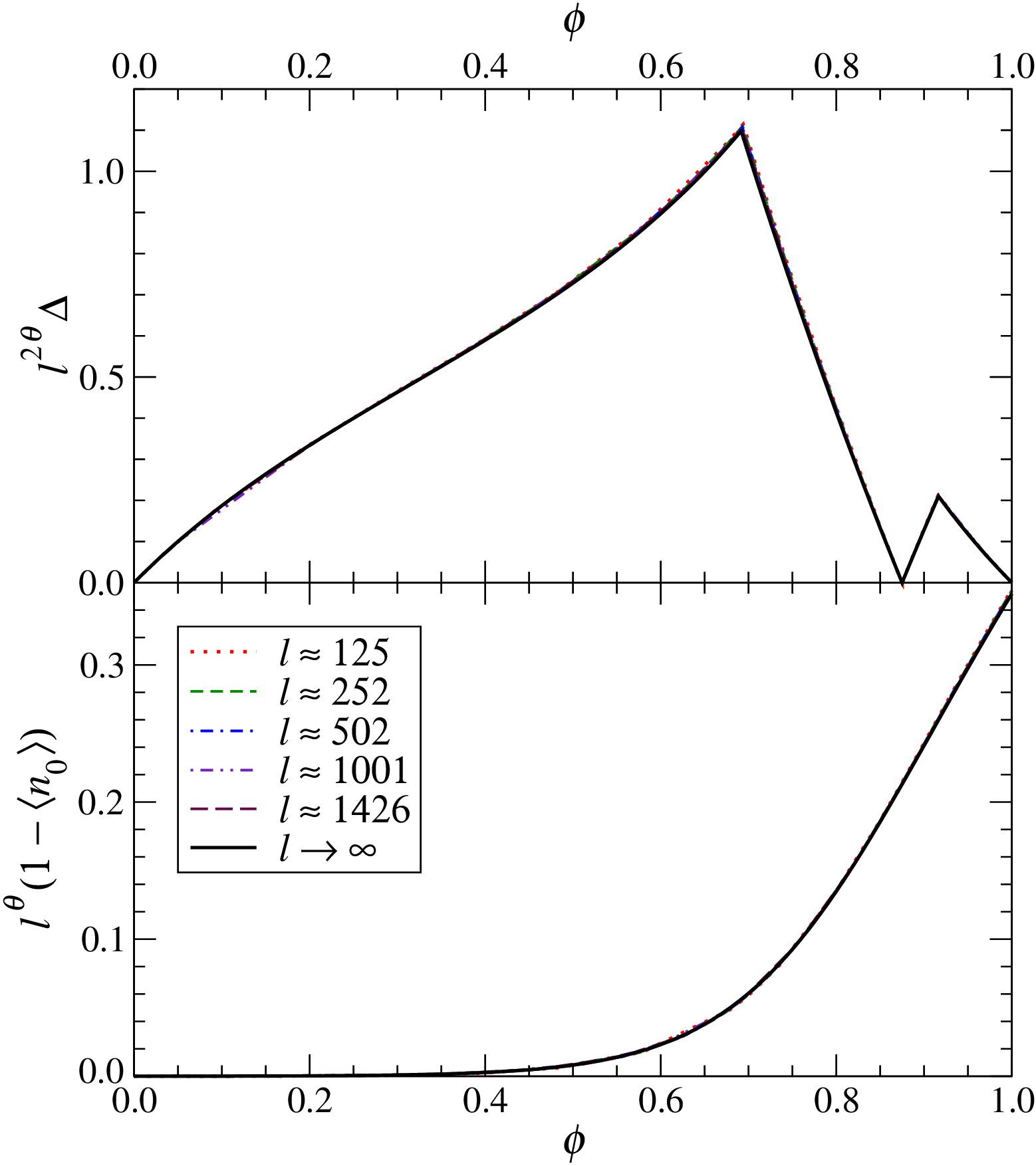}
\caption{
  $l^{2\theta}\Delta$ (above) and $l^{\theta}(1-\langle n_0\rangle)$ (below)
  vs.\ $\phi$, defined in Eq.~(\ref{phildef}), for $p=6$ ($\theta=3/4$).}
\label{gapplp6}
\end{center}
\end{figure}

In the large-$l$ limit we find that the interval between two even zeroes
approaches a constant value, i.e.,
\begin{equation}
P_l^{(k)} \equiv l_0^{(2k+2)}-l_0^{(2k)} =  P_l^* + O(l^{-2}). 
\label{pl}
\end{equation}
The large-$l$ limit is estimated to be 
\begin{eqnarray}
&&P_l^*\cong1.11072073 \quad {\rm for} \;\;p=2, \nonumber\\
&&P_l^*\cong1.10243940 \quad {\rm for} \;\;p=4, \label{plest} \\ 
&&P_l^*\cong1.08184087 \quad {\rm for} \;\;p=6. \nonumber
\end{eqnarray}
Analogously to the trap-size dependence in the superfluid region, see
Sec.~\ref{entanglementsuperfluid}, the asymptotic interval $P_l^*$ can be
estimated using the LDA of the total particle number, cf.\ Eq.~(\ref{ntlda}).
We find again identical results, showing that the LDA of the total particle
density is asymptotically exact in the large-$l$ limit also at the $n=1$ Mott
transition.  The $p\to\infty$ limit of $P_l^*$ can be easily computed using
the LDA, obtaining $P_l^*\to 1$ for $p\to\infty$.  Corrections to the LDA of
the particle density are discussed below.

The asymptotic periodic properties clearly emerge from the results shown in
Figs.\ \ref{gapplp2}, \ref{gapplp4} and \ref{gapplp6}, where
$l^{2\theta}\Delta$ and $l^{\theta}(1-\langle n_0\rangle)$, for $p=2,\,4,\,6$
respectively, are plotted versus the phase-like variable
\begin{equation}
\phi = {l-l_0^{(2k)}\over  l_0^{(2k+2)}-l_0^{(2k)}}, 
\qquad l_0^{(2k)}\le l < l_0^{(2k+2)},
\label{phildef}
\end{equation}
thus $0\le\phi<1$.

The results for the gap show that the quantity $l^{2\theta}\Delta$ approaches
an asymptotic periodic function $A_\Delta(\phi)$ in the large-$l$ limit.
Therefore, they provide a clear evidence for a modulated asymptotic behavior
\begin{equation}
\Delta \approx A_\Delta(\phi) l^{-2\theta}  [1 + O(l^{-\kappa})]
\label{deltadep}
\end{equation}
with $\kappa\approx\theta/2$ (see below).  Note that, for $p=2$, the scaling
of the gap is controlled by the same exponent found in the region $|\mu|<1$,
because we have $2\theta=1$; on the other hand, for $p=4$ and $p=6$ we have
respectively $2\theta=4/3$ and $2\theta=3/2$, which are easily differentiated
from $1$.

\begin{figure}[tbp]
\begin{center}
\leavevmode
\includegraphics*[scale=\graphicthreescale]{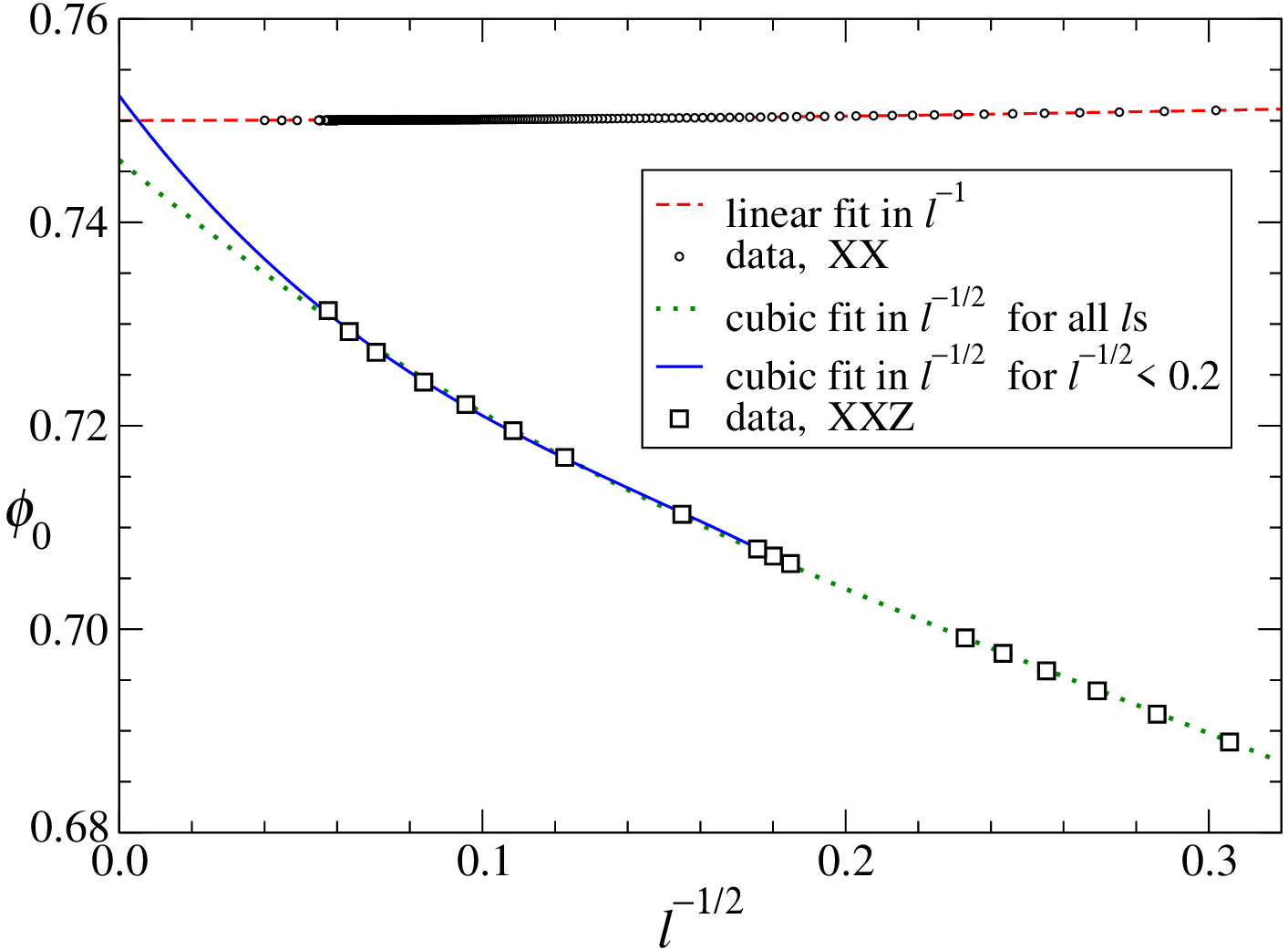}
\includegraphics*[scale=\graphicthreescale]{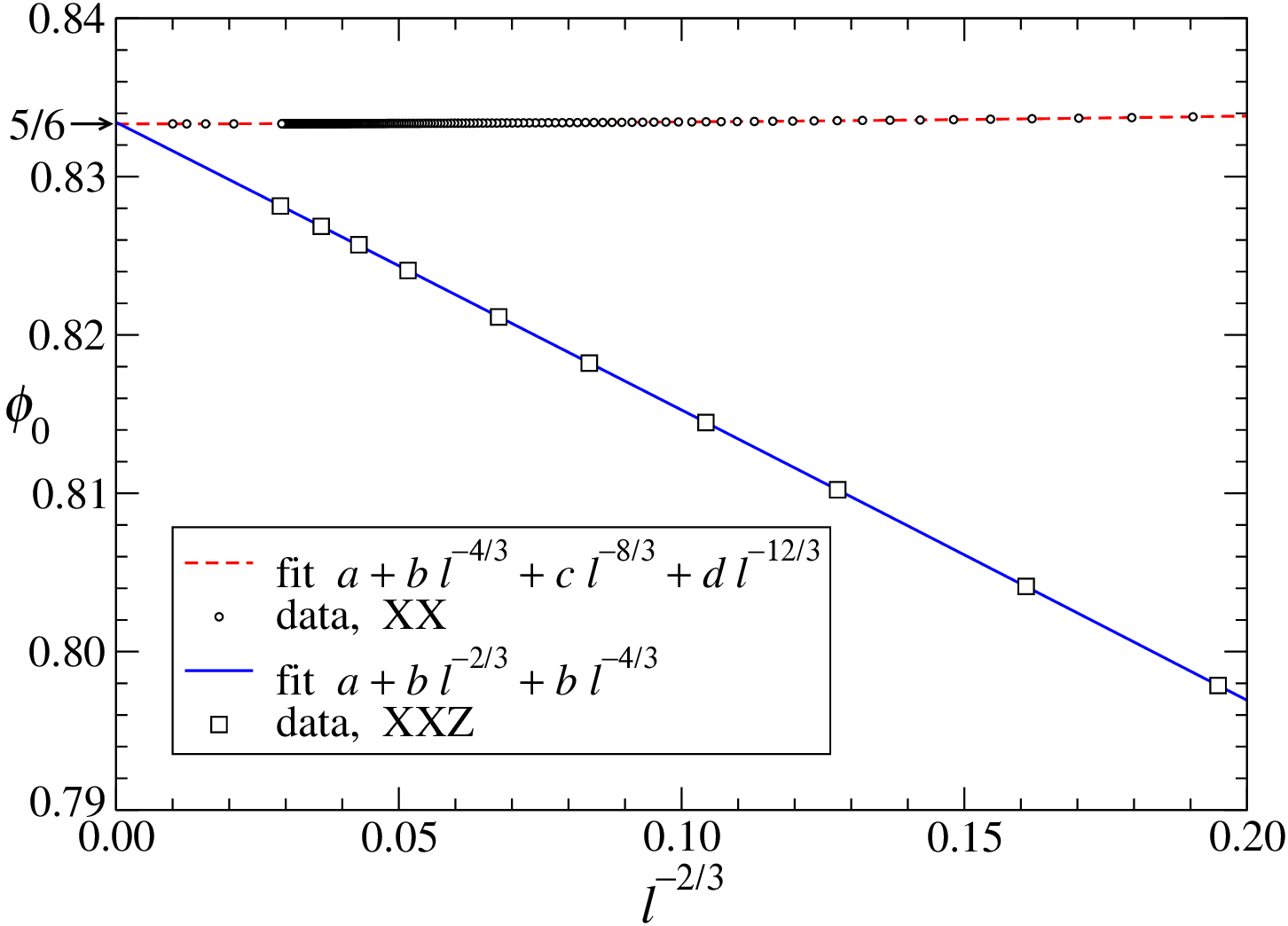}
\includegraphics*[scale=\graphicthreescale]{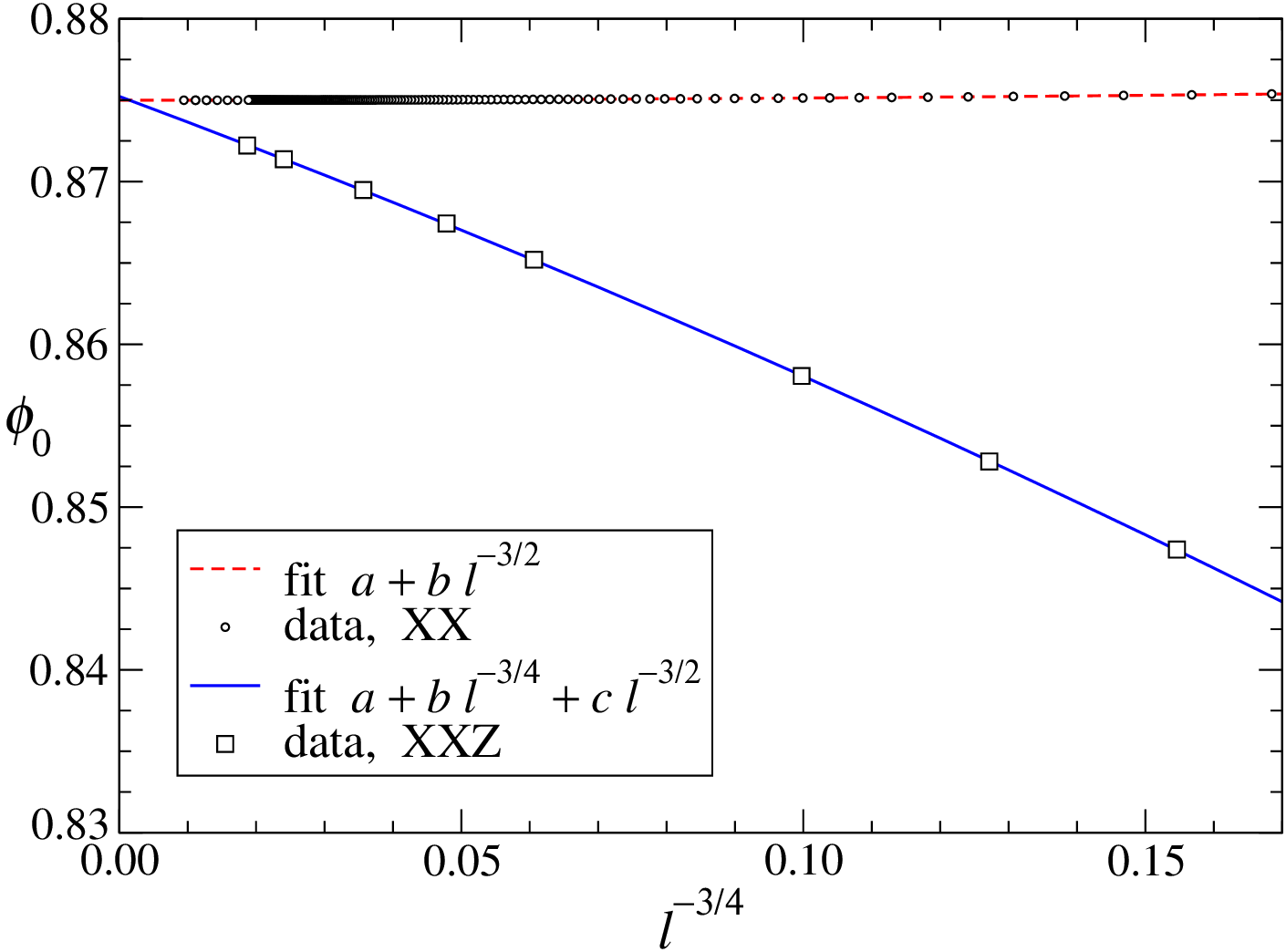}
\caption{$\phi_0$ vs.\ $l^{-\theta}$, 
  for the XX model (\ref{XX}) at $\mu=-1$ and the XXZ model (\ref{eq:H-XXZ})
  at $j_z=-1/2$ and $\mu=-1/2$, for $p=2$ (above), $p=4$ (middle), and $p=6$
  (below). }
\label{fig:XXZ,phi0}
\end{center}
\end{figure}

We should note that in these calculations the trap of size $l$ is centered in
the middle of the chain of size $L$; more precisely we consider odd $L$s to
have the center of the trap coincide with the middle site of the chain.  Even
$L$s would instead correspond to traps centered between the two central sites.
Unlike the previous cases, at $\mu=-1$ this difference must be taken into
account, but it gives only rise to an interchange of the role of the even and
odd zeroes of the gap. Therefore, in the case of even $L$s, one may simply
redefine the phase-like variable as
\begin{equation}
\phi = {l-l_0^{(2k-1)}\over  l_0^{(2k+1)}-l_0^{(2k-1)}}, 
\qquad l_0^{(2k-1)}\le l < l_0^{(2k+1)}.
\label{philevdef}
\end{equation}
The two different definitions (\ref{phildef}) and (\ref{philevdef}) remove the
dependence on the parity of $L$ (cf.\ App.~\ref{Sec:numerics}) from the
modulation functions $A(\phi)$ of the asymptotic behaviors of the observables,
such as $A_\Delta(\phi)$.  Using, e.g., definition (\ref{phildef}) for all
$L$s would lead to 
$A(\phi)|_{{\rm even}\ L} = A(\{\phi+\phi_0\})|_{{\rm odd}\ L}$,
where $\{x\}\equiv x - \lfloor x\rfloor$ is the fractional part of
$x$.  Note that centering the trap at an arbitrary distance $q$ from
the nearest site of the chain would modify significantly the modulation.

By definition, $A_\Delta(0)=A_\Delta(1)=0$. $A_\Delta$ has another zero
$\phi_0$ corresponding to the odd level crossings.  Computations of $\phi_0$
for the XX model and $p=2,\,4,\,6$ are shown in
Fig.~\ref{fig:XXZ,phi0}, where they are plotted vs $L^{-\theta}$.  They
asymptotically approaches the value $\phi_0\cong0.750000$ for $p=2$,
$\phi_0\cong0.833333$ for $p=4$, $\phi_0\cong0.875000$ for $p=6$, and
$\phi_0\cong0.916667$ for $p=10$, with $O(l^{-2\theta})$ corrections.  Note
that these values are compatible with the simple formula
\begin{equation}
\phi_0=(p+1)/(p+2).
\label{phi0guess}
\end{equation}

Other features of the modulation function $A_\Delta(\phi)$, such as the
location of the large and small peaks, $\phi_l$ and $\phi_s$ respectively, and
its peak values $A_\Delta(\phi_{l,s})$, are approached with power-law scaling
corrections $O(l^{-\kappa})$.  We find $\kappa\approx 1/4$ for $p=2$,
$\kappa\approx 1/3$ for $p=4$, and $\kappa\approx 3/8$ for $p=6$, which are
consistent with $\kappa=\theta/2$.  See, e.g., Figs.~\ref{fig:peaks,phi,p=2}
and ~\ref{fig:peaks,phi,p=4}.  Note that $\kappa=\theta/2$ should be
considered as a phenomenological result, because we do not have theoretical
arguments to derive it.
\begin{figure}[tbp]
\begin{center}
\leavevmode
\includegraphics*[scale=\graphicscale]{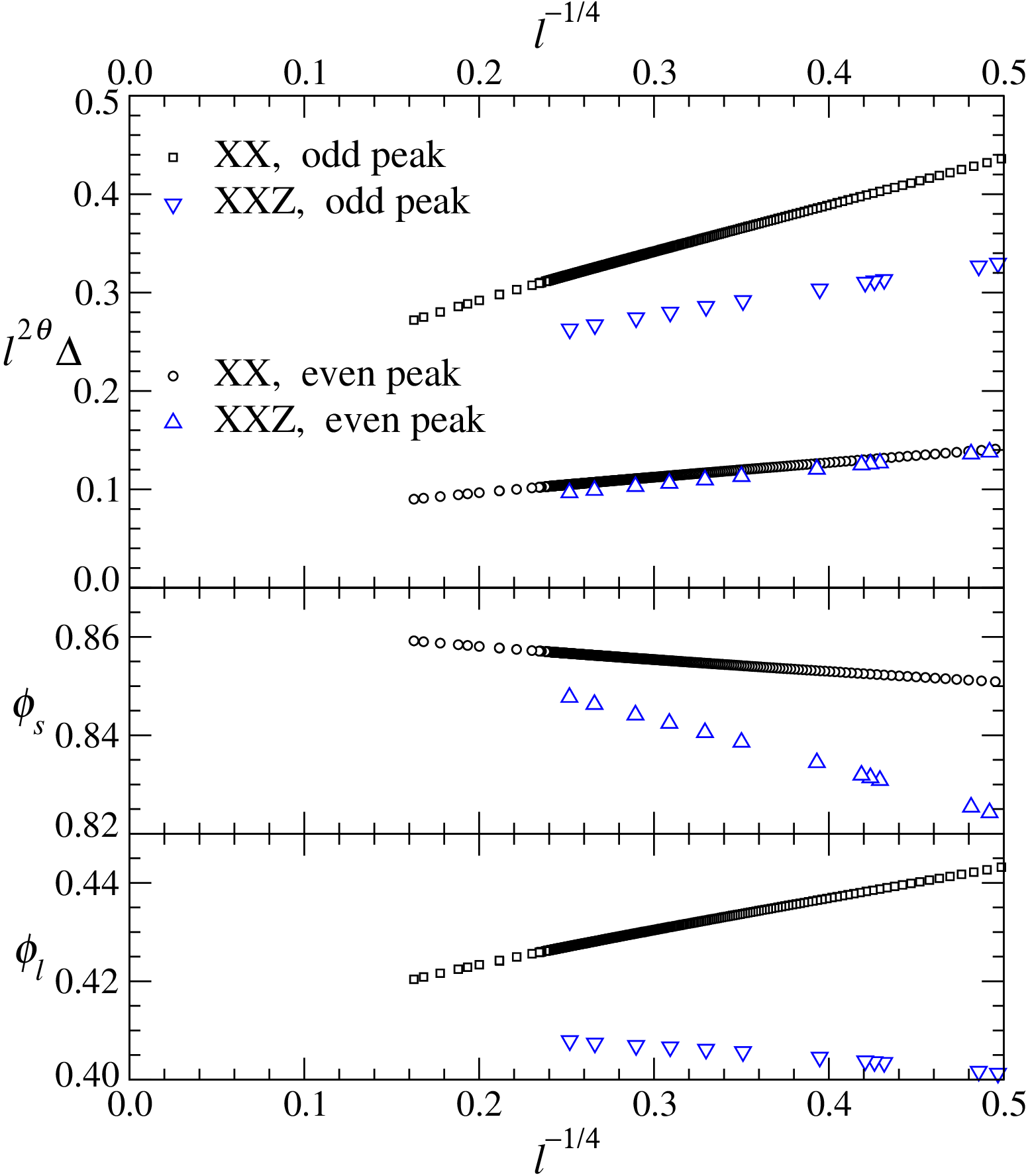} 
\caption{
Scaling of the large (odd) and small (even) peaks of the gap and their location
  $\phi_l$ and $\phi_s$ for $p=2$, for the XX model (\ref{XX}) and
the XXZ model (\ref{eq:H-XXZ}) at $j_z=-1/2$.}
\label{fig:peaks,phi,p=2}
\end{center}
\end{figure}
\begin{figure}[tbp]
\begin{center}
\leavevmode
\includegraphics*[scale=\graphicscale]{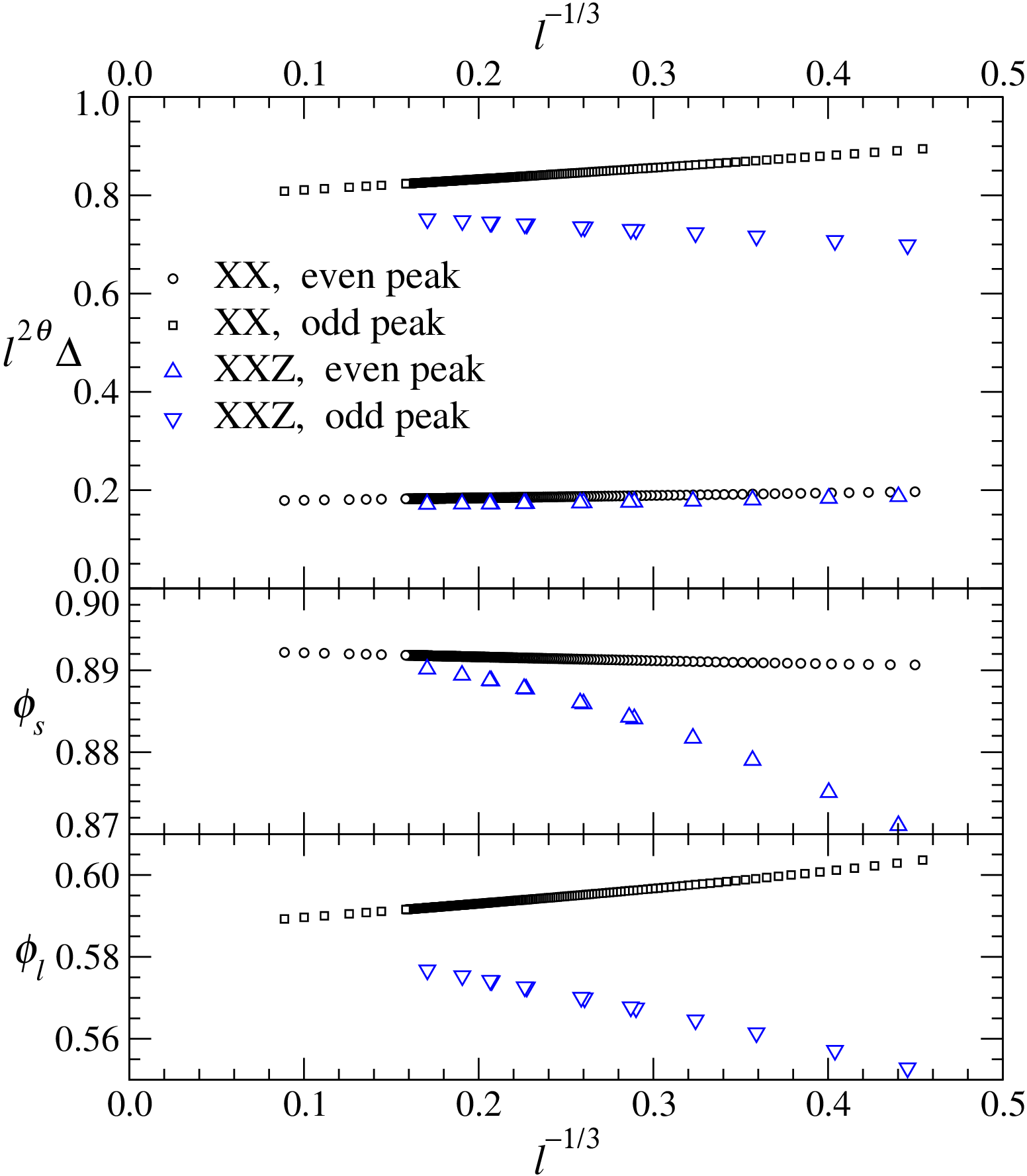} 
\caption{
Scaling of the large (odd) and small (even) peaks of the gap and their location
  $\phi_l$ and $\phi_s$ for $p=4$, for the XX model (\ref{XX}) and
the XXZ model (\ref{eq:H-XXZ}) at $j_z=-1/2$.}
\label{fig:peaks,phi,p=4}
\end{center}
\end{figure}

A faster approach to scaling is found for the ratio $\Delta_l/\Delta_s$
between subsequent large and small peaks of the gap, which is expected to be
universal (essentially because it is independent of normalizations), see
Fig.~\ref{fig:Dratio,XXZ,mu=-1}.  (We compute the ratio at $l$s corresponding
to $\Delta_s$, obtaining $\Delta_l$ by cubic spline interpolation.)  We find a
behavior compatible with
\begin{equation}
\Delta_l/\Delta_s = a + b l^{-\theta} + O(l^{-2\theta}).
\label{rdeapp}
\end{equation}
\begin{figure}[tbp]
\begin{center}
\leavevmode
\includegraphics*[scale=\graphicthreescale]{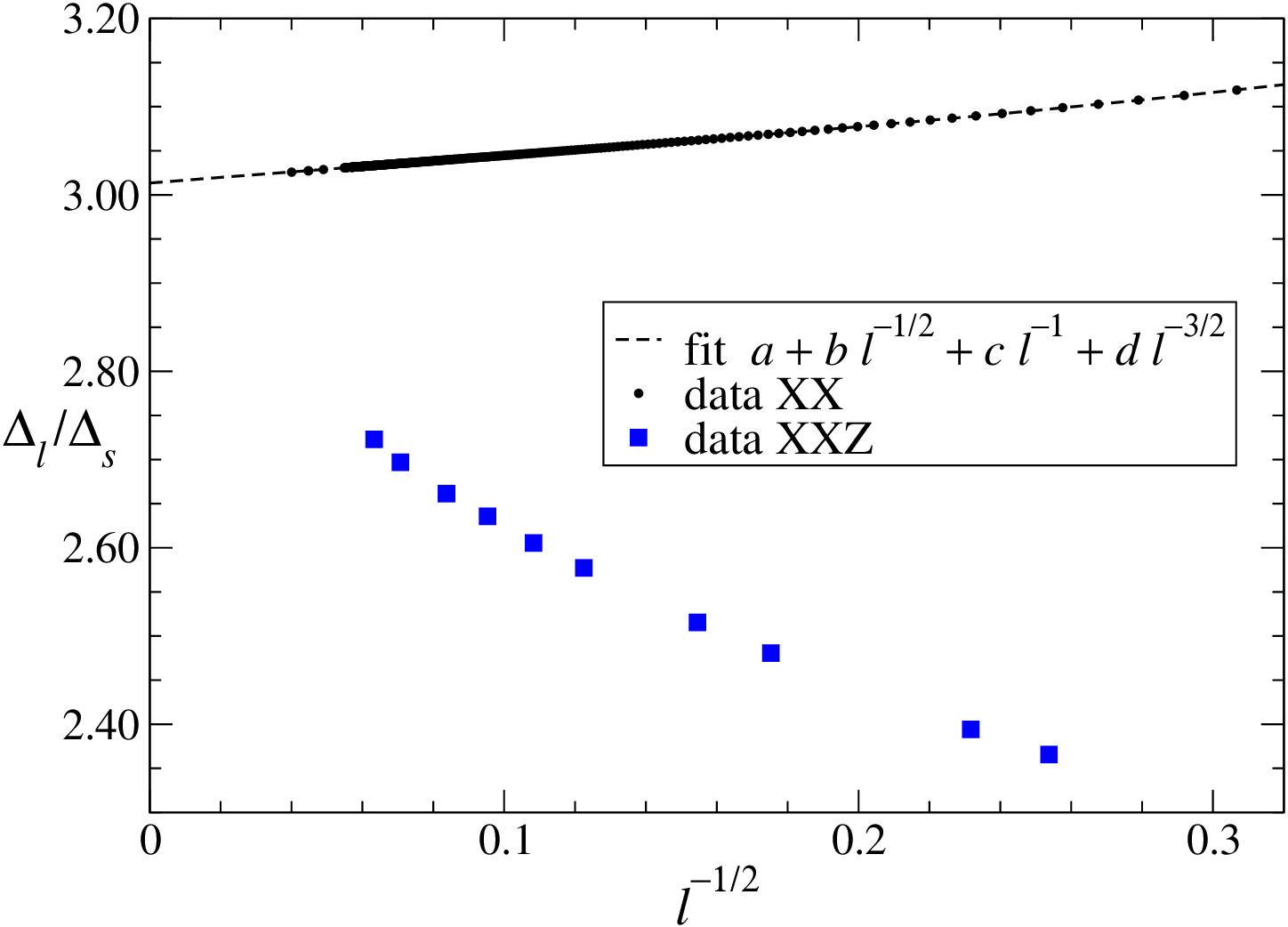}
\includegraphics*[scale=\graphicthreescale]{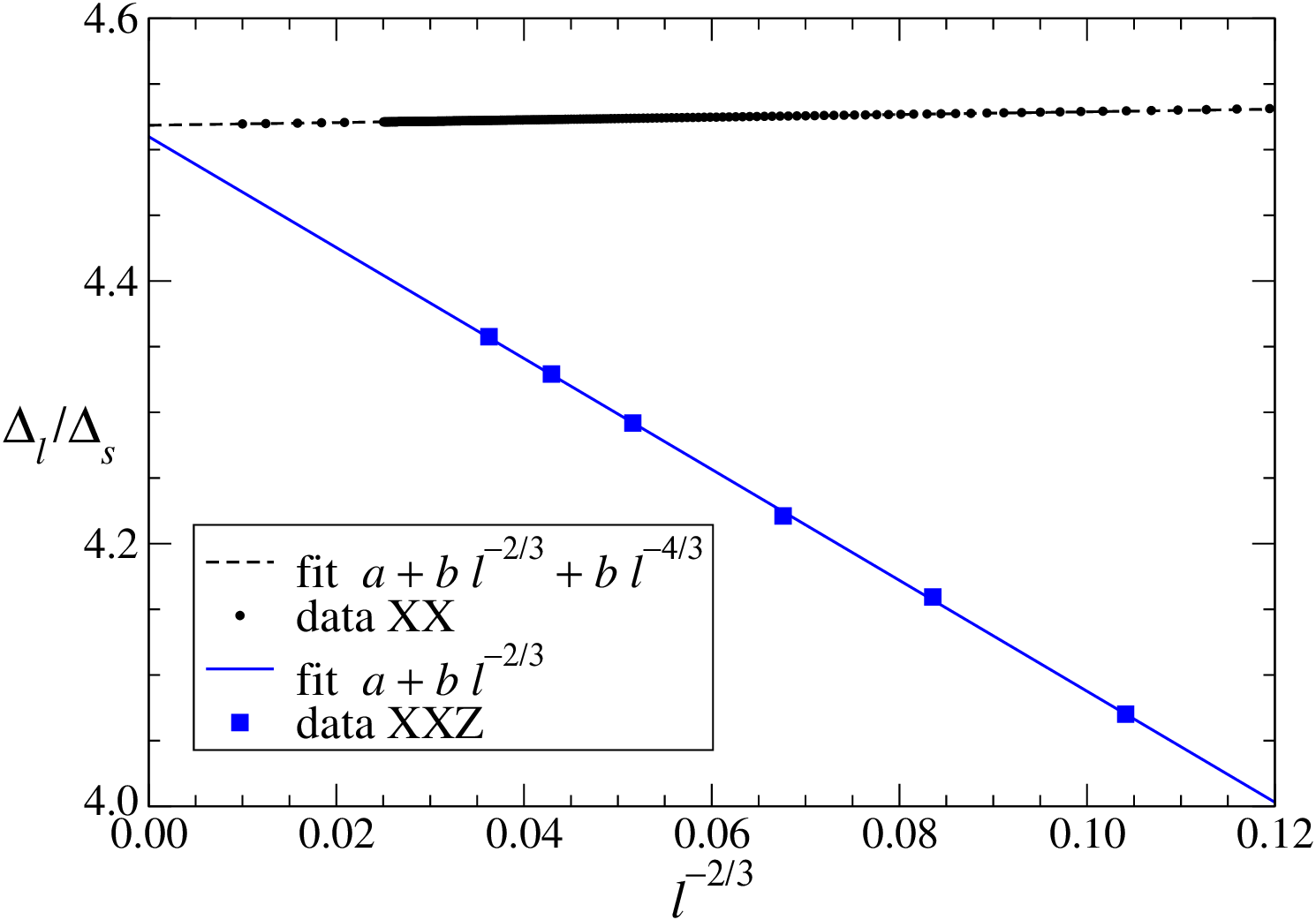}
\includegraphics*[scale=\graphicthreescale]{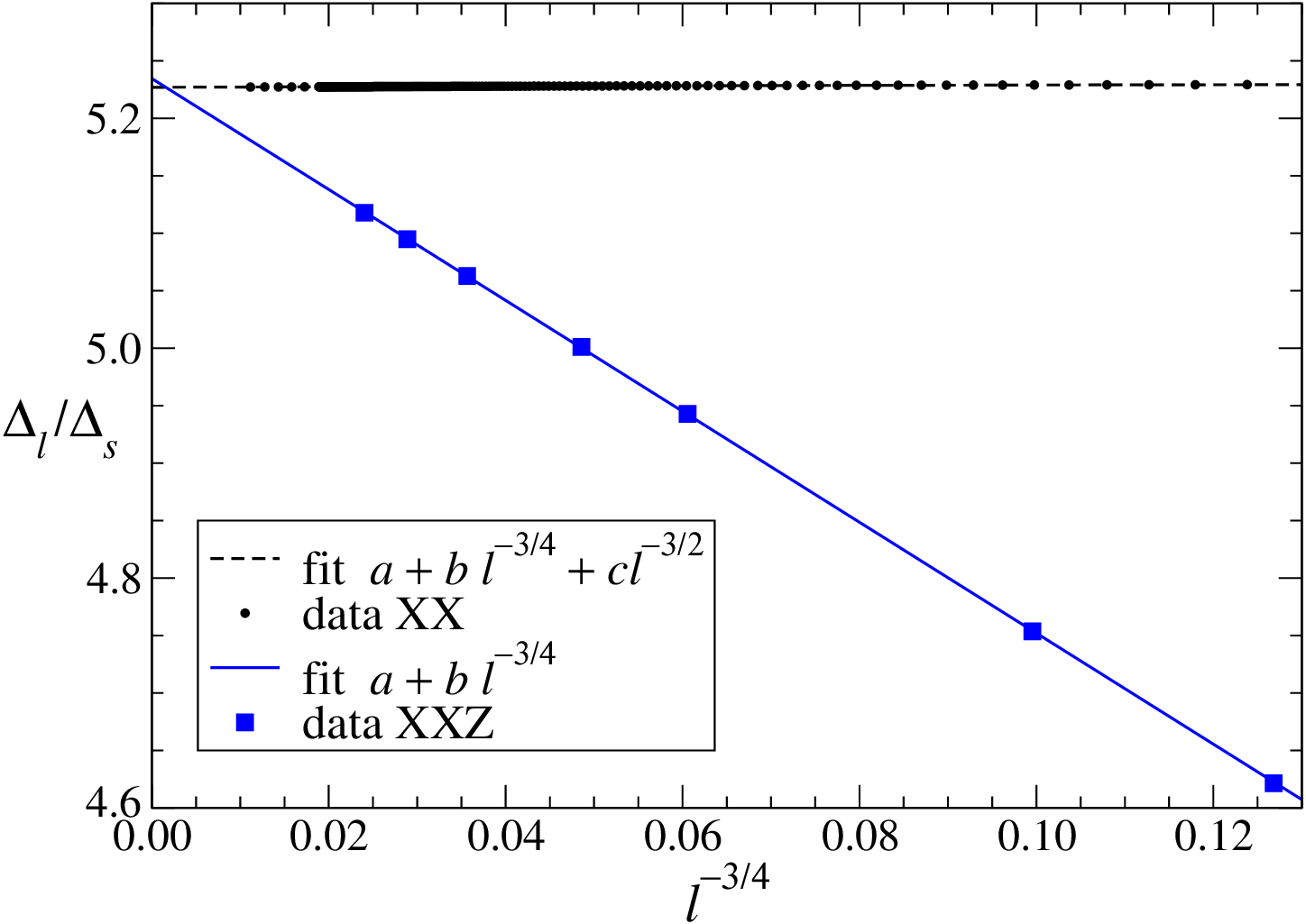}
\caption{
$\Delta_l/\Delta_s$ vs.\ $l^{-\theta}$ for 
the XX model and for the XXZ model at 
$j_z=-1/2$, for $p=2$ (above), $p=4$ (middle) 
and $p=6$ (below).}
\label{fig:Dratio,XXZ,mu=-1}
\end{center}
\end{figure}

It is worth noting that, in the case $p=2$, $A_\Delta(\phi)$ has apparently a
simple form: it is approximately formed by two reflected similar triangles, as
shown in Fig.~\ref{gapplp2}. Indeed, this shape would require
$\Delta_l/\Delta_s=\phi_0/(1-\phi_0) = 3$, while we obtain a very close value
$\Delta_l/\Delta_s = 3.013$ from the $l\to\infty$
extrapolation of the results at fixed $l$.

The behavior in the large-$p$ limit can be guessed from the results obtained
at finite $p$ shown above. Note that they do not apparently approach the
finite-size behavior of a homogeneous BH model on a chain of size $L=2l$ with
open boundary conditions, as one may naively expect.  For example, the gap of
the homogeneous BH model at $\mu=-1$ behaves as $\Delta=O(1/L^2)$ without
showing level crossings.~\footnote{ This result can be easily derived using
  the particle-hole exchange symmetry, and the corresponding result at the
  low-density Mott transition, at $\mu=1$, see, e.g., Eq.~(\ref{eq:pinfeig}).}
As a consequence, the phase-like variable does not have any corresponding
quantity in the BH model without trap. Thus the relations between the
large-$p$ limit and the finite-size behavior of the homogeneous BH model are
not straightforward.

Finally, let us mention that the gap at $\mu<-1$ shows nontrivial trap-size
dependence as well, because the phenomenon of the level crossings persists.
The numerical results at $\mu=-1.2$ show again a periodic structure of the
gap, which gets suppressed as $1/l$ for any $p$. They show large and small
peaks, but the smallest one gets rapidly suppressed in the large-$l$ limit.
We observe that the large peak turns out to scale as $\Delta_l\sim 1/l$ for
$p=2$ and $p=4$, while the small peak decreases as 
$\Delta_s\sim \exp(-a l)$; therefore, the ratio $\Delta_s/\Delta_l$
vanishes as $l\to\infty$; in the same limit, $\phi_0\to1$.
Although for $\mu<-1$ the homogeneous system
without trap has a gap proportional to $\mu_s\equiv -\mu-1$, see, e.g.,
Ref.~\cite{Sachdev-book}, here we find that the periodic trap-size dependence
of the gap tends to be suppressed as $1/l$ in the large-$l$ limit.  This is
related to the fact that, in the presence of the trap, including its large-$l$
limit, we have always some critical regions, for example the {\em
  superfluid} regions between the $\langle n_x \rangle=1$ and $\langle
n_x \rangle=0$ plateaus, for 
\begin{equation}
(-1-\mu)^{1/p} \lesssim |x|/l\lesssim (1-\mu)^{1/p}, 
\label{superflreg}
\end{equation}
which becomes larger and larger with increasing $l$. An analogous behavior is
found for the trap-size dependence of the XY chain when the middle of the trap
is in the quantum ferromagnetic phase~\cite{CV-10}.

\subsection{The particle density and its correlators}
\label{mtsspd}

\begin{figure}[tbp]
\begin{center}
\leavevmode
\includegraphics*[scale=\graphicscale]{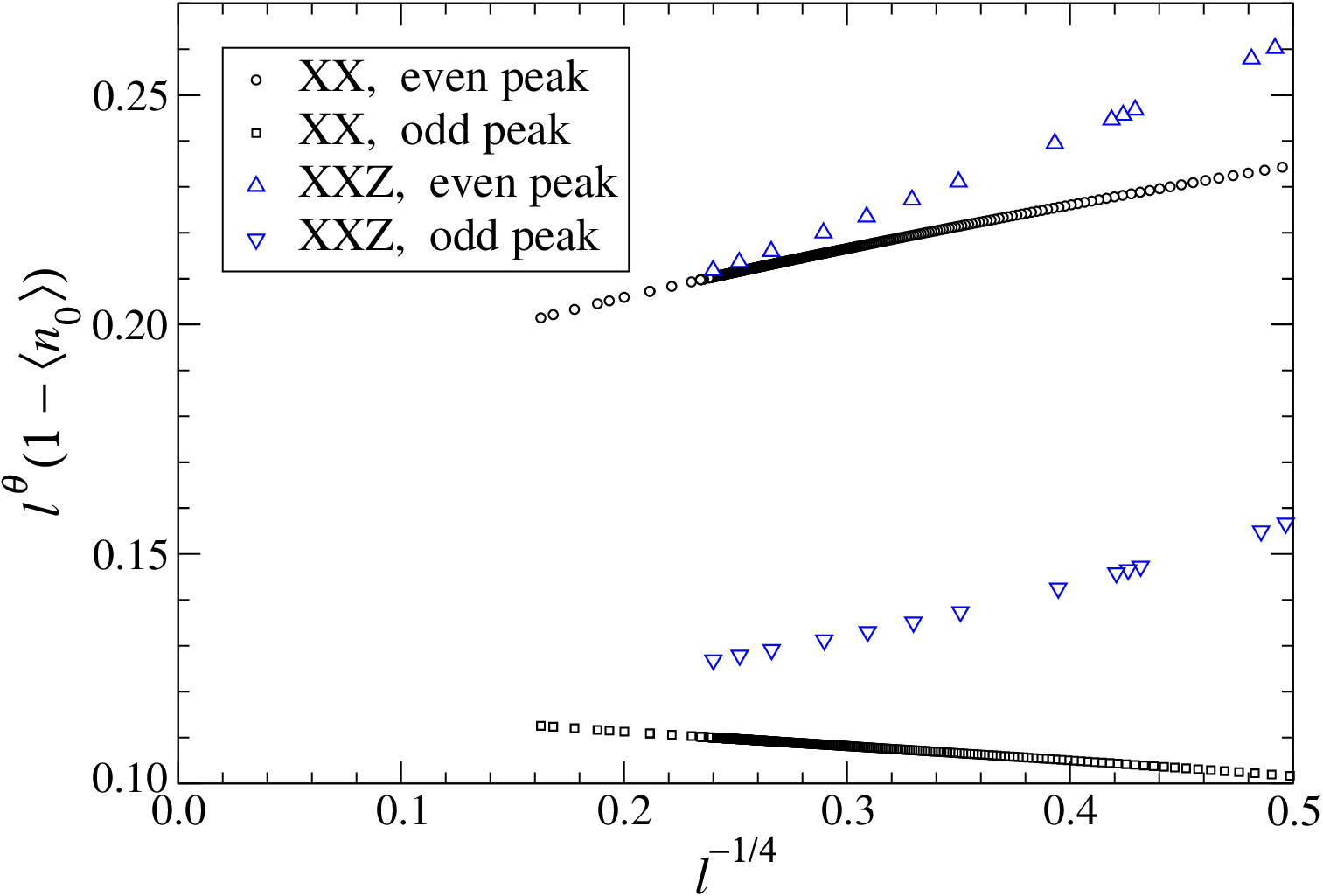}
\includegraphics*[scale=\graphicscale]{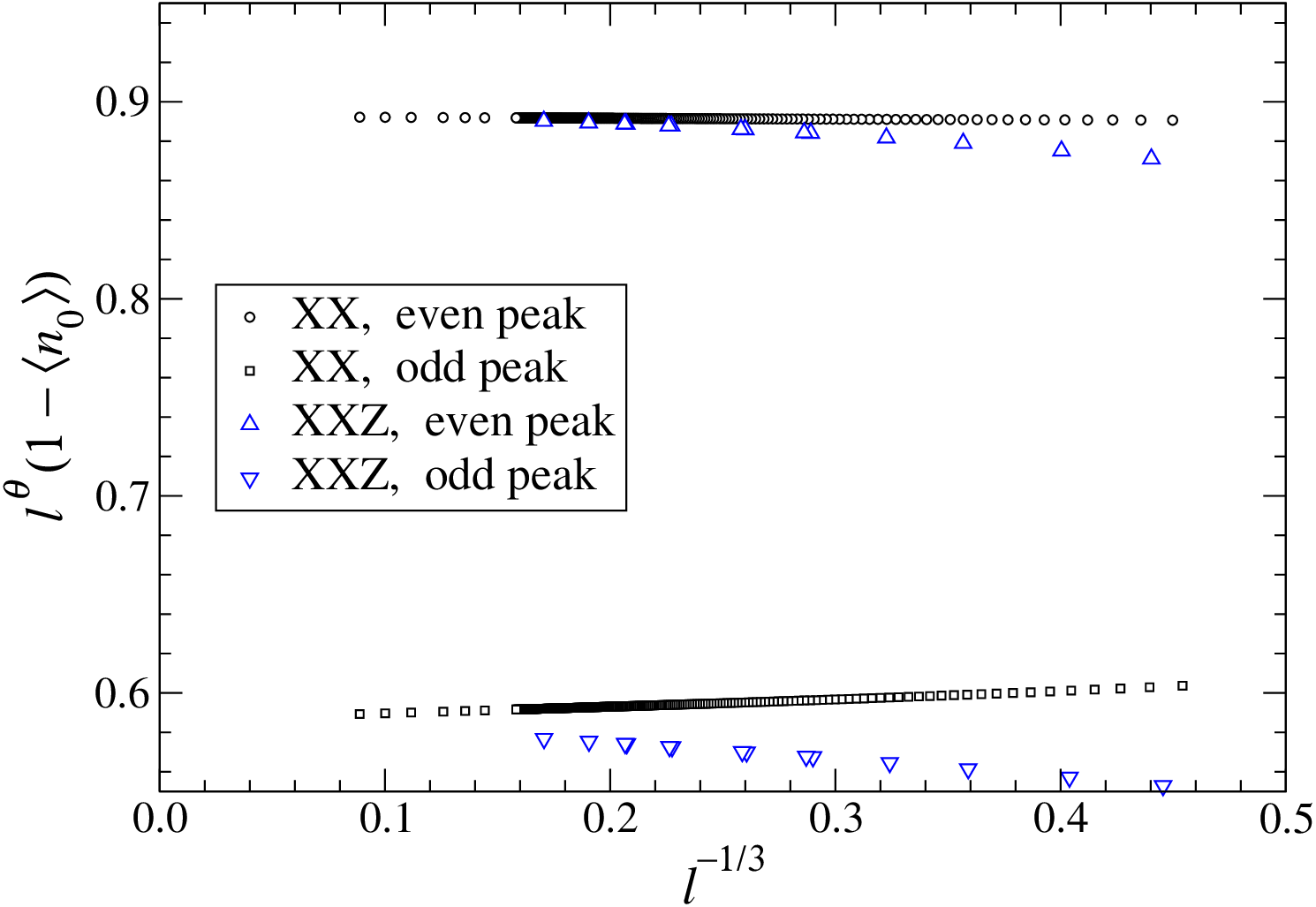}
\caption{
  Scaling of the particle density at the peaks of the gap for $p=2$ (above)
  and $p=4$ (below), for the XX model and the XXZ model (\ref{eq:H-XXZ}) at
  $j_z=-1/2$.}
\label{fig:rho0,p=2and4}
\end{center}
\end{figure}

Concerning the particle density, we recall that $\langle n_x\rangle=1$ in the
absence of the trap and the RG dimension of the particle density is $y_n=1$ at
the Mott transition. Thus, TSS predicts the scaling behavior $\langle
n_0\rangle - 1\sim l^{-\theta}$ for the particle density at the middle of the
trap and $\mu=-1$.  Like the gap, we find that the trap-size dependence of
$\langle n_0\rangle$ is described by a modulated TSS, i.e.,
\begin{equation}
\langle n_0\rangle - 1 \approx  l^{-\theta}  {\cal D}_{0}(\phi), 
\label{n0scalm-1}
\end{equation}
where $\phi$ is the same phase-like variable defined for the scaling of the
gap, cf.\ Eq.~(\ref{phildef}).  Figs.~\ref{gapplp2}, \ref{gapplp4} and
\ref{gapplp6} show $l^{\theta}(1-\langle n_0\rangle)$ vs.\ $\phi$ for
$p=2,\,4,\,6$ respectively.  The approach to the asymptotic behavior is
apparently characterized by power-law $O(l^{-\kappa})$ corrections with
$\kappa\approx \theta/2$, analogously to the behavior of the gap.  Notice
that, unlike the superfluid case, the leading term has a nonzero average over
the period.  Results for the particle density at the peaks of the gap are
shown in Figs.~\ref{fig:rho0,p=2and4} for $p=2$ and $p=4$.

\begin{figure}[tbp]
\begin{center}
\leavevmode
\includegraphics*[scale=\graphicscale]{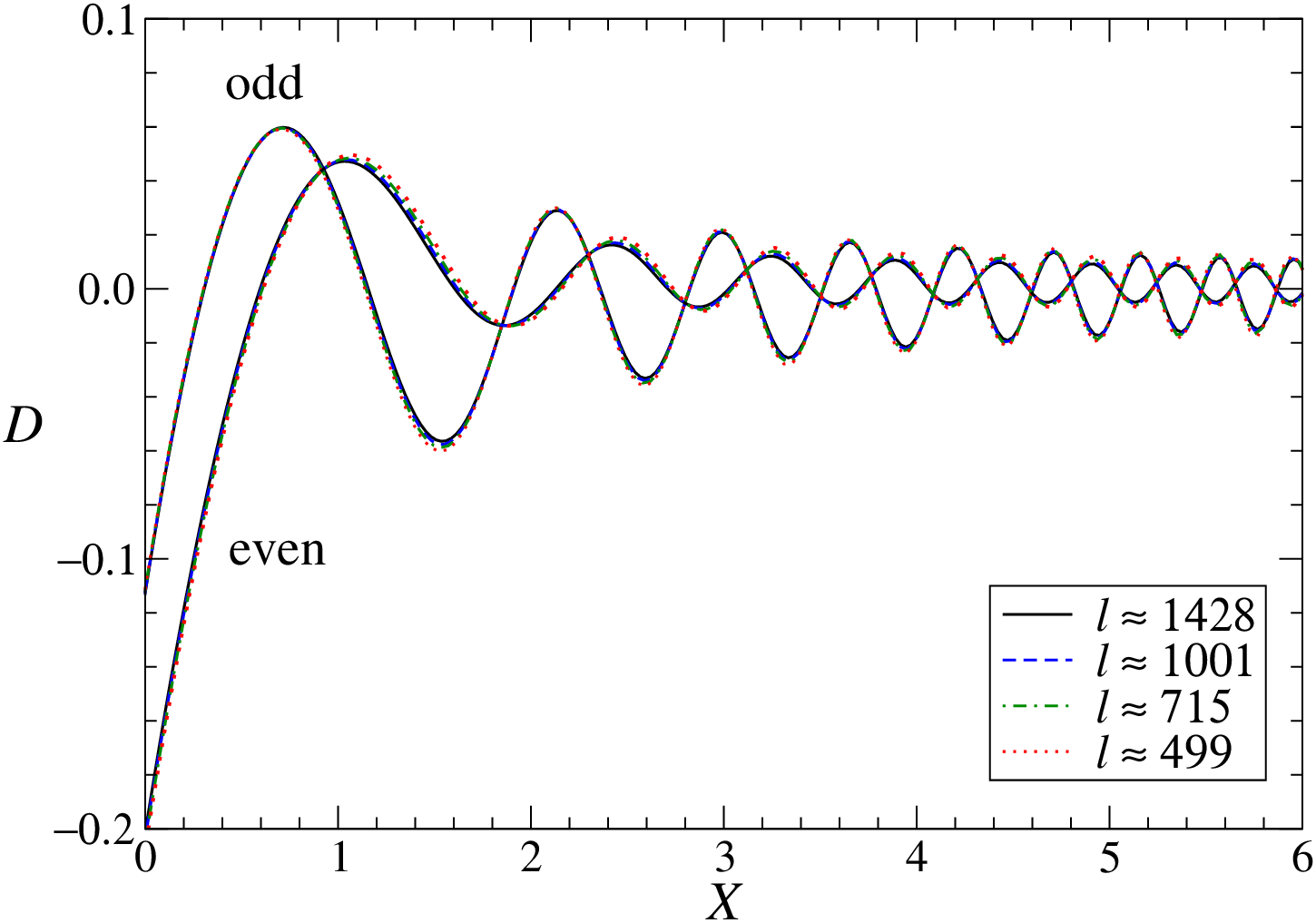}
\includegraphics*[scale=\graphicscale]{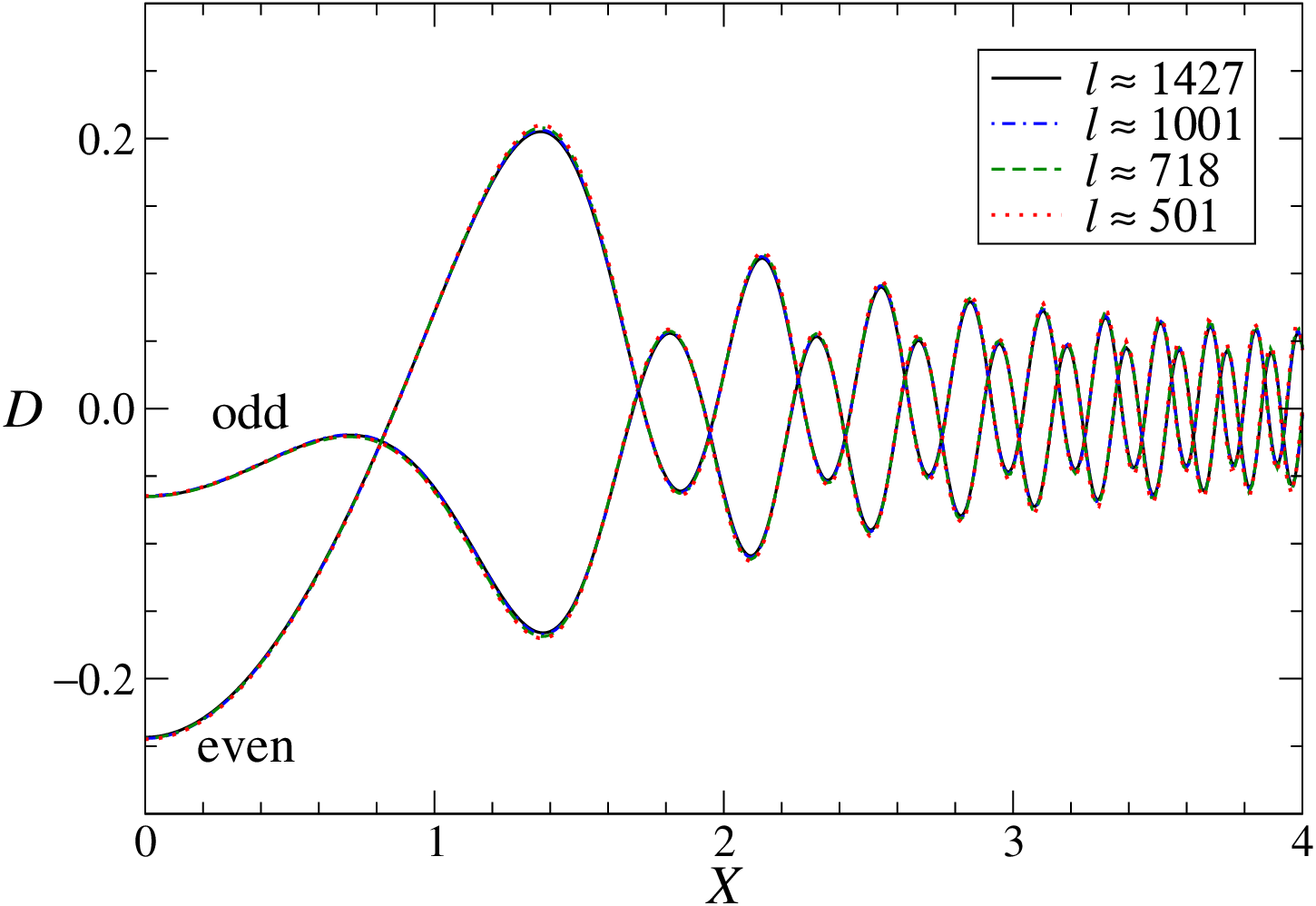}
\caption{
The function ${\cal D}$
 vs.\ $X=l^{-\theta} x$ at odd and even peaks of $\Delta$ at $\mu=-1$,
for $p=2$ (above) and $p=4$ (below).
}
\label{fig:XXrhoX,p=2and4,mu=-1}
\end{center}
\end{figure}

\begin{figure}[tbp]
\begin{center}
\leavevmode
\includegraphics*[scale=\graphicscale]{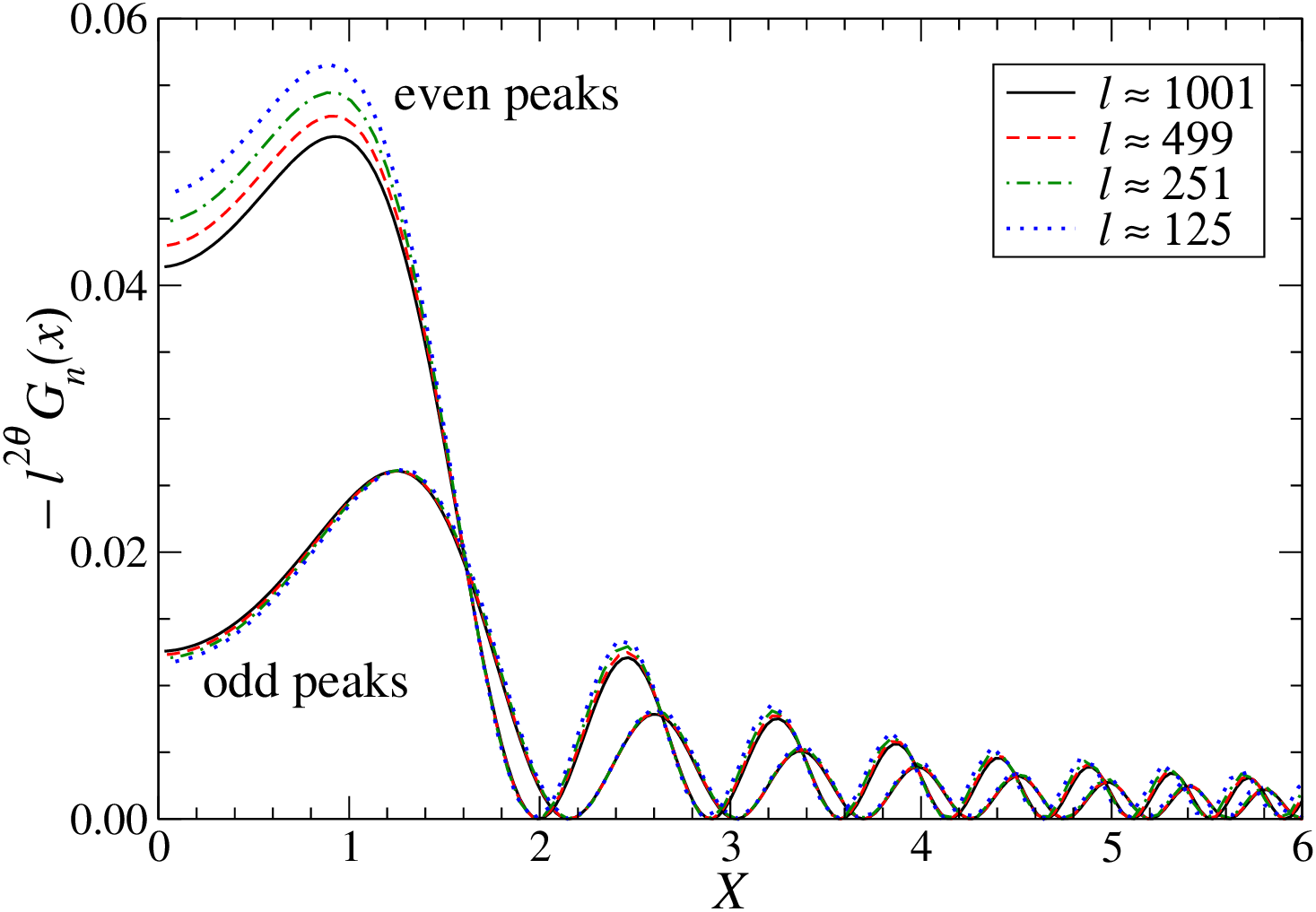}
\caption{
The rescaled connected correlation $l^{2\theta} G_n(x)$ vs.\ $X\equiv
  x l^{-\theta}$ for $p=2$, $\mu=-1$, and $l$s corresponding to even and odd
  peaks of $\Delta$.  The data at different trap sizes clearly approach a TSS
  function when increasing $l$.}
\label{fig:rhocc,mu=-1,p=2}
\end{center}
\end{figure}

The spatial dependence of the particle density at large trap size turns out to
be described by the following scaling behavior
\begin{equation}
\langle n_x\rangle \approx  \rho_{\rm lda}(x/l)
+ l^{-\theta}  {\cal D}(X,\phi), \qquad X = l^{-\theta} x.
\label{nxscalm-1}
\end{equation}
As already found in the superfluid region, the term depending on $x/l$ is
given by the LDA of the particle density, cf.\ Eq.\ 
(\ref{nxlda}). It plays the role of an analytical contribution which must be
subtracted to observe scaling in the expectation value of the density
operators at phase transitions~\cite{CV-09,CV-10}.
Some results for the scaling function ${\cal D}(X,\phi)$ at the peaks of the
gap are shown in Fig.\ \ref{fig:XXrhoX,p=2and4,mu=-1}.

As already shown by the data at $\mu=-1$ reported in
Fig.~\ref{XXpartdensnum1}, the particle density is quantitatively dominated by
its LDA at large trap size, which scales as $x/l$.

Assuming a modulated TSS, we expect that the density correlation behaves as
\begin{equation}
G_n(x) \equiv
\langle n_0 n_x\rangle_c =  l^{-2\theta}  {\cal G}_n(X,\phi).
\label{gnxm1}
\end{equation}
This is confirmed by the numerical results. For example, Fig.\
\ref{fig:rhocc,mu=-1,p=2} shows $l^{2\theta}G_n(x)$ vs.\ $X$ for $p=2$
and several values of $l$ corresponding to peaks of $\Delta$ (i.e.,
$\phi=\phi_l$ or $\phi_s$ asymptotically).  The approach to scaling is
clearly observed.

\subsection{The one-particle density matrix}
\label{mtssopdm}

\begin{figure}[tbp]
\begin{center}
\leavevmode
\includegraphics*[scale=\graphicscale]{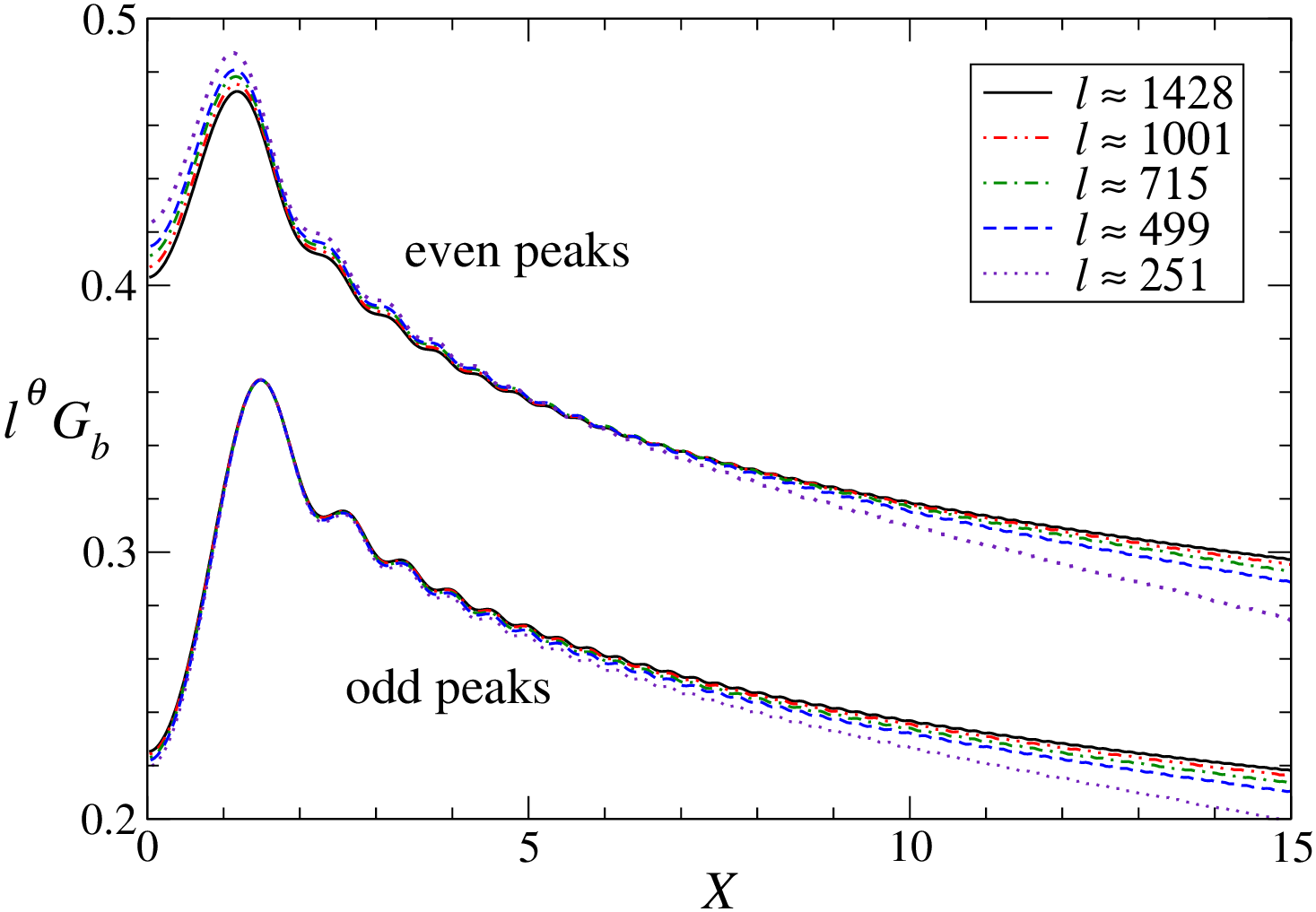}
\includegraphics*[scale=\graphicscale]{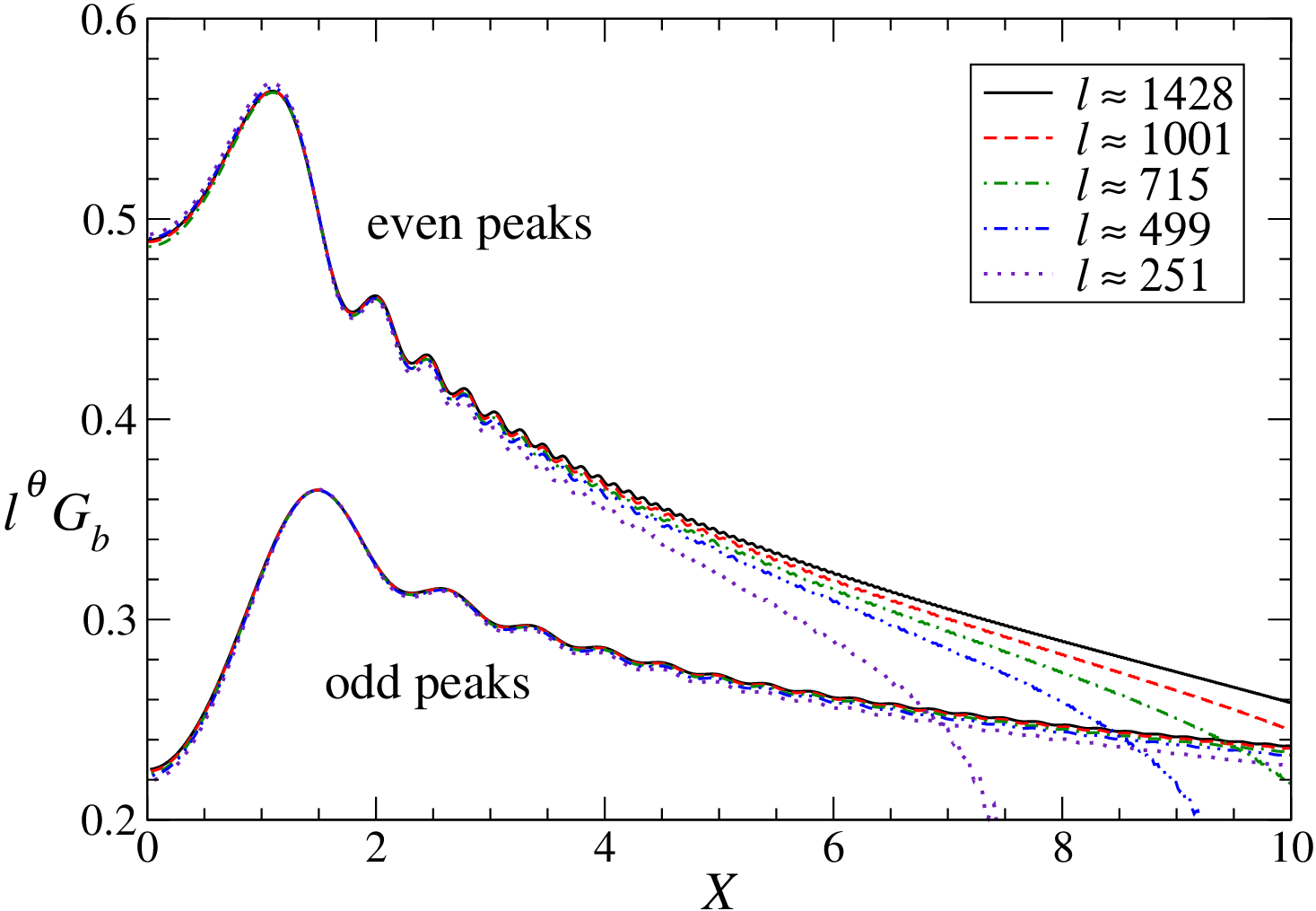}
\caption{
  The rescaled one-particle density matrix
  $l^\theta G_b(x)$ vs.\ $X\equiv x l^{-\theta}$ 
  for $p=2$ (above) and $p=4$ (below), at $\mu=-1$,
  and for $l$s corresponding to odd and even peaks of $\Delta$.}
\label{fig:cxx,mu=-1,p=2and4}
\end{center}
\end{figure}

The modulated TSS also applies to the critical behavior of the one-particle
density matrix.  We recall that the RG dimension of the bosonic field $b_i$ is
$y_b=1/2$ at the Mott transitions. Consistently, the numerical results show
the behavior
\begin{equation}
G_b(x,0)  \equiv \langle b_{0}^\dagger b_{x} \rangle
= \langle \sigma^-_{0} \sigma^+_{x} \rangle \approx
l^{-\theta} {\cal M}(X,\phi).
\label{gb0m-1a}
\end{equation}
In Figs.\ \ref{fig:cxx,mu=-1,p=2and4} we plot $l^\theta G_b(x,0)$ vs.\ $X$
for $p=2,\,4$, $\mu=-1$, and several values of $l$ corresponding to peaks of
$\Delta$ (i.e., $\phi=\phi_l$ or $\phi=\phi_s$ asymptotically).  

More generally, we may extend the modulated TSS including the dependence on
$\hat{\mu} \equiv \mu+1$ around $\hat{\mu}=0$. Setting $\mu_s = l^{2\theta}
\hat{\mu}$, we expect
\begin{equation}
G_b(x,0) \approx l^{-\theta} {\cal M}(\mu_s,X,\phi).
\label{gb0m-1}
\end{equation}

\begin{figure}[tbp]
\begin{center}
\leavevmode
\includegraphics*[scale=\graphicscale]{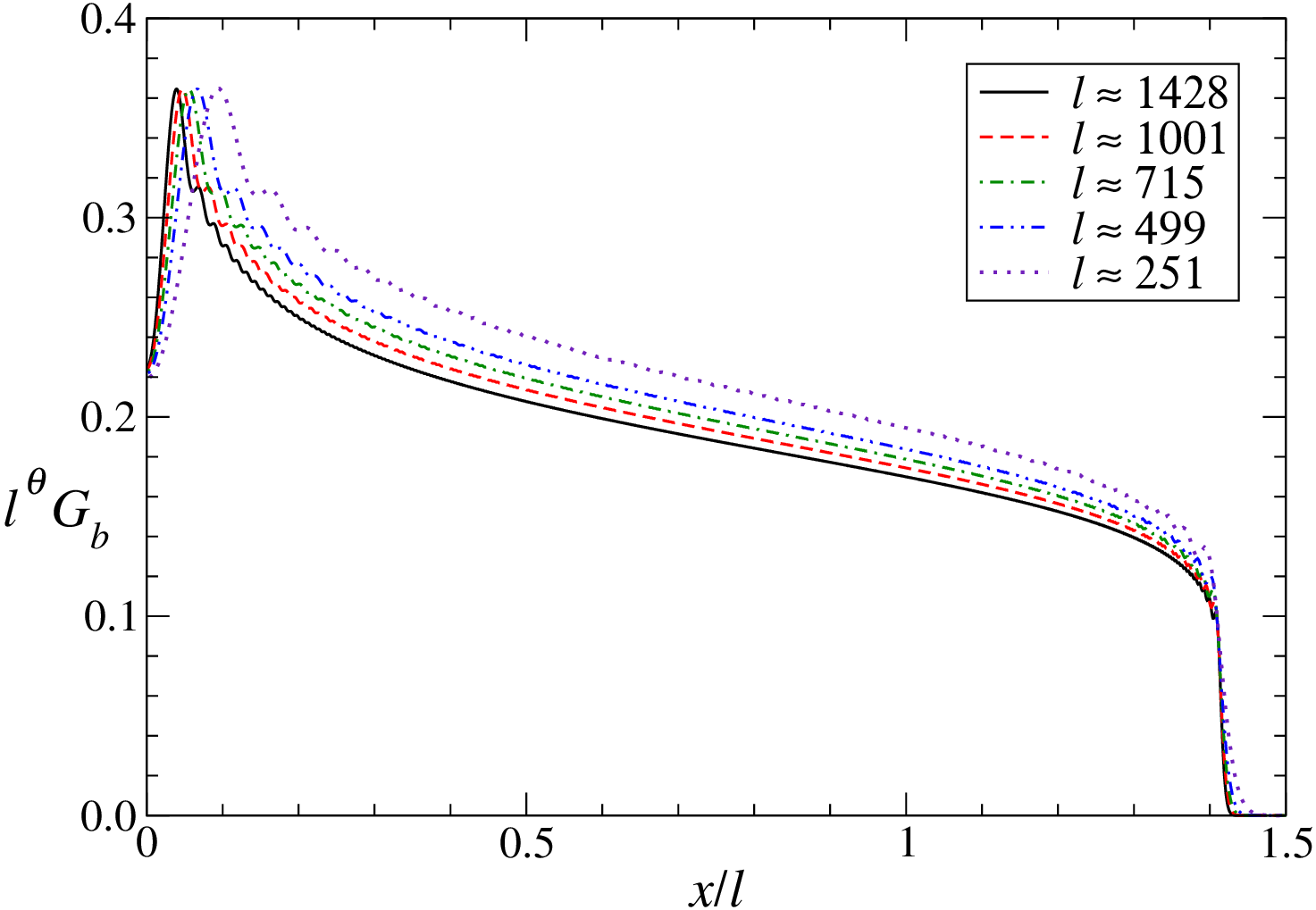}
\caption{
$l^{1/2}G_b(x,0)$ vs.\ $x/l$ for $p=2$ and $\mu=-1$.
  Data taken at the odd peaks of the gap.}
\label{fig:cxxvsl,mu=-1,p=2}
\end{center}
\end{figure}

Note that the modulated TSS is observed at fixed $X=x/l^\theta$, thus, since
$\theta<1$, the region of $x\approx l$ gets hidden in the limit
$X\to\infty$.  Fig.\ \ref{fig:cxxvsl,mu=-1,p=2} shows numerical results for
$p=2$ versus $x/l$, where we clearly observe that the behavior for $x\gtrsim
l$ is approximately scaling as $x/l$. This is essentially related to the
simple scaling of the effective chemical potential, cf.\ Eq.~(\ref{mueff}).
However, a more careful analysis shows another interesting scaling behavior
around the region 
\begin{equation}
x\approx x_c\equiv2^{1/p}l,
\label{xceq}
\end{equation}
where the spatial dependence of the effective chemical gives rise to a
transition from the superfluid phase to the $n=0$ Mott phase.  In this region
the critical behavior should be governed by the linearized potential at $x_c$,
cf.\ Eq.~(\ref{linV}), and therefore by the corresponding RG scaling.
Analogously to what observed at $\mu=0$, see Subs.~\ref{opdmsufl}, the
transition region around $x_c$ is expected to enlarge as $\Delta x \sim
l^{1/3}$ independently of $p$.  This RG prediction is fully supported by the
same numerical analysis outlined at the end of Subs.~\ref{opdmsufl}, for
several values of $p$.

\subsection{The von Neumann entanglement entropy}
\label{mum1qe}

\begin{figure}[tbp]
\begin{center}
\leavevmode
\includegraphics*[scale=\graphicscale]{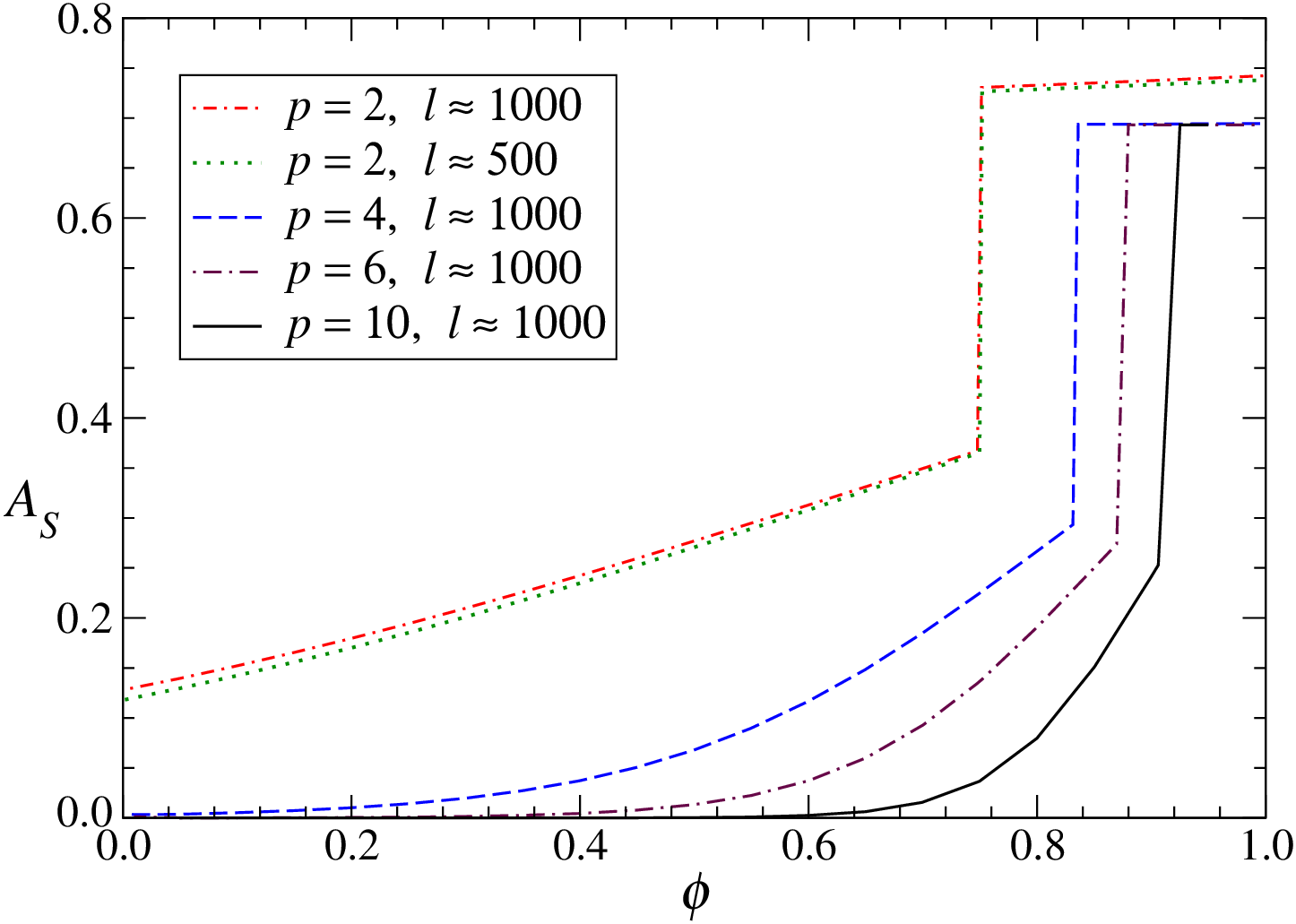}
\caption{
  The half-lattice von Neumann entanglement entropy $S_{1/2}$ at $\mu=-1$, for
  $p=2,\,4,\,6,\,10$. For $p=4$, $p=6$, and $p=10$, the plotted data 
 are already very close to the large-$l$ limit.}
\label{qesm1fig}
\end{center}
\end{figure}

Finally, we consider the von Neumann entanglement entropy, and, specifically,
the half-lattice entanglement entropy defined in Eq.~(\ref{sdef1o2}) for
chains with even $L$.  We recall that the half-lattice entanglement entropy
$S(L/2;L)$ of the homogeneous BH model vanishes at $\mu=-1$, as it does at the
low-density Mott transition point $\mu=1$, where it also vanishes in the
presence of the trap, see Sec.~\ref{hlentmu1}. We instead find that at the
$n=1$ Mott transition point $\mu=-1$ the presence of the trap gives rise to
nonzero values depending of the phase-like variable $\phi$, i.e.,
\begin{equation}
S_{1/2} \approx A_S(\phi) + O(l^{-\kappa}),
\label{qesm1}
\end{equation}
with $\kappa$ roughly compatible with $\theta/2$.
Results for $p=2,\,4,\,6,\,10$ are shown in Fig.~\ref{qesm1fig}.
Notice that the discontinuity in $A_s$ occurs at $\phi=\phi_0=(p+1)/(p+2)$,
cf.\ Eq.\ (\ref{phi0guess}).  The data suggest $A_S(\phi)\to0$ as $p\to\infty$
(for $\phi<1$), consistently with the ``naive'' $p\to\infty$ limit of the
homogeneous model with open boundary conditions.

\subsection{Universality of the modulated TSS}
\label{tssunivmum1}

\begin{figure}[tbp]
\begin{center}
\leavevmode
\includegraphics*[scale=\graphicscale]{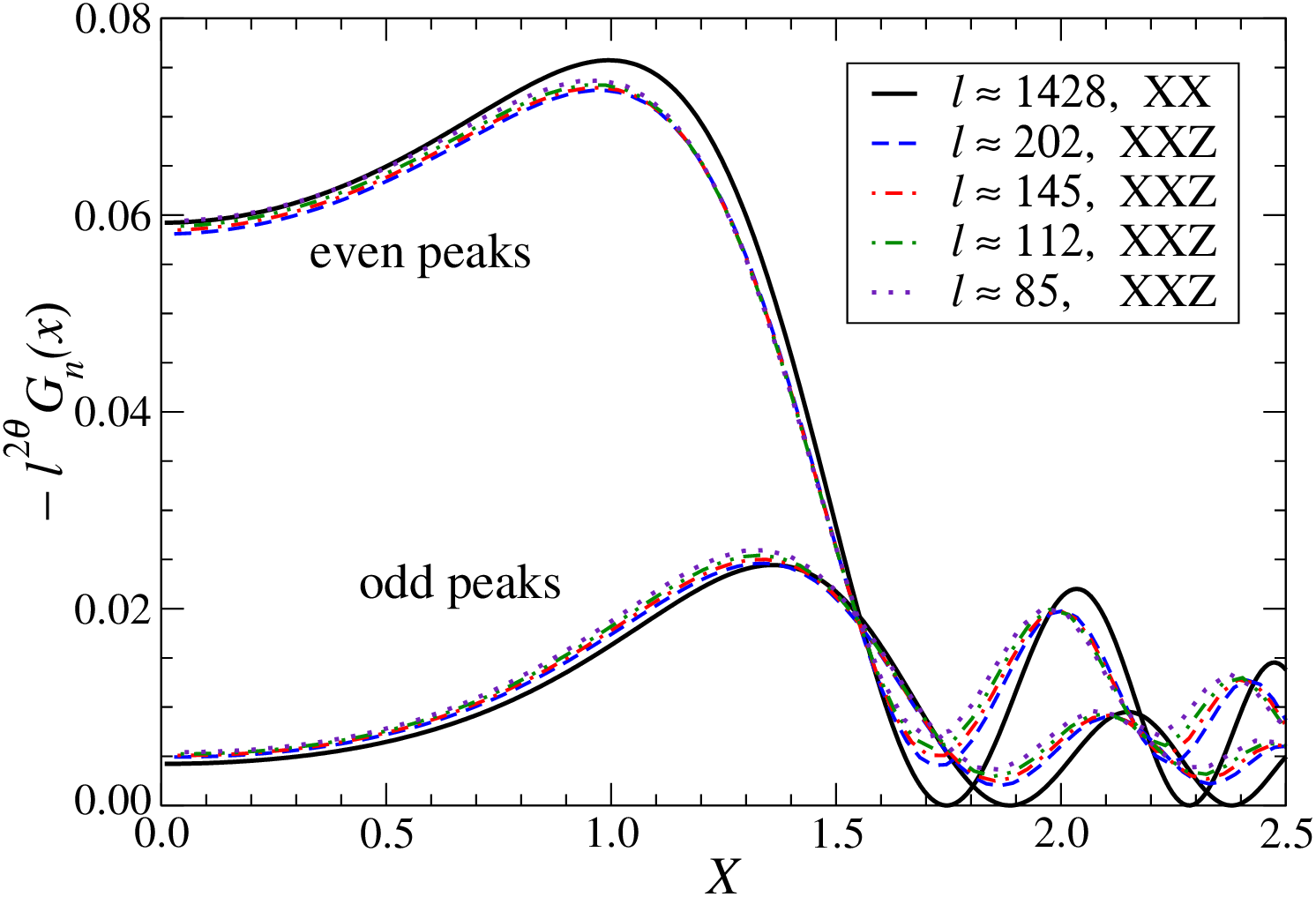}
\includegraphics*[scale=\graphicscale]{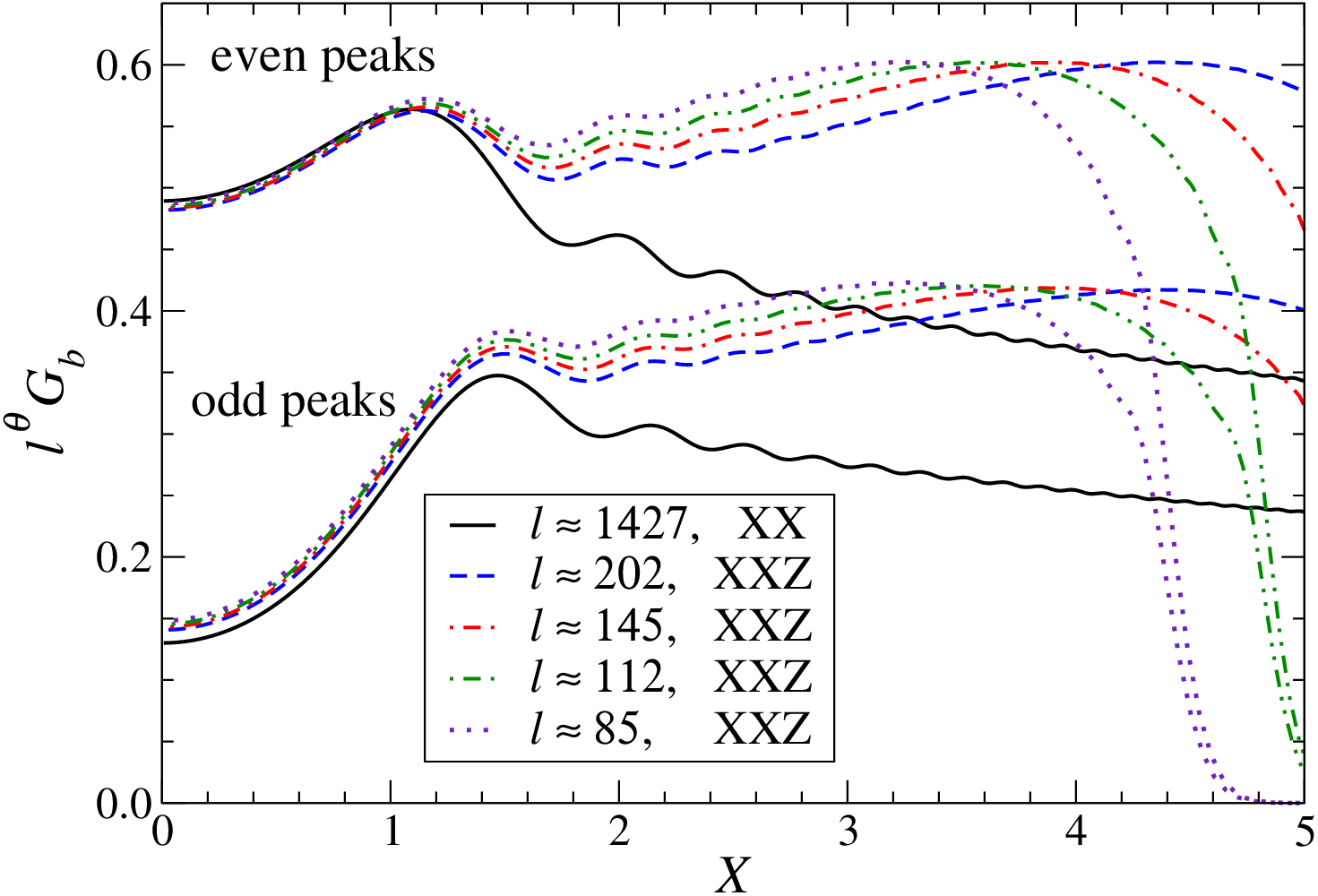}
\caption{The rescaled connected correlation $l^{2\theta}
  G_n(x)$ (above) and the rescaled one-particle density matrix $l^\theta
  G_b(x)$ (below) vs.\ $X\equiv x l^{-\theta}$ for the XXZ model at $j_z=-1/2$
  and $\mu=\mu_c$, $p=4$ and $l$s corresponding to peaks of $\Delta$. For
  comparison, we also show results for the XX model for $\mu=-1$, $p=4$ and a
  large trap size, see also Sec.\ \ref{mtsspd} and C.}
\label{fig:XXZG}
\end{center}
\end{figure}

We have shown that the BH model (\ref{bhm}) is characterized by a modulated
TSS at the $n=1$ Mott transition. A natural question concerns its
universality, i.e., whether, beside the critical exponents, also the
modulation is universal.  

In order to investigate this issue, we again consider the XXZ model
(\ref{eq:H-XXZ}), and present results obtained by DMRG calculations for
$j_z=-1/2$, $\mu=\mu_c=-1/2$ which corresponds to the $n=1$ Mott transitions,
and various power laws of the confining potential.  The values of the trap
size we could reach, $l\lesssim 300$, are significantly smaller than those for
the XX chain, which are $O(10^3)$.

The numerical results show periodic structures of the gap and the particle
density analogous to those of the XX chain, see Fig.\ \ref{gapanddensityp2}.
We find again level crossings whose interval tends to a constant in the
asymptotic large trap regime.  The asymptotic interval between even level
crossings, cf.\ Eq.~(\ref{pl}), is given by $P_l^*=1.550$ for $p=2$,
$P_l^*=1.293$ for $p=4$, and $P_l^*=1.200$ for $p=6$.  Note that these values
differ from those found for the XX model, see Eq.~(\ref{plest}).

After introducing a variable $\phi$ defined as in Eq.~(\ref{phildef}), the gap
turns out to behave analogously to the XX model, see Eq.~(\ref{deltadep}).
Results for the position $\phi_0$ of the odd zero are shown in
Fig.~\ref{fig:XXZ,phi0}: they are consistent with the asymptotic values found
for the XX model, although they show a slower approach, which is likely
$O(l^{-\theta})$ (for the XX model it was $O(l^{-2\theta})$ for this
particular quantity).  Results for the large and small peaks of the gap, and
their locations, are shown in Figs.~\ref{fig:peaks,phi,p=2} and
\ref{fig:peaks,phi,p=4}, respectively for $p=2$ and $p=4$.  They appear to
converge toward the same asymptotic values of the XX model, with similar
$O(l^{-\theta/2})$ corrections.  The ratio $\Delta_l/\Delta_s$, which is
independent of normalizations like $\phi_0$, appears to scale according to
Eq.\ (\ref{rdeapp}), and the large-$l$ extrapolated values are consistent with
those found for the XX chain, see Fig.\ \ref{fig:Dratio,XXZ,mu=-1}.  The
agreement for $p=4$ and $p=6$ is very good; for $p=2$, the data are clearly
approaching the XX value, but the values of $l$ considered are not
sufficiently large to provide a precise extrapolation.

The DMRG results for the particle density at the origin, are consistent with
the scaling behavior (\ref{n0scalm-1}) with a universal scaling function
${\cal D}_{0}(\phi)$, as also shown by the results at the peaks of the gap
reported in Fig.~\ref{fig:rho0,p=2and4}.

The behaviors of the density-density correlations function and of the
one-particle density matrix also support the universality of the modulated
TSS.  Examples are shown in Fig.\ \ref{fig:XXZG} for $p=4$.  With increasing
$l$ the data appear to approach, although slowly, the same scaling curves
found for the XX model. Note that the asymptotic curves of different models
are expected to be universal apart from normalizations.  This implies that
they should match after an appropriate rescaling of the axes.  Assuming
universality, the results of Figs.~\ref{fig:XXZG} indicate that such
rescalings are quite small for $p=4$.

In conclusion, the DMRG results for the XXZ model at $j_z=-1/2$ show a
modulated TSS analogous to that observed for the XX model, thus supporting
universality.

\section{Conclusions}
\label{conclusions}

In this paper we have studied the quantum critical behaviors of confined
particle systems described by the 1D BH model (\ref{bhm}) in the presence of a
power-law confining potential $V(r) = (r/l)^p$, at the transitions between the
Mott and superfluid phases, and within the gapless superfluid phase.  
We have considered the hard-core limit, $U\to\infty$, of the 1D BH
model, which allows us to study the effects of the confining potential by
exact and very accurate numerical results.  We have analyzed the various
zero-temperature quantum critical behaviors within the TSS
framework~\cite{CV-10,CV-09}.  In the following we list our main results.

(i) At the low-density Mott transition, the TSS limit can be analytically
derived within the quadratic spinless fermion representation, leading to a
Schr\"odinger-like equation for the lowest states.  The RG scaling arguments
leading to the TSS Ansatz (\ref{freee}) and (\ref{gapscalgen}), with critical
exponents $z=2$, $\nu=1/2$, and $\theta= p/(p+2)$, are fully confirmed.  We
have computed the TSS of several observables extending the results presented
in Ref.~\cite{CV-10}, and checked universality by DMRG calculations within the
XXZ model (\ref{eq:H-XXZ}), 
which corresponds to adding nearest-neighbor density-density
interactions in the BH model (\ref{bhm}).  The TSS functions show peculiar
behaviors, whose main features, like the discontinuities in the scaling
particle density (see, e.g., Fig.~\ref{fig:Drho0,p=2} for $p=2$), are clearly
related to the quantum nature of the transition.

(ii) The trap-size dependence turns out to be more subtle in the region of
parameters where the filling $f$ of the corresponding homogeneous system is
nonzero. This is essentially due to the presence of an infinite number of
level crossings of the lowest states when increasing the trap size.
Nevertheless, the particle density of the 1D hard-core BH model approaches its
LDA in the large-$l$ limit, i.e., the value of the particle density of the
homogeneous system at the effective chemical potential $\mu_{\rm eff}(x)
\equiv \mu + (x/l)^p$.  Corrections are suppressed by powers of the trap size,
and show a nontrivial scaling behavior.

(iii) The level-crossing phenomenon gives rise to a new interesting scenario
at the $n=1$ Mott transition, requiring a revision of the simplest TSS Ansatz
(\ref{freee}) and (\ref{gapscalgen}) into a {\em modulated} TSS: the TSS is
still controlled by the trap-size exponent $\theta= p/(p+2)$, as in the case
of the low-density Mott transition, but it gets modulated by periodic
functions of the trap size.  Indeed, the gap turns out to behave as $\Delta
\approx A_\Delta(\phi) l^{-2\theta} [1 + O(l^{-\theta/2})]$, where the
amplitude $A_\Delta(\phi)$ is a periodic function of the trap size $l$,
through the phase-like variable $\phi$ measuring the distance from the closest
even level crossing, see Figs.~\ref{gapplp2}, \ref{gapplp4} and \ref{gapplp6}
for $p=2,\,4,\,6$ respectively. Modulated TSS is also observed for other
observables, like the particle density and its correlators, and the
one-particle density matrix.  For example, the particle density $\langle
n_x\rangle$ at a distance $x$ from the middle of the trap shows the asymptotic
behavior $\langle n_x\rangle \approx \rho_{\rm lda}(x/l) + l^{-\theta} {\cal
  D}(X,\phi)$ where $X = l^{-\theta} x$, and $\rho_{\rm lda}(x/l)$ is the LDA
of the particle density, cf.\ Eq.\ (\ref{nxlda}).  DMRG computations for the
XXZ model (\ref{eq:H-XXZ}) at $j_z=-1/2$ and at the $n=1$ Mott transition show
an analogous modulated TSS, supporting its universality.  The modulated TSS
shows another peculiar aspect of the quantum nature of the Mott transitions.

(iv) We have also studied the trap-size dependence in the gapless superfluid
phase, whose corresponding continuum theory is a conformal field theory with
$z=1$.  In this region the asymptotic trap-size dependence turns out to be
characterized by two length scales with different power-law divergence in the
large trap-size limit.  One of them scales as $\xi\sim l$ and describes the
behavior of observables related to smooth modes, such as the half-lattice
entanglement; the other one scales as $\xi\sim l^\zeta$ with $\zeta=p/(p+1)$
and it is found in observables involving the modes at the Fermi scale $k_F=\pi
f$, where $f$ is the filling of the homogeneous system.  Moreover, the
asymptotic power law behaviors appear modulated by periodic functions of the
trap size. For example, the gap behaves as $\Delta \sim t(\phi)l^{-1}$ where
$t(\phi)$ is the triangle function (\ref{pertresde2}) and $\phi$ is a
phase-like variable measuring the distance from the periodic level crossings,
cf.\ Eq.~(\ref{phildefmu0pert}).  Some notable relations are found for the
behavior of observables related to smooth modes, such us the gap, the density
at the origin, and the half-lattice von Neumann entanglement entropy. Their
asymptotic behavior in the presence of a confining potential can be derived
from that of the homogeneous BH model with open boundary conditions which
represents the $p\to\infty$ limit, by replacing the lattice size $L$ with the
entanglement length scale defined from the von Neumann entanglement entropy.

The main features of the trap-size dependence reported in this paper should
not be restricted to the hard-core limit, i.e., the limit $U\to\infty$ in the
BH Hamiltonian (\ref{bhm}).  Specifically, the phenomenon of the level
crossings at finite trap size should persists at finite values of $U$, because
the total particle number is conserved by the confining potential even at
finite $U$.  A study of the trap-size dependence at the Mott transitions of
the 1D BH model at finite $U$ would be important to further assess the
extension of the universality of the modulated TSS observed in the hard-core
limit.  The presence of level crossings should also characterize the trap-size
dependence in higher dimensions. Therefore, modulated TSS may be also found at
higher dimensions.

\medskip Helpful discussions with P. Calabrese and M. Mintchev are gratefully
acknowledged.

\appendix

\section{Some details on numerical methods}
\label{Sec:numerics}

The numerical diagonalization of the Hamiltonian (\ref{fermmod}) is
straightforward.  Using {\tt lapack}, $L=5000$ chains can be diagonalized in a
few minutes on a desktop PC.  We consider chains of size $L$ with open
boundary conditions. The trap of size $l$ is centered in the middle of the
chain. The size $L$ of the chain is taken sufficiently large to obtain results
consistent with the infinite-size limit to machine precision.  

As it is clear from the discussion of the previous sections, and specifically
in Sec.~\ref{ldasec}, the density $\langle n_x\rangle$ decreases very rapidly
for $x>x_{\rm lda}\equiv l (-1-\mu)^{1/p}$: it is sufficient to choose $L$
slightly larger than $2x_{\rm lda}$ to have completely negligible finite-size
effects. This allows us to obtain results for quite large $O(10^3)$ trap
sizes.  A few examples of values of $L$ giving boundary effects
$\lesssim10^{-15}$ for $p=2$ are: $\mu=0.999$, $l=10000$: $L=1200$; $\mu=0.9$,
$l=1000$: $L=800$; $\mu=0$, $l=500$: $L=1100$; $\mu=-1$, $l=500$: $L=1600$.
For larger values of $p$ boundary effects are smaller.  Therefore, finite-$L$
effects at fixed trap size are under complete control.~\cite{QSS-10}

In the presence of the trap, most calculations are performed for odd $L$s to
have the trap centered in the middle site of the chain. We choose even $L$s
only to compute the trap-size dependence of the half-lattice von Neumann
entanglement entropy.  Notice that even and odd $L$s may yield different
results in the infinite-size limit, since the trap is centered between two
chain sites and on a chain site respectively.  For $\mu>-1$, this dependence
on the parity on $L$ disappears very rapidly with increasing $l$ and it is
totally negligible for $l\ge20$ at $\mu=0$, for $l\ge30$ at $\mu=-0.75$ and
for $l\ge120$ at $\mu=-0.9$.  For $\mu=-1$, it must be taken into account,
essentially because the role of the even and odd level crossings get
interchanged, cf.\ Eqs.\ (\ref{phildef}) and (\ref{philevdef}) 
in Sect.\ \ref{mtssgap}.

Numerical results for the XXZ model are obtained by finite-volume DMRG; the
Hamiltonian is not translation-invariant, therefore the initialization of the
procedure is slightly nonstandard.  The number of states $M$ kept in the
truncation procedure is chosen to have negligible truncation errors; for the
largest chains we considered ($L\cong1000$), we run with $M$ up to
$140$, with discarded weights $<3{\times}10^{-9}$.
Running at $L=999$, $p=2$, and $\mu=-1$ for
a cycle of $\phi$, determining $l_0^{(554)}$, $l_0^{(555)}$, $l_0^{(556)}$,
$l_{\rm peak}^{(555)}$, and $l_{\rm peak}^{(556)}$, cf.\ Subs.\ 
\ref{sect:superfluid,smallmu}, required about 100 runs for different $l,N$
combinations, each lasting about 8 hours.

\section{Finite-size behavior of
  the homogeneous 1D hard-core BH model with open boundary conditions}
\label{muintpinf}

In this section we report some exact results for the finite-size behavior of
the homogeneous 1D hard-core BH model, or equivalently for the homogeneous XX
chain, with open boundary conditions. This formally corresponds to the
$p\to\infty$ limit of the confining potential, which becomes equivalent to a
homogeneous chain of size $L=2l$ with open boundary conditions.  More
precisely, the $p\to\infty$ limit of the BH model with the trap corresponds to
a chain with an odd $L=2\lfloor l\rfloor +1$ ($\lfloor x \rfloor$ is the
largest integer not greater than $x$) when the center of the trap coincides
with the middle site of the chain, and to an even $L=2\lfloor l\rfloor$ when
the center is in the middle between two sites.

In the infinite-size limit $L\to\infty$, the filling $f$ is given
by~\cite{Sachdev-book} $f \equiv \langle n_i\rangle = (1/\pi)\arccos\mu$, thus
in the range $1>\mu>-1$ we have $0<f<1$.  Let us now consider a homogeneous
system of finite size $L$ with open boundary conditions.  The excitation
number $N$ for $|\mu|<1$ is exactly given by
\begin{equation}
N = \lfloor(L+1)f\rfloor,
\label{eq:N-pinf}
\end{equation}
without any finite-$L$ correction.  For integer $(L+1)f$, the ground state is
degenerate ($\Delta=0$); the lowest-energy simultaneous eigenvectors of the
Hamiltonian and the particle number give $N=(L+1)f$ and $N=(L+1)f-1$.

For $f=1/s$ with integer $s$, for every value of $N$ we have a vacuum
degeneracy when $L+1=Ns$, i.e., $\Delta=0$ for $L+1\equiv0\ (\mod s)$.  For
$f=r/s$ with integer $r$ and $s$, again $\Delta=0$ for $L+1\equiv0\ (\mod s)$,
but we can satisfy $L+1=Ns/r$ only for $N\equiv0\ (\mod r)$.  For irrational
$f$, $\Delta$ never vanishes for integer $L$.

Note that in the limit $\mu\to -1$, thus $f\to 1$, Eq.~(\ref{eq:N-pinf}) gives
$N=L$ without vacuum degeneration; this is the expected result for $\mu\le
-1$.  Analogously, Eq.~(\ref{eq:N-pinf}) gives $N=0$ for $\mu\to 1$ without
vacuum degeneration, which is the expected result for $\mu\ge 1$.

Eq.\ (\ref{eq:N-pinf}) suggests us to define 
\begin{equation} 
\phi \equiv \{(L+1)f\}, 
\label{phideffss}
\end{equation}
where $\{x\}\equiv x - \lfloor x\rfloor$ is the fractional part of $x$ (i.e.,
the sawtooth function).  For integer $(L+1)f$, it is useful to label
the two degenerate vacua with $\phi=0\,,1$ according to 
$N+\phi = (L+1)f$.

For each value of $\mu$, we observe that $L\Delta$ vs.\ $\phi$ collapses on a
curve proportional to the triangle function $t(\phi)$ defined in Eq.\ 
(\ref{pertresde2}), with $O(1/L)$ corrections.  Note that in the case of
rational filling $\phi$ takes only a discrete set of values; for
$\mu=0$, e.g., $\phi$ takes the values $0$ and $1/2$, corresponding to
odd and even $L$ respectively.  In
Fig.~\ref{fig:Dscal,mu=0.75,pinf} we plot data of $(L+1)\Delta$ for several
values of $\mu$, corresponding to rational and irrational fillings (we use
$(L+1)\Delta$ rather than $L\Delta$, which gives smaller, but comparable,
finite-$L$ corrections).
\begin{figure}[tbp]
\begin{center}
\leavevmode
\includegraphics*[scale=\graphicscale]{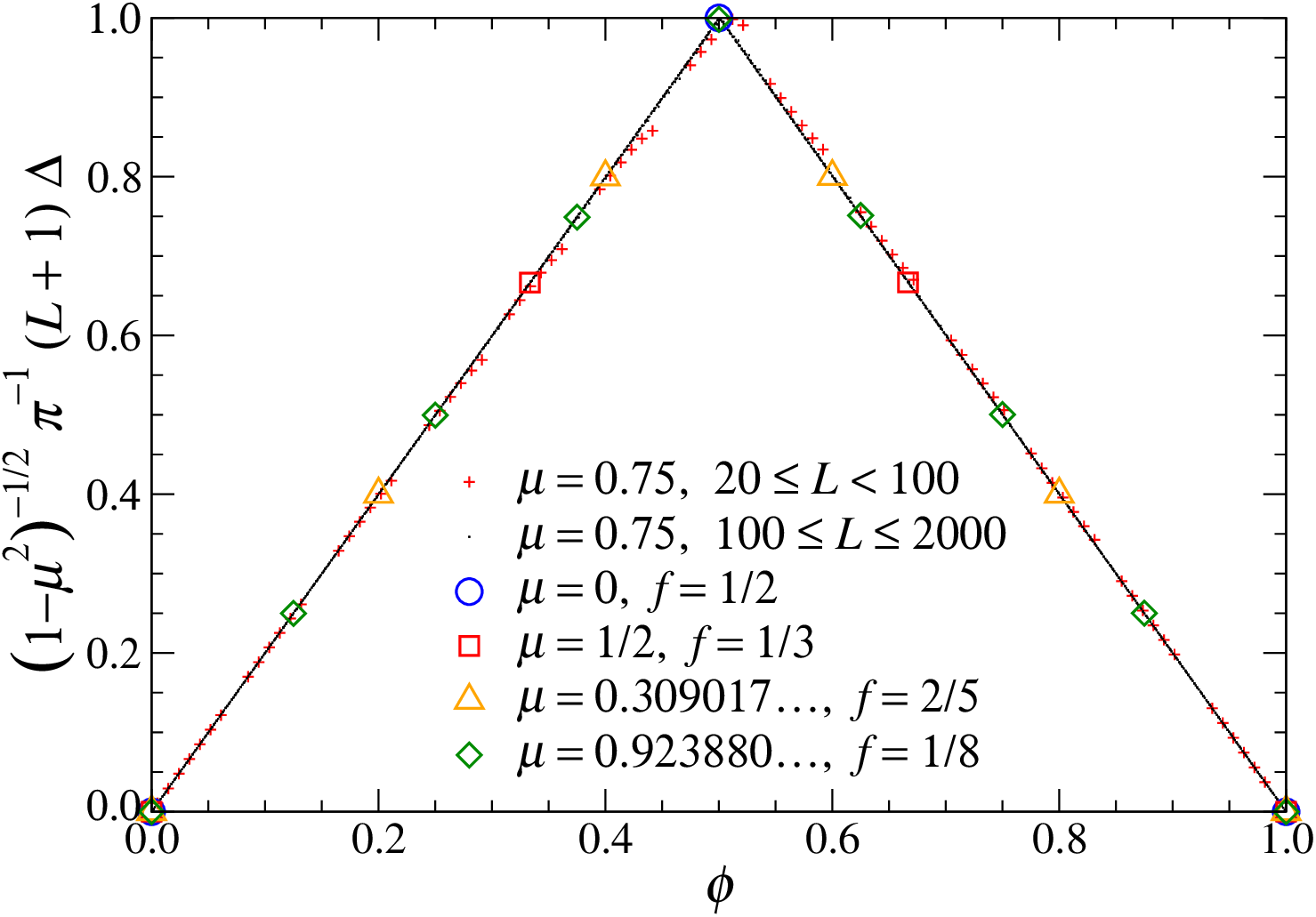}
\caption{
The rescaled gap \hbox{$(1-\mu^2)^{-1/2}\pi^{-1}(L+1)\Delta$} vs.\ $\phi$,
defined in Eq.~(\ref{phideffss}), for the homogeneous system with open
boundary conditions, for several values of $\mu$.  The apparent solid
line is due to the accumulation of data points for $\mu=3/4$,
corresponding to an irrational $f$.  Data for rational $f$ are taken
at $L\approx1000$.}
\label{fig:Dscal,mu=0.75,pinf}
\end{center}
\end{figure}
Thus we have the asymptotic behavior
\begin{equation}
L \Delta = a_\Delta \, t(\phi)+ O(1/L), \label{deltafss}
\end{equation}
This corresponds to an asymptotic periodicity of the $L$-dependence of
$L\Delta$ with period $1/f$.  The numerical results for $a_\Delta$ turn out to
be perfectly reproduced by the simple formula
\begin{equation}
a_\Delta = \pi (1-\mu^2)^{1/2}.
\label{adeltaguess}
\end{equation}

Note that $a_\Delta=0$ at $\mu=\pm 1$, indeed $\Delta=O(1/L^2)$ at $\mu=\pm 1$
without level crossings, consistently with the fact that the corresponding
continuum theory has $z=2$.  For $|\mu|>1$ we instead have $\Delta =
2(|\mu|-1) + O(L^{-2})$.

We have also studied the particle density in the middle of the chain 
$\langle n_0\rangle$, which is only defined for odd $L$.  We define
\begin{equation}
\bar\phi \equiv 2\{[(L+1)f+1]/2\},
\label{barphifss}
\end{equation}
$0\le \bar\phi \le 2$; note that either $\bar\phi=\phi$ or
$\bar\phi=\phi+1$. Then we find that, at fixed $\mu$,
\begin{equation}
(L+1)(\langle n_0\rangle-f) = 1-\bar\phi,
\label{eq:n0,allmus,pinf}
\end{equation}
without any finite-$L$ correction.  We plot in Fig.\ 
\ref{fig:rho0scal,allmus,pinf} the r.h.s.\ of Eq.\ (\ref{eq:n0,allmus,pinf})
vs.\ $\bar\phi$, for several values of $\mu$; note that, for $f=r/s$ with
integer $r$ and $s$, $\phi$ is limited to a discrete set of $s$ values:
$\{(2i-1)/s,\;i=1,\,\dots,\,s\}$ for odd $s$ and
$\{(2i-2)/s,\;i=1,\,\dots,\,s\}$ for even $s$ (in this case, we also use
$\bar\phi=2$, which is equivalent to $\bar\phi=0$, to distinguish the two
vacua: $\bar\phi=2$ for the $N=(L+1)f$ vacuum and $\bar\phi=0$ for the
$N=(L+1)f-1$ vacuum.  On the other hand, for irrational $f$s the values of
$\phi$ are distributed all over the interval $(0,2)$.
\begin{figure}[tbp]
\begin{center}
\leavevmode
\includegraphics*[scale=\graphicscale]{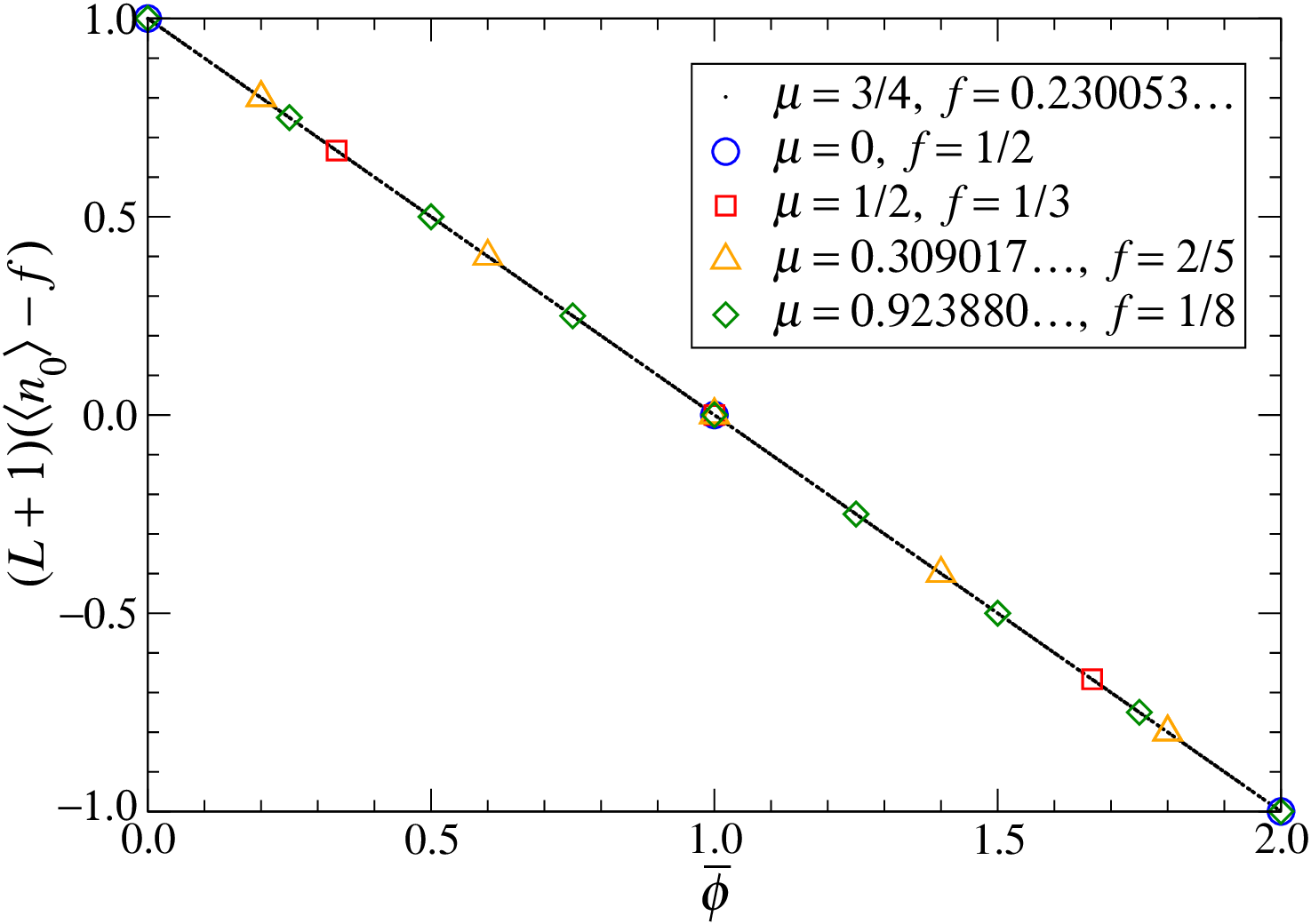}
\caption{
The l.h.s.\ of Eq.\ (\ref{eq:n0,allmus,pinf}) vs.\ $\bar\phi$,
  defined in Eq.~(\ref{barphifss}), for odd $L$s in the range $3\le L\le999$.
  }
\label{fig:rho0scal,allmus,pinf}
\end{center}
\end{figure}

For rational $f$, Eq.\ (\ref{eq:n0,allmus,pinf}) implies a very simple
rational form for $\langle n_0\rangle$.  For $\mu=0$, we have
\begin{equation}
(L+1)(\langle n_0\rangle-\half) = 
\Biggl\{\begin{array}{ll} \pm 1, & L\equiv1\ (\mod 4), \\
0, & L\equiv3\ (\mod 4), \\ \end{array}
\label{n0,mu=0,pinf}
\end{equation}
where the $\pm$ signs apply to the two degenerate lowest states with particle
number $N=(L+1)/2$ and $N=(L-1)/2$ respectively.  For $\mu=1/2$, we have
\begin{equation}
(L+1)(\langle n_0\rangle-\txf13) = 
\left\{\begin{array}{ll}
\txf23,  & L\equiv0\ (\mod 3), \\
-\txf23, & L\equiv1\ (\mod 3), \\
      0, & L\equiv2\ (\mod 3). \\ \end{array}\right.
\label{n0,mu=1/2,pinf}
\end{equation}
Eqs.\ (\ref{n0,mu=0,pinf}) and (\ref{n0,mu=1/2,pinf}) are indeed verified for
all $L$s to machine precision.  For other $\mu$s giving rational $f$, we
obtain comparable results.

For $\mu=0$, $\langle n_i\rangle$ is given by a very simple expression
for all points $i=x+\half(L+1)$, $i=1,\dots,L$:
\begin{equation}
(L+1)(\langle n_i\rangle-\half) = 
\left\{\begin{array}{ll}
 0, & \hbox{even $L$}, \\
 0, & \hbox{odd $L$, even $i$}, \\
 \pm 1, & \hbox{odd $L$, odd $i$}. \\
\end{array}\right.
\label{ni,mu=0,pinf}
\end{equation}
where again the $\pm$ signs apply to the two degenerate lowest analogously to
Eq.~(\ref{n0,mu=0,pinf}).  Eq.\ (\ref{ni,mu=0,pinf}) for $i=\half(L+1)$ gives
back Eq.\ (\ref{n0,mu=0,pinf}).

Note that all the above formulae are invariant under the particle-hole
exchange $n_i\to1-n_i$, which implies $N\to L-N$, $f\to1-f$,
and $\mu\to-\mu$.

\begin{figure}[tbp]
\begin{center}
\leavevmode
\includegraphics*[scale=\graphicscale]{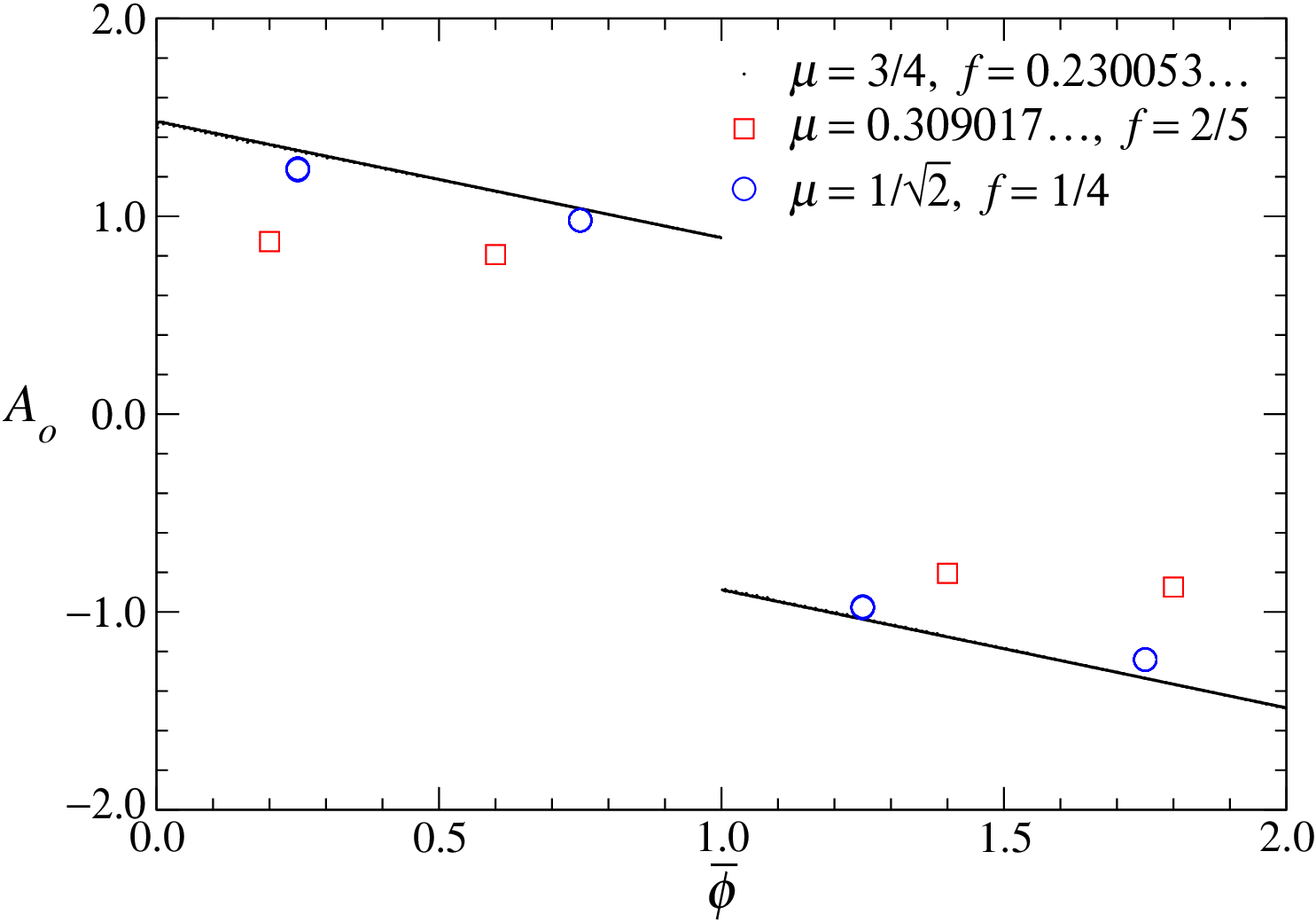}
\caption{
  The function $A_o(\mu,\bar\phi)$ as derived from data of the quantity
  reported in Eq.~(\ref{subqua}), vs.\ $\bar\phi$, cf.\ Eq.~(\ref{barphifss}),
  for several values of $\mu$.  For $\mu=3/4$, all chains with even
  $200\le L\le2000$ are plotted.  These results show that in all cases
$A_o$ can be parametrized as in Eq.\ (\ref{eq:Ae2}).} 
\label{Seoscill}
\end{center}
\end{figure}

We now consider the half-lattice von Neumann entanglement entropy 
$S(L/2;L)$ for even $L$ and open boundary conditions in the superfluid
phase, whose asymptotic large-$L$ behavior is~\cite{CC-04}
\begin{equation}
S(L/2;L) = {1\over6} \ln L + E(\mu) + O(1/L).
\label{eq:Se-E2b}
\end{equation}
The $O(1)$ term $E(\mu)$ depends on $\mu$; using the results reported in
Refs.~\cite{JK-04,ZBFS-06}, we derive
\begin{eqnarray}
E(\mu) &=& {1\over 12} \ln (1-\mu^2) + {1\over 6} \ln (4/\pi)
\label{emufunc} \\
&+& \int_0^\infty dt\,\left(
{ {\rm cosh}(t/2)\over 4\sinh^3(t/2)} 
-{e^{-t}\over 6t} - {1\over 2t\sinh^2(t/2)} \right).
 \nonumber
\end{eqnarray}
We recall that the von Neumann entanglement entropy $S(L/2;L)$ vanishes at
$\mu=\pm 1$.

An accurate numerical analysis using data up to $L=O(10^4)$ shows that, for
any $|\mu|<1$ including those corresponding to irrational filling, the
entanglement entropy of the half-lattice behaves as
\begin{equation}
S(L/2;L) = {1\over6} \ln (L+1) + E(\mu) + {1\over L} A_o(\mu,\bar\phi) 
+ O(1/L^2),
\label{eq:Se-E2a}
\end{equation}
where $\bar\phi$ is defined in Eq.~(\ref{barphifss}), and
$A_o(\mu,\bar\phi)$ has zero average over values of $\bar\phi$.
$A_o$ can be written as
\begin{equation}
A_o(\mu;\bar\phi) = \Biggl\{
\begin{array}{cl}
b + u \bar\phi &
\quad {\rm for} \quad 0 < \bar\phi < 1, \\
-b - u (2-\bar\phi) &
\quad {\rm for} \quad 1 < \bar\phi < 2, \\
\end{array}
\label{eq:Ae2}
\end{equation}
where $b$ and $u$ depend only on $\mu$.
Indeed, in the case of irrational filling $f$,
we find that, for sufficiently large $L$, the quantity
\begin{equation}
L[S(L/2;L) - {1\over6} \ln (L+1) - E(\mu)]
\label{subqua}
\end{equation}
collapses on a single curve $A_o(\mu,\bar\phi)$ at fixed $\mu$, given by
Eq.~(\ref{eq:Ae2}), with $O(1/L)$ corrections.  In the case of rational $f$,
we have a discrete set of possible values of $\bar\phi$, and the data
accumulate at points located along two lines as described by
Eq.~(\ref{eq:Ae2}).  Some results are shown in Fig.~\ref{Seoscill}.  We find
$b\cong1.48$, $u\cong-0.59$ for $\mu=3/4$; $b\cong1.37$, $u\cong-0.52$ for
$\mu=1/\sqrt{2}$; $b\cong0.91$, $u\cong-0.17$ for
$\mu=\cos(2\pi/5)\cong0.309017$.

We mention that subleading oscillations in the behavior of entanglement
entropies have been also reported in other studies, see, e.g.,
Refs.~\cite{LSCA-06,CCEN-10}.

In the case $\mu=0$, the values of $\bar\phi$ corresponding to even $L$ are
restricted to $\bar\phi=1/2,\,3/2$, which implies $A_o\sim (-1)^{L/2}$. More
precisely, our numerical results are accurately reproduced by the formula
\begin{eqnarray}
S(L/2;L) &=& {1\over6} \ln (L+1) + E(0) - (-1)^{L/2} {\pi\over 4(L+1)}
\nonumber\\
&&\ +\, O(1/L^2), \label{eq:Se-E2mu0}
\end{eqnarray}
with $E(0)=0.28776969994598...$, cf.\ Eq.~(\ref{emufunc}).  This expression for
the half-lattice entanglement entropy is consistent with an analogous formula
reported in Ref.~\cite{LSCA-06} for the XX model at $\mu=0$.
The $O(1/L^2)$ term remaining in Eq.\ (\ref{eq:Se-E2mu0}) is very
small ($\sim 10^{-5}L^{-2}$) and without oscillations within numerical
precision [$\sim 10^{-15}$ on $S(L/2;L)$].

\end{document}